\documentclass[12pt]{report}
\usepackage[dvips]{graphicx}
\usepackage{amssymb,hyperref}

\parskip=\medskipamount                         
\thicklines                                     
\def\non{\nonumber}
\def\beq{\begin{equation}}
\def\eeq{\end{equation}}
\def\beqa{\begin{eqnarray*}}
\def\eeqa{\end{eqnarray*}}
\def\beqs{\begin{eqnarray}}
\def\eeqs{\end{eqnarray}}
\def\n{\global\advance \eqnumber by 1\eqno(\the\eqnumber)}
\def\puteqno{\global\advance \eqnumber by 1 (\the\eqnumber)}

\def\qbq{\left\langle\overline{q}q\right\rangle}
\def\qbqo{\left\langle\overline{q}_0q_0\right\rangle}
\def\w3m{\left\langle\left|W_3\right|\right\rangle}
\def\openone{\leavevmode\hbox{\small1\normalsize\kern-.33em1}}

\begin{document}

\pagestyle{empty}


\begin{centering}
\mbox{}
\vfill
\huge \bf Finite Temperature QCD with Domain Wall Fermions \\
\vfill
\Large George Tamminga Fleming \\
\vfill
\large
Submitted in partial fulfillment of the \\
requirements for the degree \\
of Doctor of Philosophy \\
in the Graduate School of Arts and Sciences \\
\vfill
COLUMBIA UNIVERSITY \\
2001 \\
\end{centering}
\clearpage


\clearpage
\begin{centering}
\large
ABSTRACT \\
\vspace*{4ex}
Finite Temperature QCD with Domain Wall Fermions\\
\vspace*{4ex}
GEORGE TAMMINGA FLEMING\\
\vspace*{4ex}
\end{centering}

Domain wall fermions are a new lattice fermion formulation which preserves
the full chiral symmetry of the continuum at finite lattice spacing,
up to terms exponentially small in an extra parameter.  We discuss
the main features of the formulation and its application to study of QCD
with two light fermions of equal mass.  We also present numerical studies
of the two flavor QCD thermodynamics with $a T = 1 / 4$.

\newpage

\pagestyle{plain}
\pagenumbering{roman}
\setcounter{page}{1}

\tableofcontents

\clearpage

\listoftables

\clearpage

\listoffigures

\clearpage


\chapter*{Acknowledgments}
\addcontentsline{toc}{chapter}{Acknowledgments}

First, I would like to thank Martin Perl and Jeff Greensite.
Without their excellent guidance, I certainly would not have come to Columbia.
I would also like to thank Norman Christ and Bob Mawhinney
for their patient support and a multitude of interesting discussions.

I cannot properly express the deep debt of gratitude I owe to Pavlos Vranas.
Before I met him, I'd never heard of domain wall fermions
and the available thesis topics seemed pretty dull and uninteresting.
He opened this whole new field to me and made this project possible.
Now I leave Columbia brimming full of wonderful and exciting ideas,
thanks to him and the Columbia lattice group as a whole.

Finally, I would like acknowledge the wonderful love and support of my wife
and friend, Bonnie Tamminga Fleming.  I only hope I can live up to her example
as she writes her own dissertation.

\clearpage

\pagestyle{myheadings}
\markright{}
\pagenumbering{arabic}
\setcounter{page}{1}


\chapter*{Introductory remarks}
\addcontentsline{toc}{chapter}{Introductory remarks}

\thispagestyle{myheadings}
\markright{}

For the last fifty years, quantum field theories have been amazingly
successful at predicting observable physical phenomena involving
elementary particles.  A class of quantum field theories called
{\it gauge theories}, because each contains a special internal symmetry
called gauge symmetry, accurately describes the interactions,
or forces, among elementary particles.  Three of the four known fundamental
forces ({\it strong}, {\it weak}, and {\it electromagnetic}) are best described
by gauge theories.  Furthermore, nearly all promising new theories are based
on the principle of invariance under local symmetry transformations.

{\it Quantum Chromodynamics}, or QCD, is the gauge
theory of the strong force responsible for binding protons and
neutrons together to form nuclei. As a practical matter, although QCD is
the underlying microscopic theory responsible for nuclear physics,
theorists do not use QCD to study the formation of nuclei.
The calculation is possible in principle, but the large number of degrees
of freedom render it intractable.
However, when particles interact at high energies, the strength of the
strong interaction becomes paradoxically weak through the phenomenon
called {\it asymptotic freedom}.  Only in this perturbative regime
can expansions of QCD in Feynman diagrams be used to successfully
describe observed physical processes.

Another approach to understanding QCD employs the formal mathematical
equivalence between statistical mechanics and quantum field theories
expressed in the Euclidean path integral formalism.
Systems described by statistical mechanics also contain vast numbers
of degrees of freedom yet numerical and analytic techniques have been
developed to handle them.  When QCD is expressed in the Euclidean path
integral formalism, the analogy with statistical mechanics becomes
manifest with the QCD Lagrangian as the Boltzmann weight.
Unfortunately, the path integral over a continuous spacetime
is divergent unless the integrals are properly regularized.

Lattice QCD provides this regularization scheme by discretizing
the Euclidean spacetime manifold while exactly preserving the internal
gauge symmetry.  Unfortunately, the continuous Lorentz symmetries
of translation and rotation are explicitly broken down to discrete subgroups
in the process.  The goal is to study QCD on as fine a lattice as possible
to understand the continuum limit of infinitesimal lattice spacing
where Lorentz symmetries are restored.
Due to asymptotic freedom, the ratios of physical observables will become
independent of the lattice spacing when the spacing is small enough.
For the lattice QCD theorist, accomplishing a few steps toward
the continuum limit is an immense undertaking involving
large supercomputers.  The reward for this effort is the ability to
study non-perturbative regimes of QCD.

QCD at finite temperature is a non-perturbative regime
where lattice QCD has emerged as an ideal tool for direct study of
the thermodynamics.  Theorists have long speculated that QCD
at sufficiently high temperature or density undergoes
a deconfining phase transition where quarks and gluons are no longer
confined within meson and baryon bound states. In the limits of either
infinite or zero mass quarks, the phase transition is characterized
by the behavior of a physical observable called an {\it order parameter}.
In nature, though all the quarks have finite mass, the {\it up} and {\it down}
quarks are much lighter, as evidenced by the light pion masses.
So, it is likely that QCD in our world resembles QCD with two massless quarks.
When the quarks are massless, the deconfinement order parameter
is associated with the spontaneous breakdown of chiral symmetry, the mechanism
that generates massless pions.

In this work, we use a recently developed fermion formulation,
{\it domain wall fermions}, to study aspects of QCD that are
strongly influenced by chiral symmetry.
Using fermion formulations that were available in the past, the full chiral
symmetry of QCD is broken at finite lattice spacing along with the Lorentz
symmetry and can be restored only in the continuum limit.  With domain wall
fermions, chiral symmetry can be effectively restored
at finite lattice spacing, {\it i.e.}\ before taking the continuum limit,
albeit at some increased computational cost.

In chapter \ref{ch:QCD_with_domain_wall_fermions},
after a brief overview of continuum and lattice QCD formulations
intended primarily to fix the notation used in the rest of the work, we
then discuss the domain wall fermion formulation and the related overlap
fermion formulation.  In particular, we develop a framework, based on the
Gell-Mann, Oakes, and Renner (GMOR) relation, to discuss how domain wall
fermions restore chiral symmetry on the lattice.  Finally we discuss in
general terms the physics underlying the deconfinement phase transition of
QCD.

In chapter \ref{ch:numerical_implementation},
we discuss some specific technical issues regarding the
efficient implementation of domain wall fermions for numerical simulation
with dynamical fermions.  We discuss the even-odd preconditioning of the
fermion matrix.  We also derive the molecular dynamics equations of motion
for the fermion action for use in the hybrid Monte Carlo (HMC) algorithm.

In chapter \ref{ch:numerical_simulations}, we present the results
of numerical simulations of $N_f$=2 QCD with light degenerate quarks.
Because domain wall fermions are computationally ${\cal O}(10)$--${\cal O}(100)$
times more expensive than standard lattice fermion formulations,
we focus our efforts on studying the physics
of the finite temperature transition region where the role of chiral symmetry
should be of primary importance.  Our initial studies
on $8^3\!\times\!4$ lattices located the transition region.  Further studies
on $16^3\!\times\!4$ lattices searched for conclusive evidence
of critical behavior in the transition region which might indicate the order
of the transition.  Finally, we make a quantitative estimate of the degree
of chiral symmetry restoration in the broken phase using the GMOR framework
introduced in chapter \ref{ch:QCD_with_domain_wall_fermions}.

We use the following conventions throughout this work.  All lattice fields
and bare parameters are to be dimensionless.  To compare these
dimensionless lattice numbers with their physical dimensionful analogues,
we employ the dimensionful lattice spacing.  For example, the lattice pion
mass $m_\pi$ at a given lattice spacing $a$ may be compared to the
physical pion mass $m_\pi^{(phys)}$ by $m_\pi^{(phys)} = m_\pi / a$.

\clearpage


\chapter{QCD with domain wall fermions}
\label{ch:QCD_with_domain_wall_fermions}

\thispagestyle{myheadings}
\markright{}

\section{Continuum description}
\label{sec:continuum_QCD}

QCD, or Quantum Chromodynamics, is the current best description of
the strong nuclear force.  The theory contains
six types, or {\it flavors}, of fermions called {\it quarks}.
Analogous to the theory of electrodynamics, where fermions
called {\it electrons} carry electric charge,
quarks carry one of three types of color charge.
As electrons interact through an electromagnetic gauge field,
these color-charged quarks interact through color-electric
and color-magnetic gauge fields following a logical extension
of Maxwell's field equations.  The complex charge structure of QCD yields 
a far richer theory with quark confinement and gauge field self-interactions
not found in the simpler electrodynamics.

QCD is described schematically by the Lagrangian density
\beq
{\cal L} = {\cal L}_{\rm quarks} + {\cal L}_{\rm gluons}
\eeq
which yields the action when integrated over space-time.  
By treating the time coordinate as a complex parameter, we can equivalently
define the Lagrangian either on the real time axis ({\it Minkowski} space-time)
or the imaginary time axis ({\it Euclidean} space-time), as the theory defined
on one can be analytically continued to the other.
We choose the Euclidean prescription because it allows the interpretation
of the action as a Boltzmann weight in an equivalent statistical mechanics
framework.

The first term in the Lagrangian density includes a description of how quarks
interact with {\it gluons}, the bosons that constitute gauge fields
\beq
\label{eq:cont_quark_lagrangian}
{\cal L}_{\rm quarks} = \sum_{j=1}^{N_f} \sum_{\mu=1}^4
\overline{q}_j \left[
  \gamma_\mu \left(
    \partial_\mu - i g A_\mu
  \right) + m_j
\right] q_j .
\eeq
The $\gamma_\mu$ are Hermitian matrices which satisfy the anticommutation
relations
$\left\{\gamma_\mu, \gamma_\nu \right\} = 2 \delta_{\mu,\nu}$
appropriate for Euclidean space-time.
$N_f$ is the number of quark flavors (six), the gauge fields $A_\mu$
are elements of the algebra of the Lie group ${\rm SU}(N_c)$
\beq
\label{eq:cont_gauge_field}
A_\mu = \sum_{a=1}^{N_c^2-1} A_\mu^a \frac{\lambda^a}{2}
\eeq
and $N_c$ is the number of color charges (three).
The $\lambda^a$ are the generators of the algebra of ${\rm SU}(N_c)$
and are normalized
\beq
{\rm tr} \left( \lambda^a \lambda^b \right) = 2 \delta^{ab} .
\eeq

The second part of the Lagrangian includes a description
of the self-interaction of the gluonic fields in the absence of quarks
\beqs
\label{eq:cont_gauge_lagrangian}
& {\cal L}_{\rm gluons} = \sum_{\mu,\nu=1}^4
\frac{1}{2} {\rm Tr} \left[ F_{\mu\nu} F_{\mu\nu} \right]. & \\*
\label{eq:YM_field_strength}
& F_{\mu\nu} = \sum_{a=1}^{N_c^2-1} F_{\mu\nu}^a \frac{\lambda^a}{2} = \left(
  \partial_\mu A_\nu^a - \partial_\nu A_\mu^a + g f^{abc} A_\mu^b A_\nu^c
\right) \frac{\lambda^a}{2} &
\eeqs
where $F_{\mu\nu}$ is the Yang-Mills field strength tensor.
Note that the summation convention is implied in the right hand side
of (\ref{eq:YM_field_strength}).
$f^{abc}$ are the structure functions of ${\rm SU}(N_c)$
and satisfy the commutation relations
\beq
\left[ \lambda^a , \lambda^b \right] = 2 i f^{abc} \lambda^c .
\eeq

The action is the Euclidean spacetime integral of the Lagrangian density
\beq
S\left( A, \overline{q}, q \right)
= \int d^4x \left( {\cal L}_{\rm quarks} + {\cal L}_{\rm gluons} \right) .
\eeq
The partition function is given by the path integral over all possible
field configurations
\beq
\label{eq:cont_part_fn}
Z = \int [{\cal D}A][{\cal D}\overline{q}][{\cal D}q]~
e^{-S(A,\overline{q},q)} .
\eeq
Physical quantities are computed as a weighted average of appropriately
chosen observables ${\cal O}$ over all field configurations,
normalized by the partition function
\beq
\label{eq:cont_expect_val}
\left\langle{\cal O}(A,\overline{q},q)\right\rangle
= \frac{1}{Z} \int [{\cal D}A][{\cal D}\overline{q}][{\cal D}q]
~{\cal O}(A,\overline{q},q)~e^{-S(A,\overline{q},q)} .
\eeq
As written, this partition function is formally divergent because it 
includes integration over an infinite number of gauge configurations which are
related by gauge transformations, thus of equal weight in the path integral.
Restricting the path integral to sample only one element of each set
of gauge-equivalent configurations will resolve this problem.
For example, the method of Faddeev and Popov can be used
\cite{Montvay:1994cy,Rothe:1997}.
Even still, there are further divergences which must be carefully
regularized to ensure the partition function is well defined.

In addition to invariance under local ${\rm SU}(N_c)$ gauge transformations,
when the $m_j$=0, the action is also invariant under independent global
transformations of the left and right chiral components of the fermion fields.
We define the chiral matrix $\gamma_5$ as the last remaining linearly
independent matrix that anticommutes with the $\gamma_\mu$ matrices:
$\left\{ \gamma_5, \gamma_\mu \right\}$=0.
We then define the chiral projection operators
\beq
\label{eq:chiral_projectors}
P_\pm = \frac{1\pm\gamma_5}{2}
\eeq
and the right and left components are given by
$q_R = P_+ q$ and $q_L = P_- q$.  Now the
transformations explicitly are
\beqa
q_R & \to & e^{i \alpha_R \frac{\lambda^a}{2}} q_R , \\
q_L & \to & e^{i \alpha_L \frac{\lambda^a}{2}} q_L , \\
q_R & \to & e^{i \theta_R} q_R , \\
q_L & \to & e^{i \theta_L} q_L .
\eeqa
Formally, these transformations generate the symmetry group
$ {\rm SU}_L(N_f) \otimes {\rm SU}_R(N_f)
  \otimes {\rm U}_L(1) \otimes {\rm U}_R(1) $.
Furthermore, we stress that these are the symmetries of the classical action
only.  The full theory will only respect a subgroup of these symmetries
due to quantum interactions.

The Noether currents commonly associated with these
symmetries are the {\it non-singlet vector current}
\beq
\label{eq:nonsinglet_vector_current}
j_\mu^a(x) = \overline{q}(x) \gamma_\mu \frac{\lambda^a}{2} q(x) ,
\eeq
the {\it non-singlet axial vector current}
\beq
\label{eq:nonsinglet_axial_current}
j_{\mu 5}^a(x) = \overline{q}(x) \gamma_\mu \gamma_5 \frac{\lambda^a}{2} q(x) ,
\eeq
the {\it singlet vector current}
\beq
\label{eq:singlet_vector_current}
J_\mu(x) = \overline{q}(x) \gamma_\mu q(x) ,
\eeq
and the {\it singlet axial vector current}
\beq
\label{eq:singlet_axial_current}
J_{\mu 5}(x) = \overline{q}(x) \gamma_\mu \gamma_5 q(x) .
\eeq
The conservation of the singlet axial vector current
(\ref{eq:singlet_axial_current}) is broken by the Adler--Bell--Jackiw
(ABJ) anomaly.
The conserved charge associated with the singlet vector current
(\ref{eq:singlet_vector_current}) is the {\it baryon number} and is conserved
even when the bare quark masses are non-zero.

For massless quarks, the remaining non-singlet chiral symmetry may be
spontaneously broken by the ground state:
${\rm SU}_L(N_f) \otimes {\rm SU}_R(N_f) \to {\rm SU}_V(N_f)$.
This implies that the non-singlet vector current
(\ref{eq:nonsinglet_vector_current}) is still conserved.
It remains conserved even for non-zero quark masses,
provided they are degenerate, and the associated conserved charge
is called {\it isospin}.  The implied breaking of the non-singlet
axial vector current is responsible for the appearance of the
$N_f^2 - 1$ Nambu-Goldstone bosons.  Though quark mass terms explicitly break
this current, for small masses spontaneous and explicit symmetry breaking
effects work together to produce light Nambu-Goldstone bosons whose mass
is proportional to the square root of the quark mass.

Another interesting consequence of chiral symmetry is the role it plays
in the {\it spectral representation} of the Dirac operator.
From equation (\ref{eq:cont_quark_lagrangian}), we define the continuum Dirac
operator
\beq
\label{eq:dirac_matrix}
\!\not\!{\rm D} = \sum_\mu \gamma_\mu \left( \partial_\mu - i g A_\mu \right) .
\eeq
This anti-Hermitian operator has a continuous imaginary eigenvalue spectrum:
$\!\not\!{\rm D} q_\lambda = i \lambda q_\lambda$, where each eigenvalue
occurs with the density $\rho\left(\lambda\right)$, normalized so
$\int d\lambda\ \rho\left(\lambda\right) = 1$. 
Because $\!\not\!{\rm D}$ anticommutes with $\gamma_5$, then for each
non-zero eigenvector $q_\lambda$ we can generate a new eigenvector
$\gamma_5 q_{\lambda} = q_{-\lambda}$ whose eigenvalue is the negative
of the original
\beq
\label{eq:cont_evalue_pairing}
\!\not\!{\rm D} \left( \gamma_5 q_\lambda \right)
= - \gamma_5 \left( \!\not\!{\rm D} q_\lambda \right)
= - i \lambda \left( \gamma_5 q_\lambda \right) 
= - i \lambda q_{-\lambda}.
\eeq
Thus, non-zero eigenvalues always occur in pairs of opposite sign:
$\left( i\lambda, -i\lambda \right)$.

A simple application of the spectral representation is to derive the spectral
identity that relates the chiral condensate to the pseudoscalar susceptibility
in QCD with degenerate quarks.  The chiral condensate in a volume $V$ is
\beqs
\label{eq:cont_qbq}
\qbq & \equiv & - \frac{1}{4 N_c V}
\int_V d^4x\ \left\langle \overline{q}_x q_x \right\rangle \non \\*
& = & \frac{1}{4 N_c V} {\rm Tr} \left\langle
  \left( \!\not\!{\rm D} + m \right)^{-1}
\right\rangle \non \\*
& = & \int_{-\infty}^{+\infty} d\lambda\ 
\frac{\overline\rho(\lambda)}{m+i\lambda}
= 2 \int_0^\infty d\lambda\ \frac{\overline\rho(\lambda)}{m+i\lambda}
\eeqs
where we have inserted an extra minus sign so that the symbol $\qbq$
always represents a positive number for positive $m$.  We have chosen
the normalization such that $\qbq \sim 1/m$ as $m \to \infty$.
$\overline\rho\left(\lambda\right)$, an even function, is the quantum average
of the eigenvalue density so it depends implicitly on the bare masses
and couplings.

The pseudoscalar, or pion, susceptibility is (no sum on $a$)
\beqs
\label{eq:cont_chi_pi_1}
\chi_{\pi^a} & = & - \frac{2}{4 N_c V} \int_V d^4x d^4y \left\langle
  \overline{q}_y \gamma_5 \frac{\tau^a}{2} q_y
  \overline{q}_x \gamma_5 \frac{\tau^a}{2} q_x
\right\rangle \non \\*
& = & \frac{1}{4 N_c V} {\rm Tr} \left\langle
  \gamma_5 \left( \!\not\!{\rm D} + m \right)^{-1}
  \gamma_5 \left( \!\not\!{\rm D} + m \right)^{-1}
\right\rangle 
\eeqs
where we have implicitly assumed that $N_f \ge 2$.
We can anticommute $\!\not\!{\rm D}^{-1}$ and $\gamma_5$
\beqs
\label{eq:cont_chi_pi_2}
\chi_{\pi^a} & = & \frac{1}{4 N_c V} {\rm Tr} \left\langle
  \left( m - \!\not\!{\rm D} \right)^{-1}
  \left( m + \!\not\!{\rm D} \right)^{-1}
\right\rangle \non \\*
& = & \int_{-\infty}^{+\infty} d\lambda\
\frac{\overline\rho(\lambda)}{(m-i\lambda)(m+i\lambda)} .
\eeqs
If we multiply the right hand side of (\ref{eq:cont_qbq}) by the
identity factor $\left(m-i\lambda\right)/\left(m-i\lambda\right)$
and note that $i \lambda \overline\rho(\lambda)$ is an odd function,
then (\ref{eq:cont_qbq}) and (\ref{eq:cont_chi_pi_2}) give the identity
\beq
\label{eq:cont_spect_ident}
\qbq = m \chi_{\pi^a} .
\eeq
Finally, we emphasize it is the chiral symmetry, manifested
in (\ref{eq:cont_evalue_pairing}), that is responsible for this identity.
Not surprisingly, (\ref{eq:cont_spect_ident}) can also be derived as the
integral form of the flavor non-singlet axial Ward-Takahashi identity,
another consequence of chiral symmetry.  In section \ref{sec:finite_ls},
we will show that (\ref{eq:cont_spect_ident}), within the context of the
spectral representation, provides a useful starting point for understanding
chiral symmetry breaking for lattice fermions.

\section{Lattice actions for gauge fields}
\label{sec:lattice_gauge_actions}

Lattice gauge theories provide a regularization scheme for the divergences of
the path integral by discretizing the Euclidean spacetime manifold.  Not
surprisingly, this discretization process destroys the continuous Lorentz
symmetries, leaving only a hypercubic remnant.  Other symmetries might also
be broken if we make a poor choice when transcribing our continuum action
into the lattice scheme.  Wilson demonstrated \cite{Wilson:1974sk} that the
local gauge symmetry, at least, can always be exactly preserved on the lattice
and other symmetries are restored in the limit of vanishing lattice spacing:
$a \to 0$.

We can define the lattice partition function as the integral over all field
configurations of the lattice action
\beq
\label{eq:lat_part_fn_1}
Z = \int \left[ {\cal D} U_\mu \right] \left[ {\cal D} \overline{q} \right]
\left[ {\cal D} q \right] e^{-\left( S_G + S_F \right)}.
\eeq
Purely gluonic contributions to the action are contained in $S_G$.
Fermionic contributions are included in $S_F$ and will be discussed
in section \ref{sec:sym_dwf_action}.

Starting from the $A_\mu(x)$ fields of (\ref{eq:cont_gauge_field}),
we define the {\it gauge links} $U_\mu(x)$ as the integral of the path ordered
exponential of the $A_\mu(x)$ along the link connecting the site $x$
to the neighboring site one lattice spacing away in the $\mu$ direction:
$x+\hat\mu$.  As the $A_\mu(x)$ are elements of the algebra of the gauge group
${\rm SU}(N_c)$, the gauge links are elements of the group itself.
Additionally, the $U_\mu(x)$ are used to parallel transport quark fields
from the site $x+\hat\mu$ to the site $x$.

The Wilson action for ${\rm SU}(N_c)$ gauge theory
in $d$ dimensions is \cite{Wilson:1974sk}
\beq
\label{eq:wilson_gauge_action}
S_G^{(W)} = \beta \sum_x \sum_{\mu < \nu} \left[
  1 - \frac{1}{N_c} {\rm Re Tr} U_P(x,\mu,\nu)
\right]
\eeq
where $\mu, \nu \in [1,d]$
and the plaquettes $U_P(x,\mu,\nu)$ are
\beq
\label{eq:plaquette}
U_P(x,\mu,\nu) =
U_\mu(x) U_\nu(x+\hat\mu) U_\mu^\dagger(x+\hat\nu) U_\nu^\dagger(x) .
\eeq
When the plaquette is expanded in powers of the lattice spacing, the leading
term appearing at ${\cal O}(a^2)$ corresponds to the continuum gauge field
Lagrangian density (\ref{eq:cont_gauge_lagrangian}).  Terms in the expansion
which appear at higher powers of $a$ are called {\it irrelevant} because
their influence vanishes in the continuum limit.  Although the analogy with
common statistical mechanics notation is intentional, here $\beta$ is a bare
parameter of the lattice theory which corresponds to the gauge coupling $g$
in the continuum theory: $\beta = 2 N_c / g^2$ as $a \to 0$.

Wilson later proposed \cite{Wilson:1979wp} introducing irrelevant operators
to the gauge action to improve the scaling of Monte Carlo simulations.
Here we consider adding only the rectangle operator
\beq
\label{eq:rect_gauge_action}
S_G^{(R)} = \beta \sum_x \left\{
  \sum_{\mu < \nu} \left[
    1 - \frac{c_0}{N_c} {\rm Re Tr} U_P(x,\mu,\nu)
  \right] + \sum_{\mu \ne \nu} \left[
    1 - \frac{c_1}{N_c} {\rm Re Tr} U_R(x,\mu,\nu)
  \right]
\right\}.
\eeq
We require that the coefficients satisfy the constraint
$c_0 + 8 c_1 = 1$, as predicted in weak coupling lattice 
perturbation theory.  The rectangles $U_R(x,\mu,\nu)$
are
\beq
\label{eq:rectangle}
U_R(x,\mu,\nu) =
U_\mu(x) U_\mu(x+\hat\mu) U_\nu(x+2\hat\mu)
U_\mu^\dagger(x+\hat\mu+\hat\nu) U_\mu^\dagger(x+\hat\nu) U_\nu^\dagger(x) .
\eeq
In chapter \ref{ch:numerical_simulations}, we will compare numerical
simulations that include a different choice of gauge action looking for
behavior which appears closer to the continuum limit.

\section{Domain wall fermions on lattice boundaries}
\label{sec:sym_dwf_action}

The standard approach to discretizing the continuum quark Lagrangian density
(\ref{eq:cont_quark_lagrangian}) is to define the quark fields on lattice sites
and replace derivatives with finite differences. For example, we can define
the gauge covariant lattice central difference operator corresponding
to the covariant derivative in the continuum
\beq
\label{eq:1st_ord_diff}
{\nabla(\mu)}_{x,x^\prime}  \equiv \frac{1}{2a} \left[
  U_\mu(x) \delta_{x+\hat\mu,x^\prime}
  - U^\dagger_\mu(x-\hat\mu) \delta_{x-\hat\mu,x^\prime}
\right] .
\eeq
As we shall see below, this naive prescription leads to the well known problem
of species doubling: on a $d$ dimensional finite lattice the discretized
action describes $2^d$ identical species, one for each corner of the Brillouin
zone.  Nielsen and Ninomiya \cite{Nielsen:1981rz,Nielsen:1981xu} showed
when the lattice fermion action is local, Hermitian
and translationally invariant then either the fermions are doubled
or chiral symmetry of the continuum action is broken on the lattice.

Wilson \cite{Wilson:1975id} proposed giving all but one of the 
doubled fermions a mass proportional to the lattice cutoff 
by adding an irrelevant term to the fermionic action.
Using the lattice central difference operator corresponding
to the Laplacian in the continuum
\beq
\label{eq:2nd_ord_diff}
{\Delta(\mu)}_{x,x^\prime} \equiv \frac{1}{a^2} \left[
  U_\mu(x)\delta_{x+\hat\mu,x^\prime}
  - 2\delta_{x,x^\prime}
  + U^\dagger_\mu(x-\hat\mu)\delta_{x-\hat\mu,x^\prime}
\right],
\eeq
the lattice action for Wilson fermions is
\beq
\label{eq:frm_w}
S_F^{(W)} = a^d \frac{\overline\psi_x}{a^{(d-1)/2}} \left\{
  \frac{m_0}{a} \delta_{x,x^\prime}
  + \sum_{\mu=1}^d \left[
    \gamma_\mu {\nabla(\mu)}_{x,x^\prime}
    - \frac{ar}{2} {\Delta(\mu)}_{x,x^\prime}
  \right]
\right\} \frac{\psi_{x^\prime}}{a^{(d-1)/2}}.
\eeq
We suppress spin and flavor indices and the summation convention
implies summing $x, x^\prime$ over all sites on the lattice.
The bare fermion fields $\overline\psi$ and  $\psi$, the bare mass $m_0$
and coefficient of the irrelevant term $r$ are all dimensionless quantities
whereas $\nabla(\mu)$ and $\Delta(\mu)$ are dimensionful according to the
powers of the lattice spacing in their definitions.
Rearranging terms, we put the action into the standard form
of a fermion matrix bilinear
\beq
S_F^{(W)} = \overline\psi_x M_{x,x^\prime}^{(W)} \psi_{x^\prime}
\eeq
where $M_{x,x^\prime}^{(W)}$ is the Wilson-Dirac fermion matrix
\beq
\label{eq:frm_w_mat}
M_{x,x^\prime}^{(W)} = \left(m_0 + r d \right) \delta_{x,x^\prime}
- \sum_{\mu=1}^d \frac{1}{2} \left[
  \left(r-\gamma_\mu\right) U_\mu(x) \delta_{x+\hat\mu,x^\prime}
  + \left(r+\gamma_\mu\right)
  U_\mu^\dagger(x-\hat\mu) \delta_{x-\hat\mu,x^\prime}
\right].
\eeq
For numerical simulation, $r$ is usually set to one to take advantage of 
the simplified spin structure of the spin projection matrices:
$P_{\pm\mu} = \left( 1 \pm \gamma_\mu \right) / 2$.

The free Wilson fermion propagator is obtained by inverting
the fermion matrix with all $U_\mu(x)$ set to the identity.
We block-diagonalize the fermion matrix by Fourier transform to give
\beq
\label{eq:frm_w_mat_ft}
\widetilde{M}_p^{(W)} = m_0 - \sum_{\mu=1}^d \left[
  i \gamma_\mu \sin p_\mu + r \left(\cos p_\mu - 1\right)
\right],
\eeq
where the $p_\mu$ are dimensionless momenta that reside on the
reciprocal lattice and range over $-\pi < p_\mu \le \pi$.
In the massless limit $r \to 0$ corresponding to the naive case above,
we see that the propagator has a massless pole when each $p_\mu = 0$ or $\pi$.
These are the corners of the Brillouin zone mentioned above and each pole is
associated with one doubler species.  When $0 < r \le 1$ only the fermion
species with $p_\mu=0$ remains massless and the others acquire a mass
on the order of the lattice cutoff.  However, even though we have removed
the doublers with the irrelevant Wilson term, chiral symmetry is not restored
because the Wilson term commutes with $\gamma_5$.

The domain wall fermion formulation, originally proposed by Kaplan
\cite{Kaplan:1992bt}, solves the doubling problem while conserving the
${\rm SU}_L(N_f) \otimes {\rm SU}_R(N_f)$ chiral symmetry of massless quarks
on the lattice through the introduction of an infinite internal dimension
which we label with coordinate $s$.  To demonstrate the construction, we start
with an odd $d+1$ dimensional action of free lattice Wilson fermions, see
(\ref{eq:frm_w}), where the bare mass parameter $m_0$ is a function
of one direction, the $s$-direction.  Although $m_0(s)$ can be
any monotonically increasing function that crosses zero, for simplicity
we discuss only step function mass terms
\beq
\label{eq:m0_step_func}
m_0(s) = \left\{
  \begin{array}{cc}
    m_+, & s \ge 0 \\
    m_-, & s < 0
  \end{array}
\right. .
\eeq
Despite the presence of a mass term and the lack of chiral symmetry in odd
dimensions, there is a normalizable, chiral basis of states that solve
the equation
\beq
M^{(K)} \Psi^\pm_p = \left[
  \sum_{\mu=1}^d i \gamma_\mu \sin p_\mu
\right] \Psi^+_p
\eeq
where $M^{(K)}$ is Kaplan's Dirac matrix and $p_\mu$ is a $d$ dimensional
momentum.  The existence of such a basis indicates the spectrum contains states
that propagate as massless fermions in the even $d$ dimensions.
Kaplan's solution
\beq
\Psi^+_p = (\pm 1)^s e^{i p \cdot x} e^{-\mu_0 |s|},
\quad \sinh \mu_0 \equiv m_0
\eeq
indicates the massless modes are exponentially localized on the domain wall,
where the mass changes sign.  Furthermore, when the extent $L_s$ along
the $s$-direction is infinite, only one chirality of the massless modes is
normalizable.  So, even when the lattice spacing is finite, the action
describes a single massless chiral fermion flavor at scales well below
the lattice cutoff in the $L_s \to \infty$ limit.

On a lattice with a finite $L_s$, the solutions of the opposite chirality
become normalizable and localized on the boundaries.  In essence, the
boundaries become anti-domain walls and at low energies the effective theory
is naturally vector-like.  As the separation between the opposing walls
is increased by increasing $L_s$, terms that break chiral symmetry will
eventually become negligible due to the exponentially suppressed overlap
between modes of opposite chirality on opposite walls.  So, domain wall
fermions at large $L_s$ are ideally suited for simulating the massless quarks
of a vector-like theory like QCD.

If we reinterpret the $s$-direction as an internal flavor space,
then gauge fields are coupled to the fermions in the natural way
\cite{Kaplan:1992sg,Frolov:1993ck,Narayanan:1993wx}.
We write this as the set of constraints on the gauge field
\beq
U_{\mu \le d}(x,s) = U_\mu(x) ; \qquad U_{d+1}(x,s) = \openone .
\eeq
From this perspective, domain wall fermions are identical to heavy Wilson
fermions in $d$ dimensions that carry an additional flavor index $s$
and derivative terms along this $s$ direction are now interpreted
as flavor mixings in a sophisticated mass matrix.

Shamir proposed \cite{Shamir:1993zy} a variant of domain wall fermions where
free boundaries at both ends of the extra direction serve as the domain
wall/anti-wall.  This variant naturally arises when considering the step
function mass term (\ref{eq:m0_step_func}) in the limit $m_- \to -\infty$,
provided we use some care in handling the boundaries.  From a practical view,
this reduces by half the number of lattice sites necessary to separate
the walls a fixed distance.  The action is a fermion bilinear
\beq
\label{eq:frm_d_action}
S_F^{\rm (dwf)} =
\overline\Psi_{x,s} M_{x,s;x^\prime,s^\prime}^{\it(dwf)}
\Psi_{x^\prime,s^\prime} .
\eeq
Starting with the Wilson fermion matrix (\ref{eq:frm_w_mat}) we build
the flavor mixing matrix from first and second order difference
operators and the chiral matrix $\Gamma_5 \equiv \gamma_{d+1}$ of the effective
even $d$ dimensional theory
\beq
\label{eq:frm_d_mat_1}
M_{x,s;x^\prime,s^\prime}^{\it(dwf)} = 
M_{x,x^\prime}^{\it(W)}(-m_0) \delta_{s,s^\prime} + \left[
  \Gamma_5 \hat\nabla_{s,s^\prime} - \frac{r_5}{2} \hat\Delta_{s,s^\prime}
\right] \delta_{x,x^\prime} .
\eeq
The hat notation on the difference operators denotes that the operators
are ungauged and dimensionless because both act on a flavor space:
\beqs
\hat\nabla_{s,s^\prime} & = & \frac{1}{2} \left(
  \delta_{s+1,s^\prime} - \delta_{s-1,s^\prime}
\right) , \\
\hat\Delta_{s,s^\prime} & = & \delta_{s+1,s^\prime} - 2 \delta_{s,s^\prime}
+ \delta_{s-1,s^\prime} .
\eeqs

Following the literature, we collect terms in the action that are diagonal
in the $s$-direction into an operator that acts parallel
to the $d$ dimensional spacetime
\beq
\label{eq:frm_d_mat_2}
M_{x,s;x^\prime,s^\prime}^{\it(dwf)}
= D_{x,x^\prime}^\parallel \delta_{s,s^\prime}
+ \delta_{x,x^\prime} D_{s,s^\prime}^\perp .
\eeq
The parallel piece is simply a Wilson matrix with slightly different
mass term,
$
D_{x,x^\prime}^\parallel = M_{x,x^\prime}^{\it(W)}(-m_0)
+ r_5 \delta_{x,x^\prime},
$
which we write out explicitly for completeness
\beqs
\label{eq:frm_d_dslash_parallel}
D_{x,x^\prime}^\parallel & = & 
\left( r d + r_5 - m_0 \right) \delta_{x,x^\prime} \non \\*
& & - \sum_{\mu=1}^d \frac{1}{2} \left[
  \left(r-\gamma_\mu\right) U_\mu(x) \delta_{x+\hat\mu,x^\prime}
  + \left(r+\gamma_\mu\right) U_\mu^\dagger(x-\hat\mu)
    \delta_{x-\hat\mu,x^\prime}
\right].
\eeqs
The second term is the flavor mixing matrix (with flavor index $s$)
\beq
\label{eq:frm_d_dslash_perp}
D_{s,s^\prime}^\perp = \left\{
  \begin{array}{llll}
    - \left( \frac{r_5-\Gamma_5}{2} \right) \delta_{1,s^\prime} &
    + m_f \left( \frac{1+\Gamma_5}{2} \right)
      \delta_{L_s-1,s^\prime} &,& s=0 \\
    - \left( \frac{r_5-\Gamma_5}{2} \right) \delta_{s+1,s^\prime} &
    - \left( \frac{r_5+\Gamma_5}{2} \right)
      \delta_{s-1,s^\prime} &,& 0<s<L_s-1 \\
    m_f \left( \frac{1-\Gamma_5}{2} \right) \delta_{0,s^\prime} &
    - \left( \frac{r_5+\Gamma_5}{2} \right)
      \delta_{L_s-2,s^\prime} &,& s=L_s-1
  \end{array}
\right. .
\eeq
We have included a new parameter $m_f$ that plays the role
of generalized boundary conditions in the $s$-direction.  For $m_f = 0$
the boundaries are free and for $m_f = \pm 1$ and $r_5 = 1$ the boundaries are
(anti)periodic.  We shall also see that $m_f$ controls the bare mass for the
effective light modes.

Until now, we have included a parameter $r_5$ analogous to the Wilson $r$
because it is possible to find massless modes in the low energy theory
for $0 < r_5 \le 1$ \cite{Kaplan:1992bt}.  While $r$ plays the physical role
of setting the lattice mass scale of the $2^d-1$ doubler states
for Wilson fermions, $r_5$ has no such physical interpretation.
Furthermore, for $r_5 \ne 1$ the massless modes are no longer eigenvectors
of $\Gamma_5$.  So, for the remainder of this work we will assume $r_5 = 1$.

As in (\ref{eq:frm_w_mat_ft}), we can diagonalize $D^\parallel$ in the free
case by Fourier transform
\beq
\label{eq:frm_d_dslash_parallel_ft}
\widetilde{D}_p^\parallel
= b(p) - \sum_{\mu=1}^d i \gamma_\mu \sin p_\mu 
\eeq
where $b(p) \equiv 1 - m_0 - \sum_{\mu=1}^d r \left( \cos p_\mu - 1 \right)$.
As in our previous discussion of Kaplan's formulation, we want to search for
a normalizable basis of states upon which the full domain fermion matrix acts
like the chirally symmetric Dirac matrix in $d$ dimensions
\beq
\widetilde{M}_{p;s,s^\prime}^{\rm(dwf)} \widetilde\Psi_{p,s^\prime}
= - \sum_{\mu=1}^d i \gamma_\mu \sin p_\mu \widetilde\Psi_{p,s} .
\eeq
Since we chose $r_5=1$, the $\widetilde{\Psi}$ should also be
eigenvectors of $\Gamma_5$: $\widetilde\Psi^\pm_{p,s} = \phi^\pm_{p,s} u^\pm$
where $u^\pm$ are constant eigenvectors of $\Gamma_5$.

We can reformulate the problem of identifying this basis into solving
the zero eigenvalue problem of a different matrix
\beq
{\cal M}_{p;s,s^\prime} \equiv \widetilde{M}_{p;s,s^\prime}^{\rm(dwf)}
+ \sum_{\mu=1}^d i \gamma_\mu \sin p_\mu \delta_{s,s^\prime}
= b(p) \delta_{s,s^\prime} + D_{s,s^\prime}^\perp .
\eeq
Note that ${\cal M}$ essentially contains all the terms
in $\widetilde{M}^{\rm(dwf)}$ that can break chiral symmetry
in the $d$ dimensional spacetime.  So, the basis of chiral states 
corresponds to the null space of ${\cal M}$
\beq
{\cal M}_{p;s,s^\prime} \widetilde\Psi_{p,s^\prime} \equiv
\left[
  b(p) \delta_{s,s^\prime}
  + D_{s,s^\prime}^\perp
\right] \left(
  P_+ \phi^+_{p,s^\prime} + P_- \phi^-_{p,s^\prime}
\right) = 0
\eeq
where $P_\pm \equiv \left( 1 \pm \Gamma_5 \right) / 2$ are chiral projection
matrices analogous to (\ref{eq:chiral_projectors}).  For $m_f$=0,
the eigenvectors which solve this equation are
\beq
\phi^\pm_{p,s} = b(p)^{\mp s} \phi^\pm_{p,0}
\eeq
%
If we can choose $m_0$ to satisfy the condition
\beq
\label{eq:free_dwf_condition}
-1 < 1 - m_0 - \sum_{\mu=1}^d \left( \cos p_\mu - 1 \right) < 1 
\eeq
then $\Psi^-_p$ is localized at the boundary $s=$ and,
as $L_s \to \infty$, it is the only normalizable solution at any finite $s$.
For the zero momentum state, $0 < m_0 < 2$ will satisfy the condition.
Similarly, for $2 r < m_0 < 2 \left( 1 + r \right)$, $d$ zero momentum doubler
states will be bound to the domain wall.  It is straightforward to derive
the other conditions for the remaining doubler states.  If we allow $L_s$
to be finite, $\Psi^+$ becomes normalizable and localized at $s=L_s-1$.

We remind the reader that the states $\widetilde\Psi_{p,s}$ are not
eigenvectors of the free domain wall fermion Dirac matrix.  Since that
Dirac matrix is non-Hermitian, it is essential to calculate the free
propagator from the Hermitian second order operator:
$M^{\dagger{\rm(dwf)}} M^{\rm(dwf)}$.  For non-zero $m_f$, Vranas
\cite{Vranas:1997tj,Vranas:1997da} analytically computed the wave functions
of the nearly massless modes and found the effective mass (for $r=1$) is
\beq
\label{eq:free_m_eff}
m_{\rm eff} = m_0 \left( 2 - m_0 \right)
\left[ m_f + \left( 1 - m_0 \right)^{L_s} \right] .
\eeq
The $L_s$-dependent contribution is exactly the overlap between
the wave functions of the right and left modes on opposite walls.

In the interacting case, the precise shape of the wave functions
of the nearly massless modes depends upon each gauge configuration.
So it is convenient to define the effective light fermion fields
through the projection
\beqs
\label{eq:dwf_4d_fermions}
q_x & = & P_- \Psi_{x,0} + P_+ \Psi_{x,L_s-1} \non \\*
\overline{q}_x & = & \overline\Psi_{x,L_s-1} P_- + \overline\Psi_{x,0} P_+ .
\eeqs
For large $L_s$, these fields have a large overlap with the nearly
massless modes and a vanishing overlap with the other heavy modes
described above.

Besides the nearly massless modes, the action also describes $2^d-1$ fermion
doublers that propagate along the wall as massive fermions with the mass
of ${\cal O}(r/a)$.  Furthermore, naive counting suggests there are 
an additional $2^d \left( L_s - 1 \right)$ massive states in the spectrum.
These states will contribute to a bulk infinity as $L_s \to \infty$.
In section $\ref{sec:large_ls_limit}$, we will identify this divergence
and introduce terms into the action to cancel it and preserve the
$L_s \to \infty$ limit.

\section{The limit of infinite domain wall separation}
\label{sec:large_ls_limit}

Using the rules for integration of Grassmann variables, we can explicitly
perform the integration over the fermionic degrees of freedom in the lattice
partition function (\ref{eq:lat_part_fn_1}).  The resulting expression now
depends explicitly on only the gauge degrees of freedom
\beq
\label{eq:lat_part_fn_2}
Z = \int \left[ {\cal D} U_\mu \right] \left( \det M \right) e^{-S_G}
= \int \left[ {\cal D} U_\mu \right] e^{-S_{\rm eff}}
\eeq
where $M$ is the lattice Dirac fermion matrix which appears in the 
fermion action, {\it e.g.\ }(\ref{eq:frm_w_mat}) or (\ref{eq:frm_d_mat_1}),
and $S_{\rm eff} = S_G - {\rm Tr} \log M$.
Since the fermionic weight in the path integral is given by determinant,
we can consider replacing $M$ with a different matrix $M^\prime$ provided
$\det M = \det M^\prime$.  As we shall see, this freedom allows us to
uncover important details about the physics of the lattice fermions.
Following the approach of Neuberger \cite{Neuberger:1998bg}
and Kikukawa and Yamada \cite{Kikukawa:1998pd}, we will construct a product
of matrices whose determinant is the same as the domain wall fermion Dirac
matrix in section \ref{sec:sym_dwf_action}, yet whose $L_s \to \infty$ limit
is manifestly apparent.

First, we choose a Weyl basis for the Dirac matrices
\beq
\label{eq:dirac_matrices}
\gamma_{\mu=1,\cdots,d} = \left(
  \begin{array}{cc}
    0 & \sigma_\mu \\
    \sigma_\mu^\dagger & 0
  \end{array}
\right),
\eeq
where each block is a $2^{\frac{d}{2}-1} \times 2^{\frac{d}{2}-1}$ matrix.
We define the $q \times q$ matrices 
\beqs
\label{eq:frm_d_mat_b}
B & = & \left( r d + 1 - m_0 \right) \delta_{x,x^\prime}
- \frac{r}{2} \sum_{\mu=1}^d \left[
  U_\mu(x) \delta_{x+\hat\mu,x^\prime}
  + U_\mu^\dagger(x-\hat\mu) \delta_{x-\hat\mu,x^\prime}
\right], \\
\label{eq:frm_d_mat_c}
C & = & \frac{1}{2} \sum_{\mu=1}^d \sigma_\mu \left[
  - U_\mu(x) \delta_{x+\hat\mu,x^\prime}
  + U_\mu^\dagger(x-\hat\mu) \delta_{x-\hat\mu,x^\prime}
\right].
\eeqs
where $q=2^{\frac{d}{2}-1} N_f N_c \Omega$ and $\Omega$ is the number
of lattice sites in the $d$ dimensional spacetime.  This block notation allows
us to write the matrix $D^\parallel$ of section \ref{sec:sym_dwf_action},
see (\ref{eq:frm_d_dslash_parallel}), in the compact form of a $2q \times 2q$
block matrix
\beq
D^\parallel = \left(
  \begin{array}{cc}
    B & -C \\
    C^\dagger & B
  \end{array}
\right)
\eeq
Continuing with the same strategy, the full domain wall fermion Dirac matrix
in (\ref{eq:frm_d_mat_2}) can be written as the block matrix
\beq
\label{eq:frm_d_mat_3}
M^{\it (dwf)} = \left(
  \begin{array}{cccc}
    \begin{array}{cc}
      B         & -C \\
      C^\dagger &  B
    \end{array} &
    \begin{array}{cc}
      0 &  0 \\
      0 & -1
    \end{array} &
    &
    \begin{array}{cc}
      m_f & 0 \\
      0   & 0
    \end{array} \\
    \begin{array}{cc}
      -1 & 0 \\
       0 & 0
    \end{array} &
    \ddots &
    \ddots &
    \\
    &
    \ddots &
    \ddots &
    \begin{array}{cc}
      0 &  0 \\
      0 & -1 
    \end{array} \\
    \begin{array}{cc}
      0 & 0   \\
      0 & m_f
    \end{array} &
    &
    \begin{array}{cc}
      -1 & 0 \\
       0 & 0
    \end{array} &
    \begin{array}{cc}
      B         & -C \\
      C^\dagger &  B
    \end{array}
  \end{array}
\right).
\eeq

It is a potential advantage of this block form that we can permute blocks
of rows and columns, leaving the determinant unchanged up to a sign.
If $q$ is even, which is always true if $d \ge 4$, then the sign is always
positive.  We choose the set of permutations to make $M^{\it (dwf)}$ as
block lower triangular as possible, leaving a single non-zero block in the
upper right corner.  This is done by swapping pairs of block columns and then
moving the first block-column on the left all the way to the right.  The new
matrix $M^{\prime{\it(dwf)}}$ in terms of $2q \times 2q$ blocks is
\beq
\label{eq:frm_d_mat_prime}
M^{\prime{\it(dwf)}} = \left(
  \begin{array}{cccc}
    \alpha_1 &        &                &  \beta_{L_s} \\
     \beta_1 & \ddots &                &              \\
             & \ddots & \alpha_{L_s-1} &              \\
             &        &  \beta_{L_s-1} & \alpha_{L_s}
  \end{array}
\right) 
\eeq
where the $\alpha_s$ and $\beta_s$ are block lower triangular
\beqs
\alpha \equiv \alpha_{1,\cdots,L_s-1} = \left(
  \begin{array}{cc}
    B         &  0 \\
    C^\dagger & -1
  \end{array}
\right), & & \alpha_{L_s} = \alpha \left(
  \begin{array}{cc}
    1 & 0 \\
    0 & -m_f
  \end{array}
\right) \\
\beta \equiv \beta_{1,\cdots,L_s-1} = \left(
  \begin{array}{cc}
    -1 & -C \\
     0 &  B
  \end{array}
\right), & & \beta_{L_s} = \left(
  \begin{array}{cc}
    -m_f & 0 \\
     0   & 1
  \end{array}
\right) \beta .
\eeqs

We use the technique suggested by Neuberger \cite{Neuberger:1998bg}
to simplify the determinant of the nearly lower triangular
$M^{\prime{\it(dwf)}}$.  First, we factor (\ref{eq:frm_d_mat_prime})
into the product of a lower triangular and an upper triangular matrix
\beq
M^{\prime{\it(dwf)}} = \left(
  \begin{array}{cccc}
    \alpha &        &        &      0       \\
     \beta & \ddots &        &              \\
           & \ddots & \alpha &              \\
           &        &  \beta & \alpha_{L_s}
  \end{array}
\right) \left(
  \begin{array}{cccc}
    1 &        &   & -v_1        \\
      & \ddots &   & \vdots      \\
      &        & 1 & -v_{L_s-1}  \\
      &        &   & (1-v_{L_s})
  \end{array}
\right)
\eeq
where the $v_s$ are given by
\beqs
& & v_s = \left( -\alpha^{-1} \beta \right)^s \left(
  \begin{array}{cc}
    -m_f & 0 \\
     0   & 1
  \end{array}
\right), \nonumber \\*
& & v_{L_s} = \left(
  \begin{array}{cc}
    1 &  0     \\
    0 & -1/m_f
  \end{array}
\right)  \left( -\alpha^{-1} \beta \right)^{L_s} \left(
  \begin{array}{cc}
    -m_f & 0 \\
     0   & 1
  \end{array}
\right).
\eeqs
As $\alpha$ is block lower triangular, it is straightforward to compute
it's inverse in block form.  Now the determinant is
\beqs
\label{eq:det_frm_d_mat_1}
& \det M^{\it(dwf)} & = \left( \det\alpha \right)^{L_s-1} \det\alpha_{L_s}
\det \left( 1 - v_{L_s} \right) \\*
& & = (-1)^{q L_s} \left( \det B \right)^{L_s} \det \left[
  \left(
    \begin{array}{cc}
      -m_f & 0 \\
       0   & 1
    \end{array}
  \right)
  - T^{-L_s}
  \left(
    \begin{array}{cc}
      1 &  0   \\
      0 & -m_f
    \end{array}
  \right)
\right]. \nonumber
\eeqs
where it simplifies what is to follow if we define the $2q \times 2q$ matrix
$T$ and its inverse
\beqs
\label{eq:dwf_xfer_mat}
& T \equiv \left(
  \begin{array}{cc}
    B^{-1} & -B^{-1} C^\dagger \\
    -C B^{-1} & C B^{-1} C^\dagger + B
  \end{array}
\right) & \\
\label{eq:dwf_xfer_mat_inv}
& T^{-1} = \left(
  \begin{array}{cc}
    C^\dagger B^{-1} C + B & C^\dagger B^{-1} \\
    B^{-1} C & B^{-1}
  \end{array}
\right) . &
\eeqs
At this point we have already made substantial progress
since the dimensionality of the matrices on the right hand side of
(\ref{eq:det_frm_d_mat_1}) no longer depend on $L_s$.

However, to fully expose the $L_s \to \infty$ limit, we should find
a way to raise $T^{-1}$ to an arbitrarily large power.  Let $\openone$
be the $2q \times 2q$ identity matrix and $\Gamma_5$ be the $2q \times 2q$
block matrix
\beq
\Gamma_5 \equiv \left(
  \begin{array}{cc}
    1 &  0   \\
    0 & -1
  \end{array}
\right) .
\eeq
Using these matrices and the identities
\beqs
\left(
  \begin{array}{cc}
    1 &  0   \\
    0 & -m_f
  \end{array}
\right) & = & \frac{1}{2} \left[
  (1 - m_f) \openone + (1 + m_f) \Gamma_5
\right] \\
\left(
  \begin{array}{cc}
    -m_f & 0 \\
     0   & 1
  \end{array}
\right) & = & \frac{1}{2} \left[
  (1 - m_f) \openone - (1 + m_f) \Gamma_5
\right]
\eeqs
we can bring $\det M^{\it(dwf)}$ into the form
\beqs
\label{eq:det_frm_d_mat_2}
\det M^{\it(dwf)} & = & (-1)^{q L_s} \left( \det B \right)^{L_s}
\det \left( \openone + T^{-L_s} \right) \det \Gamma_5 \non \\*
& & \times \det \frac{1}{2} \left[
  - ( 1 + m_f ) \openone
  - ( 1 - m_f ) \Gamma_5 \tanh \left( -\frac{L_s}{2} \log T \right)
\right] \non \\*
& = & (-1)^{q \left( L_s + 1 \right)} \left( \det B \right)^{L_s}
\det \left( \openone + e^{L_s H} \right) \non \\*
& & \times \det \left[
  \frac{ (1 + m_f) \openone + (1 - m_f) \Gamma_5 \tanh(\frac{L_s}{2}H) }{2}
\right] .
\eeqs
where in the second equation we have used $\det \Gamma_5 = (-1)^q$ and
defined the $2q \times 2q$ matrix $H \equiv -\log{T}$.

Finally, we see the $L_s \to \infty$ limit of (\ref{eq:det_frm_d_mat_2})
is divergent from the factors $\left( \det B \right)^{L_s}$ and
$\det \left( \openone + e^{L_s H} \right)$.  This should not come as a surprise
if we recall that domain wall fermions can be interpreted as a theory of $L_s$
heavy Wilson fermions that mix to create low energy chirally symmetric massless
modes.  We should remove this bulk divergence if our partition function is
to correctly model a single light fermion flavor.

A simple method to remove the contribution of the heavy modes was proposed
by Vranas \cite{Vranas:1997tj,Vranas:1997da}.  For $m_f = 1$ the last
determinant factor in (\ref{eq:det_frm_d_mat_2}) is unity. So, if we divide
the partition function for arbitrary $m_f$ by the partition function with
fixed $m_f$=1, we will cancel the bulk divergence and render finite the
$L_s \to \infty$ limit.  The technique used to place the determinant
into the denominator is to add Pauli-Villars regulator fields\footnote{
  Pauli-Villars fields, also called {\it pseudofermions,} commute like bosons
  yet have the same matrix spin structure as fermions.
} into the action with the same Dirac matrix as the fermion fields except
with fixed $m_f$=1.  In the path integral over the Pauli-Villars fields,
the integration is now Gaussian rather than Grassmannian, so the determinant
of the Pauli-Villars matrix will appear in the denominator.  Explicitly,
the proposed partition function is
\beqs
\label{eq:lat_part_fn_dwf}
Z^{\rm(dwf)} &=& \int \left[{\cal D}U\right] \left[{\cal D}\overline\Psi\right]
\left[{\cal D}\Psi\right] \left[{\cal D}\Phi_{PV}^\dagger\right]
\left[{\cal D}\Phi_{PV}\right]
e^{-\left(S_G + S_F^{\rm(dwf)} + S_{PV}^{\rm(dwf)}\right)} \nonumber \\*
&=& \int \left[{\cal D}U\right] \frac{\det M(m_f)}{\det M(m_f\!=\!1)} e^{-S_G}
\nonumber \\*
&=& \int \left[{\cal D}U\right] \det \left[
  \frac{ (1+m_f) \openone + (1-m_f) \Gamma_5 \tanh(\frac{L_s}{2}H) }{2}
\right] e^{-S_G}
\eeqs
where
\beq
\label{eq:frm_d_PV_action}
S_{PV}^{\rm (dwf)} = \Phi_{PV}^\dagger M^{\it(dwf)}(m_f\!=\!1) \Phi_{PV}
\eeq
is the Pauli-Villars action defined as in (\ref{eq:frm_d_action}).
As we shall see in chapter \ref{ch:numerical_implementation}, it is sometimes
convenient to also rewrite the fermionic path integral in terms
of pseudofermions as well
\beq
\label{eq:lat_part_fn_dwf_2}
Z^{\rm(dwf)} = \int \left[{\cal D}U\right] \left[{\cal D}\overline\Phi_F\right]
\left[{\cal D}\Phi_F\right] \left[{\cal D}\Phi_{PV}^\dagger\right]
\left[{\cal D}\Phi_{PV}\right]
e^{-\left(S_G + S_F^{\rm(dwf)} + S_{PV}^{\rm(dwf)}\right)}
\eeq
where the fermionic action is now written as
\beq
S_F^{\rm(dwf)} = \Phi_F^\dagger \left[ M^{\it(dwf)} \right]^{-1} \Phi_F .
\eeq

To take the $L_s \to \infty$ limit of (\ref{eq:lat_part_fn_dwf}), we recall
that the hyperbolic tangent function converges to the sign function
\beq
\lim_{n\to\infty} \tanh(nx) = \epsilon(x)
\equiv \left\{
  \begin{array}{rl}
     1, & x > 0 \\
    -1, & x < 0
  \end{array}
\right. .
\eeq
Then the fermion determinant for QCD with exact chiral symmetry is 
\beq
\label{eq:det_frm_o_old}
\left. \frac{\det M(m_f)}{\det M(m_f\!=\!1)} \right|_{L_s \to \infty}
= \det \left[
  \frac{ (1+m_f) \openone + (1-m_f) \Gamma_5 \epsilon(H) }{2}
\right] .
\eeq
So, we have achieved our goal of finding the contribution to the partition
function of domain wall fermions in the $L_s \to \infty$ limit.

As an aside, it is possible to construct a transfer matrix formulation of
domain wall fermions analogous to the Hamiltonian formulation of lattice
gauge theory.  The crucial difference is that the domain wall Hamiltonian
corresponds to evolution along the $s$-direction rather than in time.  Due
to the trivial gauge links in this direction, it is possible to construct
a Hamiltonian that is $s$-independent.  The matrix $T$ introduced in 
(\ref{eq:dwf_xfer_mat}) and its logarithm $H$ are the $s$-independent
transfer matrix and Hamiltonian of this formulation.  Finally, we note the
domain wall fermion transfer matrix formulation described here is just a 
specific case of the overlap formulation
\cite{Furman:1995ky,Narayanan:1994sk,Narayanan:1995gw}.

\section{Effects of finite domain wall separation}
\label{sec:finite_ls}

In the preceding section, we have described how domain wall fermions restore
chiral symmetry in the limit of infinite domain wall separation.  However,
for numerical simulations we must necessarily work at finite domain separation.
For physical quantities defined at large distances, we expect that the effect
of the chiral symmetry breaking induced by finite $L_s$ is to shift the bare
mass $m_f$ by some small additive piece $m_{\rm res}$, the residual mass.
By examining the non-singlet axial Ward-Takahashi identity, we hope to identify
and measure the residual mass in the confined phase of QCD.

The domain wall fermion action (\ref{eq:frm_d_action})
with degenerate quark masses is invariant under a global ${\rm SU}_V(N_f)$
symmetry.  Following Furman and Shamir \cite{Furman:1995ky}, we can define the
corresponding $d+1$ dimensional conserved non-singlet vector current as
\beqs
{\cal J}_{\mu=1,\cdots,d}^a(x,s) & = &
\overline\psi_{x,s} \left(\frac{1-\gamma_\mu}{2}\right) U_\mu(x)
  \frac{\lambda^a}{2} \psi_{x+\hat\mu,s} \\*
& & - \overline\psi_{x+\hat\mu,s} \left(\frac{1+\gamma_\mu}{2}\right)
  U_\mu^\dagger(x) \frac{\lambda^a}{2} \psi_{x,s} \non \\
{\cal J}_{d+1}^a(x,s) & = & \left\{
  \begin{array}{lc}
    \overline\psi_{x,s} P_- \frac{\lambda^a}{2} \psi_{x,s+1}
    - \overline\psi_{x,s+1} P_+ \frac{\lambda^a}{2} \psi_{x,s} , &
    0 \le s < L_s-1 \\*
    \overline\psi_{x,L_s-1} P_- \frac{\lambda^a}{2} \psi_{x,0}
    - \overline\psi_{x,0} P_+ \frac{\lambda^a}{2} \psi_{x,L_s-1} , &
    s = L_s-1
  \end{array}
\right. .
\eeqs
We can see immediately this current is conserved as its divergence vanishes
\beq
\label{eq:dwf_vc_divergence}
\sum_{\mu=1}^{d+1} \partial_\mu {\cal J}_\mu^a(x,s) = 0.
\eeq
However, as the low energy effective theory is $d$ dimensional QCD,
we can also define a conserved $d$ dimensional vector current.
Towards this end, we move the terms in (\ref{eq:dwf_vc_divergence}) for
$\mu=d+1$ to the right hand side to define the continuity equation
\beq
\label{eq:dwf_continuity}
\sum_{\mu=1}^{d} \partial_\mu {\cal J}_\mu^a(x,s) = \left\{
  \begin{array}{lc}
    -{\cal J}_{d+1}^a(x,0) - m_f {\cal J}_{d+1}^a(x,L_s-1) , &
    s = 0 \\*
    -{\cal J}_{d+1}^a(x,s) + {\cal J}_{d+1}^a(x,s-1) , &
    0 < s < L_s-1 \\*
    m_f {\cal J}_{d+1}^a(x,L_s-1) + {\cal J}_{d+1}^a(x,L_s-2) , &
    s = L_s-1
  \end{array}
\right. ,
\eeq
where
\(
\partial_\mu {\cal J}_\mu^a(x,s)
= {\cal J}_\mu^a(x,s) - {\cal J}_\mu^a(x-\hat\mu,s) 
\)
for $\mu = 1, \cdots, d$.  If we define the $d$ dimensional non-singlet vector
current as
\beq
\label{eq:dwf_vc}
{\cal V}_\mu^a(x) = \sum_{s=0}^{L_s-1} {\cal J}_\mu^a(x,s)
\eeq
then we can see from (\ref{eq:dwf_continuity}) that it is also divergenceless:
$\sum_{\mu=1}^d \partial_\mu {\cal V}_\mu^a(x) = 0$.
So, isospin is a symmetry of the low energy effective action of domain wall
fermions.

To proceed as above to construct a $d$ dimensional axial vector current is not
so straightforward as it cannot be uniquely defined.  Clearly, opposite chiral
rotations should be applied to the states on opposing walls.  Following Furman
and Shamir \cite{Furman:1995ky} we perform opposite chiral transformations
on each half of the fermions in the $s$-direction.  Analogous
to (\ref{eq:dwf_vc}), the corresponding $d$ dimensional non-singlet axial
vector current is
\beq
{\cal A}_\mu^a(x) = \sum_{s=0}^{L_s-1}
\epsilon \left( s - \frac{L_s - 1}{2} \right) {\cal J}_\mu^a(x,s)
\eeq
where our notation assumes, for simplicity, that $L_s$ is even.  Unlike
the vector current in (\ref{eq:dwf_vc_divergence}), the divergence of the axial
vector current has two contributions that do not trivially vanish
\beq
\sum_{\mu=1}^d \partial_\mu {\cal A}_\mu^a(x) =
2 m_f {\cal J}_{d+1}^a(x,L_s-1)
+ 2 {\cal J}_{d+1}^a(x,L_s/2) .
\eeq
When expressed in terms of the fields $q_x, \overline{q}_x$
of (\ref{eq:dwf_4d_fermions}), ${\cal J}_{d+1}^a(x,L_s-1)$ takes
the familiar form of the pseudoscalar density
\beq
{\cal J}_{d+1}^a(x,L_s-1) = \overline{q}_x \gamma_5 \frac{\lambda^a}{2} q_x .
\eeq
Thus, when $m_f \ne 0$ the first term reproduces the correct
contribution expected from a naive fermion mass term.
Although the second contribution, ${\cal J}_{d+1}^a(x,L_s/2)$,
does not vanish for finite $L_s$, we present numerical results in
section \ref{sec:GMOR_study} that indicate it is exponentially suppressed.

To better understand the consequences of working at finite $L_s$,
we will follow the role of ${\cal J}_{d+1}^a(x,L_s/2)$ as we derive
the Gell-Mann, Oakes, and Renner (GMOR) relation
\cite{Gell-Mann:1968rz} in the domain wall framework
\cite{Fleming:1998cc,Fleming:1999eq}.
We start with the axial Ward-Takahashi identity
\beqs
\label{eq:dwf_AWI}
\sum_{\mu=1}^d \partial_\mu \left\langle
  {\cal A}_\mu^a(x) ~ \overline{q}_0 \gamma_5 \frac{\lambda^a}{2} q_0
\right\rangle & = & 2 m_f \left\langle
  \overline{q}_x \gamma_5 \frac{\lambda^a}{2} q_x
  ~ \overline{q}_0 \gamma_5 \frac{\lambda^a}{2} q_0
\right\rangle \non \\*
& & + 2 \left\langle
  {\cal J}_{d+1}^a(x,L_s/2) ~ \overline{q}_0 \gamma_5 \frac{\lambda^a}{2} q_0
\right\rangle \non \\*
& & - \delta_{x,0} \qbqo
\eeqs
where the last term comes from the derivative of the Heavyside
function used for time ordering and the zero subscript of $\qbqo$
indicates the contribution to the chiral condensate from a single point,
analogous to setting $x$=0 in (\ref{eq:cont_qbq}) rather than averaging
over the whole spacetime volume.  The left hand side of (\ref{eq:dwf_AWI})
will vanish when summed over the whole lattice,
leaving us with the GMOR relation
\beq
\label{eq:dwf_GMOR}
\qbqo = m_f \chi_{\pi,0} + \Delta{\cal J}_5
\eeq
where $\chi_\pi$ is the pion (or pseudoscalar) susceptibility
\beq
\label{eq:sc_pi_d_1}
\chi_{\pi,0} \equiv - \frac{2}{4 N_c} \sum_x \left\langle
  \overline{q}_x \gamma_5 \frac{\lambda^a}{2} q_x
  ~ \overline{q}_0 \gamma_5 \frac{\lambda^a}{2} q_0
\right\rangle
\eeq
and $\Delta{\cal J}_5$ is the mid-plane contribution due to finite $L_s$
\beq
\label{eq:DeltaJ5}
\Delta{\cal J}_5 \equiv - \frac{2}{4 N_c} \sum_x \left\langle
  {\cal J}_{d+1}^a(x,L_s/2) ~ \overline{q}_0 \gamma_5 \frac{\lambda^a}{2} q_0
\right\rangle .
\eeq
As before, the zero subscript $\chi_{\pi,0}$ in (\ref{eq:sc_pi_d_1}) means this
quantity is computed only at a single lattice point, analogous to setting $x$=0
in (\ref{eq:cont_chi_pi_1}).

In the chirally broken phase, both susceptibilities should be dominated
in the massless limit by the light Goldstone degrees of freedom localized
on the walls.  As ${\cal J}_{d+1}^a(x,L_s/2)$ is located midway between
the two boundaries, its overlap with the boundary states should be
exponentially suppressed.  So, assuming an infinite lattice volume,
we will model $\chi_{\pi,0}$ as
\beq
\label{eq:chi_pi_param}
\chi_{\pi,0} = \frac{a_{-1}}{m_f + m_{\rm res}} + a_0 + {\cal O}(m_f)
\eeq
and $\Delta{\cal J}_5$ will also have a pole when $m_f\!=\!-m_{\rm res}$
although we expect that the $c_i$ will be exponentially suppressed in $L_s$
\beq
\label{eq:DeltaJ5_param}
\Delta{\cal J}_5 = \frac{c_{-1}}{m_f + m_{\rm res}} + c_0 + {\cal O}(m_f) .
\eeq
Thus, $m_\pi^2 \to 0$ as $m_f \to - m_{\rm res}$.  We also note that
the parameters used here are assumed only to be independent of
$m_f$ and can vary with $m_0$ and $L_s$ but we imagine performing
the following derivation while holding $m_0$ and $L_s$ fixed.

We write the GMOR relation as
\beq
\label{eq:GMOR_mass_1}
\frac{\qbqo}{\chi_{\pi,0}} = m_f + \frac{\Delta{\cal J}_5}{\chi_{\pi,0}}
\eeq
and substitute our parameterizations (\ref{eq:chi_pi_param}) and
(\ref{eq:DeltaJ5_param}) in the right hand side
\beq
\label{eq:GMOR_mass_2}
\frac{\qbqo}{\chi_{\pi,0}}
= m_f + \frac{c_{-1}+c_0(m_f+m_{\rm res})}{a_{-1}+a_0(m_f+m_{\rm res})} .
\eeq
As we tune $m_f$ to the pion pole, {\it i.e.\ } $m_f \to -m_{\rm res}$,
the left hand side of (\ref{eq:GMOR_mass_2}) will vanish.  On the right hand
side, this requires $m_{\rm res} = c_{-1} / a_{-1}$ which allows us
to eliminate $c_{-1}$ as a free parameter in (\ref{eq:GMOR_mass_2}).

In the continuum identity the slope of $\qbqo / \chi_{\pi,0}$ versus the bare
quark mass is always one.  However, for domain wall fermions as
$m_f \to - m_{\rm res}$ at fixed $m_0$ and $L_s$
\beq
\left.
  \frac{\partial}{\partial m_f} \left(\frac{\qbqo}{\chi_{\pi,0}}\right)
\right|_{m_f=-m_{\rm res}, m_0, L_s} = 1 + \frac{c_0 - a_0 m_{\rm res}}{a_{-1}}
\eeq
so the slope is never one except when $c_0 = a_0 m_{\rm res}$.
To clarify this situation, let us eliminate the parameter $c_0$
in favor of $b_0 = c_0 - a_0 m_{\rm res}$, the parameter which,
when non-zero, indicates the slope of $\qbqo / \chi_{\pi,0}$ is not unity
as $m_f \to -m_{\rm res}$.
In this new parameterization, we see that (\ref{eq:DeltaJ5_param})
becomes
\beq
\label{eq:DeltaJ5_new_param}
\Delta{\cal J}_5 = m_{\rm res} \chi_{\pi,0} + b_0
\eeq
and the GMOR relation (\ref{eq:GMOR_mass_1}) becomes implicitly
\beq
\label{eq:qbq_o_spc_implicit}
\qbqo = \left( m_f + m_{\rm res} \right) \chi_{\pi,0} + b_0
\eeq
and explicitly
\beq
\label{eq:qbq_o_spc_ansatz}
\frac{\qbqo}{\chi_{\pi,0}} = \left( m_f + m_{\rm res} \right) \left[
 1 + \frac{b_0}{a_{-1} + a_0 \left( m_f + m_{\rm res} \right)}
\right] .
\eeq

Finally, we recall that the chiral condensate receives contributions from
chiral symmetry breaking not just at large distances but from all energy
scales up to the lattice cutoff.  We should expect that chiral symmetry
violations of the GMOR relation due to lattice artifacts will appear in a form
that cannot be interpreted as an additional residual mass in the low energy
effective Lagrangian.  We conjecture that the non-zero $b_0$ is precisely
the cumulative contribution due to the chiral symmetry breaking of the heavy
modes at the lattice cutoff.  As such, we could absorb the heavy contribution
into a new definition of the chiral condensate, which we call
the ``subtracted'' chiral condensate
\beq
\qbqo_{\rm sub} = \qbqo - b_0 .
\eeq
One potential advantage of including this contribution in the chiral
condensate is that it guarantees the slope of the lattice spectral identity
will be unity when the spectrum becomes chiral as $m_f \to - m_{\rm res}$
at fixed $m_0$ and $L_s$
\beq
\left.
  \frac{\partial}{\partial m_f}
  \left(\frac{\qbqo_{\rm sub}}{\chi_{\pi,0}}\right)
\right|_{m_0,L_s} = 1 .
\eeq

In chapter \ref{ch:numerical_simulations}, we will use the approach of this
section to estimate the residual mass for our simulations near the critical
temperature of QCD.  For a given domain wall separation, we would like
to establish whether the dominant contribution to the effective quark mass
comes from the bare input mass, $m_f$, or the residual mixing of the chiral
surface states, $m_{\rm res}$.  Ideally, to ensure good control over the chiral
limit, we would prefer to choose $L_s$ large enough so $m_{\rm res} \ll m_f$.
Only then can we be confident that any observed critical behavior, or lack
thereof, is a consequence of the chiral symmetry of the QCD Lagrangian.

\section{The QCD finite temperature phase transition}
\label{sec:phase_transition_discussion}

An outstanding success of lattice QCD has been the demonstration of confinement
and spontaneously broken chiral symmetry at low temperatures and the existence
of a phase transition above some critical temperature, $T_c$.  A current goal
of lattice theorists is to calculate the physical value for $T_c$, as well as
other physically observable consequences of the transition, for comparison with
anticipated experimental results from the Relativistic Heavy Ion Collider
(RHIC) and other experiments designed to study the thermodynamics of QCD.

In either the quenched limit with infinitely massive quarks or the chiral limit
with massless quarks, the finite temperature phase transition is characterized
by the behavior of an observable which is not invariant under some symmetry
of the action.  In the symmetric phase, the observable has a vanishing
expectation value because the ground state of the system respects the symmetry
of the action.  In the broken phase, the observable has, in general,
a non-vanishing expectation value because the ground state of the system
no longer respects the symmetry.  The observable is an order parameter
for the transition.

For quenched QCD, the gauge actions (\ref{eq:wilson_gauge_action}) and
(\ref{eq:rect_gauge_action}) are globally symmetric under a multiplication
of all time-like gauge links originating from a given space-like hyperplane
by the same element of the center of the gauge group.  For ${\rm SU}(N_c)$,
the center is $Z_{N_c}$.  However, a closed loop of link variables
with non-trivial winding around the temporal direction of the lattice,
commonly called a Wilson (or Polyakov) line,
\beq
W_R({\bf x}) = \frac{1}{\dim(R)} {\rm tr} \left[
  \prod_{x_4=0}^{L_4-1} U_4({\bf x},x_4)
\right]
\eeq
is not invariant under the transformations described above since it involves
only one time-like link originating from each space-like hyperplane.
The Wilson line observable measures the free energy of a isolated static
fermion which couples to the gauge fields in representation $R$
of the gauge group
\cite{Polyakov:1978vu,Susskind:1979up,McLerran:1981pk,Kuti:1981gh}.
In the low temperature phase of quenched QCD, the free energy should be
infinite and the expectation value of the Wilson line zero.  Hence, the Wilson
line is the order parameter for the quenched QCD phase transition and
the symmetric phase is at low temperature.

Lattice QCD studies have firmly established that the quenched transition is
a first order phase transition. This means that a first derivative of the log
of the partition function is discontinuous in the transition region.
This discontinuity in the energy density is called the latent heat and is
the measure of the amount of energy which must be put into the symmetric state
to drive it into the broken state.

For QCD with dynamical fermions, the Wilson line is no longer an order
parameter because the global $Z_{N_c}$ symmetry of the gauge sector is not
a symmetry of the fermion action.  The physical interpretation is that once
dynamical fermion-antifermion pairs are present in the vacuum, the static
fermion can be screened by the vacuum, so the free energy must be finite.
In the low temperature phase, since one doesn't expect to see isolated fermions
the Wilson line expectation value should still be small.  Thus, the Wilson
line should undergo rapid evolution through the transition and can serve as
an approximate order parameter.

In the limit that the $N_f \ge 2$ dynamical fermions of QCD become massless,
the appearance of the ${\rm SU}_L(N_f) \otimes {\rm SU}_R(N_f)$ global chiral
symmetry discussed in sections \ref{sec:continuum_QCD} and \ref{sec:finite_ls}
suggests the possibility of a new order parameter for the transition.
The chiral condensate (\ref{eq:cont_qbq}) is not invariant under chiral
transformations.  So, if the ground state of the system is invariant under
chiral transformations then the expectation value for the chiral condensate
must vanish.  However, if the chiral symmetry of the action is spontaneously
broken then the chiral condensate will, in general, have a non-zero expectation
value.  Contrary to the quenched theory, the symmetric phase for the massless
$N_f \ge 2$ theory occurs at high temperature.

The order of the transition depends strongly on the number of massless
quark flavors.  For $N_f \ge 3$, renormalization group analysis indicates
there are no stable infrared fixed points, a necessary condition for 
the transition to be second order, {\it i.~e.}\ the second derivative
of the log of the partition function is discontinuous in the critical region
\cite{Paterson:1981fc,Pisarski:1984ms}.  Numerical lattice QCD simulations
confirm this conclusion as first order transitions have been observed
in the massless limit of QCD with $N_f$=3,4.  For $N_f$=2, the order
of the transition may be determined by the fate of the anomalously broken
${\rm U}_A(1)$ symmetry in the high temperature phase.  If the ${\rm U}_A(1)$
symmetry remains broken while chiral symmetry is restored, then $N_f$=2 QCD
is expected to be in the ${\rm O}(4)$ universality class, and models
in that class have second order phase transitions \cite{Svetitsky:1986ye}.
However, if ${\rm U}_A(1)$ symmetry is effectively restored
in the finite temperature phase transition then the universality class
changes to ${\rm O}(2)\otimes{\rm O}(4)$ and, as other models in that class
are known to have first order transitions, it is possible that the finite
temperature transition for $N_f$=2 QCD will be first order
\cite{Pisarski:1984ms,Wilczek:1992sf,Rajagopal:1993qz}.

Earlier attempts to address this question using staggered fermions
\cite{Chandrasekharan:1996ci,Gottlieb:1997ae,Bernard:1997iz,Kogut:1998rh,
      Chandrasekharan:1999yx} did not produce conclusive results.
While staggered fermions have an exact
${\rm U}_V(1)\otimes{\rm U}_{A^\prime}(1)$ chiral symmetry,
the ${\rm U}_{A^\prime}(1)$ is actually an unbroken subgroup
of the flavor non-singlet chiral symmetry and not the ${\rm U}_A(1)$
of the continuum.  At finite lattice spacing, the flavor singlet ${\rm U}_A(1)$
is actually broken at the classical level and it becomes quite
difficult to disentangle the signal of anomalous symmetry breaking
from the explicit breaking due to lattice artifacts.

As $L_s\to\infty$, ${\rm U}_A(1)$ is a symmetry of the classical massless
domain wall action.  The domain wall Dirac operator can have exact zero modes
\cite{Narayanan:1995gw} and a well-defined integer index, even for gauge fields
at strong coupling.  We demonstrated this numerically in earlier studies
\cite{Chen:1999ne,Chen:1999kg,Kaehler:1999} by observing the effects
of fermionic zero modes produced by smooth topological gauge configurations
for small bare quark masses and for $L_s\gtrsim 10$.  Further, these effects
persisted even after the addition of a moderate amount of Gaussian noise
to the classical gauge fields.  From this we conclude that domain wall fermions
are an ideal formulation to study the ${\rm U}_A(1)$ problem.  In preliminary
studies \cite{Vranas:1998vm,Christ:1998,Fleming:2000xe,Vranas:1999pm,
              Vranas:1999rz}, we have strong evidence that ${\rm U}_A(1)$
remains broken in the high temperature phase, but by such a small amount
that it may be possible that the breaking is negligible so that the symmetry
is effectively restored.  It is a primary goal of this work to address whether
numerical lattice simulations of $N_f$=2 QCD with domain wall fermions give
evidence of a first or second order transition in the massless limit.

In the real world, the fermions of QCD are neither massless nor infinitely
massive, so both the Wilson line and the chiral condensate serve only as
approximate order parameters.  As such, we expect that the values of these
observables will undergo a rapid change as we adjust the bare parameters
of the theory to move through the transition region.  However, the width
of the transition region is an important phenomenological number and can
be directly computed on the lattice.

In sections \ref{sec:beta_crit_8nt4} and \ref{sec:beta_crit_16nt4},
we will measure the Wilson line and chiral condensate in lattice simulations
of $N_f$=2 QCD with two light degenerate flavors of domain wall fermions.
We can locate the region of the finite temperature transition by searching
for simultaneous rapid changes in both approximate order parameters
as we vary the temperature.  Once we have located the critical temperature,
we will study the behavior of the chiral condensate for small quark masses
in the hope that we will see some indication of the order
of the phase transition in the chiral limit.

\clearpage


\chapter{Numerical implementation of QCD with domain wall fermions}
\label{ch:numerical_implementation}

\thispagestyle{myheadings}
\markright{}

\section{General considerations}
\label{sec:num_genl_consid}

A naive approach to numerical simulation of lattice QCD follows from our
discussion in sections \ref{sec:continuum_QCD}
and \ref{sec:lattice_gauge_actions}, particularly equations
(\ref{eq:cont_expect_val}) and (\ref{eq:lat_part_fn_1}), regarding
the computation of expectation values.  To evaluate the lattice partition
function on a computer, we approximate the continuous multidimensional
path integral over an infinite lattice volume as a sum over finite volume
lattice configurations with a finite precision representation of real numbers
\beq
\label{eq:lat_part_fn_3}
Z \to \sum_{\{U\}} e^{-S(U)}, \qquad
\left\langle {\cal O}(U) \right\rangle \to
\frac{1}{Z} \sum_{\{U\}} {\cal O}(U) e^{-S(U)}.
\eeq
Each $U$ is one possible lattice configuration and $\{U\}$ is the set of all
representable configurations.  After fixing the lattice size and precision,
only a countable number of configurations can be represented in the memory
of a computer.  So, to evaluate the expectation value
in (\ref{eq:lat_part_fn_3}), it is possible to program a computer to generate,
one by one, each configuration $U$ that can be represented, compute
the observable ${\cal O}(U)$ and the Boltzmann weight $e^{-S(U)}$
on each configuration and do the weighted average indicated.

To see this approach as completely intractable, consider a simple Ising model:
on each lattice site is a single degree of freedom which can take on one of two
possible values, which we can represent as a single bit in the memory of our
computer.  If $V$ is the total number of sites in a given lattice volume,
then the number of representable configurations that our program must evaluate
is $2^V$.  The time required to compute any expectation value grows
exponentially in the size of the configuration.  For QCD, the only difference
is the counting of representable degrees of freedom per lattice site once the
number of bits of memory per lattice site is fixed.  So, this method
is intractable for all but the smallest lattice volumes.

We recall from the statistical mechanics of systems with many degrees
of freedom, the Boltzmann distribution in configuration space is strongly
peaked around the minimum of the action.  This implies that the naive method
above is incredibly inefficient since most of the configurations generated will
have vanishingly small weight for expectation values.  To surmount this problem,
we use {\it importance sampling}, which creates a sample ensemble
of configurations drawn from the set of all representable configurations
with a probability given by the Boltzmann weight.  Configurations with large
weights will appear more frequently in our sample.  The number
of configurations in the sample should be large enough that expectation values
computed on the sample agree, within a given precision, with ones computed
on the full set.  A computer program could generate configurations at random
and compute the weight and expectation values.  However, this approach is
no better than the naive method above since most configurations will be
very unlikely to appear in a sample ensemble.  So, an importance sampling
method should provide us with an algorithm which only generates configurations
with a good chance to be added to the sample.

A classic example of an importance sampling method is the Metropolis
algorithm.  The algorithm has two parts: the {\it update process} and the
{\it acceptance test}.  The initial condition for the algorithm is to start
at any point in the configuration space, let's call it $U_1$.  The algorithm
then starts by computing the Boltzmann weight, $w_1$,
of the initial configuration and adding it to the sample ensemble.  Next, we
use the update process to generate a second configuration with known
probability $P_u\left(U_2 \leftarrow U_1\right)$ and compute its weight, $w_2$.
Then we use the acceptance test to decide if the second configuration should be
accepted into the sample with probability $P_a\left(U_2 \leftarrow U_1\right)$:
\beq
\label{eq:metropolis_acceptance}
P_a\left(U_2 \leftarrow U_1\right) =
\min \left\{
  1, \frac{
    P_u\left(U_1 \leftarrow U_2\right) w_2
  }{
    P_u\left(U_2 \leftarrow U_1\right) w_1
  }
\right\}
\eeq
The total probability that the second configuration is accepted given the
first configuration is $ P\left(U_2 \leftarrow U_1\right) =
P_a\left(U_2 \leftarrow U_1\right) P_u\left(U_2 \leftarrow U_1\right)$.
We then iterate the algorithm by applying the update process and acceptance
test to the last configuration accepted until the sample ensemble
is sufficiently large.

Two important issues must be addressed when designing update processes
for the Metropolis algorithm.  The first is whether the algorithm can
eventually construct a sample ensemble and the second is how efficiently
the algorithm can construct the ensemble.  The first issue can be resolved
if we establish two properties.  The property called {\it detailed balance},
which guarantees that the Boltzmann distribution of our sample ensemble
will converge to the distribution of the full set, requires 
\beq
\label{eq:detailed_balance}
P\left(U_j \leftarrow U_i\right) w_i = P\left(U_i \leftarrow U_j\right) w_j .
\eeq
From the preceding paragraph, it is straightforward to verify that the
Metropolis algorithm satisfies detailed balance.  The second property,
called {\it ergodicity}, requires that the update process of the Metropolis
algorithm can reach any point in the configuration space with finite
probability in a finite number of steps.  This property ensures that no
important regions of configuration space are left unsampled provided
the algorithm is run long enough.  However, it is typically very difficult
to determine analytically the efficiency of a given algorithm to construct
an ergodic ensemble.  In such cases, one attempts to address this issue through
numerical studies.

For pure gauge theories on the lattice, a typical update process is to
choose a single gauge link, propose a change to that link and compute the
difference in the action, $\Delta S$, between the two configurations.
Since the gauge action is local, $\Delta S$ is inexpensive to compute.
However, to maintain efficiency, some care must be taken in the update process
to make sure that the proposed change to the single link is still sufficiently
small to allow a reasonable chance of passing the acceptance test.  Performing
a Metropolis step at each link in the lattice once in succession is commonly
called a {\it sweep}.  One potential disadvantage of this approach is that two
configurations in the update sequence separated by only a few sweeps will
still be related, or ``nearby'' in configuration space.  So, many sweeps
are typically required to evolve the sequence between ``distant'' regions
of configuration space.

As discussed in section \ref{sec:large_ls_limit}, in lattice QCD
with dynamical fermions, the Grassmann degrees of freedom are integrated out
analytically, leaving a partition function expressed in only in terms
of gauge degrees of freedom
\beq
\label{eq:lat_part_fn_dwf_3}
Z = \int \left[{\cal D}U\right]
e^{- S_G + {\rm Tr} \log M_F - {\rm Tr} \log M_{PV}}
\eeq
or, equivalently, in terms of gauge and pseudofermion degrees of freedom
\beq
\label{eq:lat_part_fn_dwf_4}
Z=\int\left[{\cal D}U\right]
\left[{\cal D}\Phi_F^\dagger\right]\left[{\cal D}\Phi_F\right]
\left[{\cal D}\Phi_{PV}^\dagger\right]\left[{\cal D}\Phi_{PV}\right]
e^{-S_G-\Phi_F^\dagger M_F^{-1}\Phi_F-\Phi_{PV}^\dagger M_{PV}\Phi_{PV}} .
\eeq
Local update processes will no longer be efficient in either case
since the calculation of $\Delta S_{\rm eff}$ is no longer local.
While solving the sparse linear system $M_F \chi_F = \Phi_F$ is not
as computationally demanding as the full diagonalization normally required
to compute ${\rm Tr} \log M$, both are essentially non-local computations.
So, a successful fermionic update process should simultaneously update
a large number of degrees of freedom for a given calculation of the solution
to $M_F \chi_F = \Phi_F$ without a corresponding large change in the action.

One popular algorithm that accomplishes this is the hybrid Monte Carlo
algorithm.  The strategy is to treat the gauge degrees of freedom
as generalized coordinates in a Hamiltonian framework of one higher dimension
(not to be confused with the extra dimension of domain wall fermions).
We introduce conjugate momenta $H_\mu(x)$, which appear in the extended path
integral as a trivial bosonic field
\beq
\label{eq:lat_part_fn_Ham}
Z = \int \left[ {\cal D} U \right] \left[ {\cal D} H \right] e^{-\cal H}
= \int \left[ {\cal D} U \right] e^{-S_{\rm eff}}
\eeq
where the Hamiltonian
\beq
\label{eq:HMD_Hamiltonian}
{\cal H} =  S_{\rm eff}
+ \sum_{x,\mu}{\rm tr}\left[ \frac{1}{2} H_\mu(x) H_\mu(x)\right]
\eeq
now serves as the Boltzmann weight for the extended system $(U,H)$.
The update process for the momenta merely requires the generation of a Gaussian
distributed field, which is always accepted.  Hamilton's equations of motion
are then used to evolve the system along the classical trajectory.  Provided
the computational cost of the classical evolution is not prohibitive,
the result in a simultaneous update of all the degrees of freedom
between calculations of the (hopefully small) change in the Boltzmann
weight.

\section{The hybrid Monte Carlo algorithm}
\label{sec:hmc_algorithm}

The {\it hybrid Monte Carlo} (HMC) algorithm combines a non-local update
process called {\it hybrid molecular dynamics} (HMD) with the Metropolis
acceptance test described in section \ref{sec:num_genl_consid}
\cite{Gottlieb:1987mq}.  The full path integral is rewritten in terms
of the pseudofermionic fields $\Phi_F$ and $\Phi_{PV}$
as in (\ref{eq:lat_part_fn_dwf_4}).  The algorithm starts with the generation
of Gaussian distributed momenta $H_\mu(x)$ and the computation
of their contribution to the initial Hamiltonian.  For the fermionic
and Pauli-Villars contributions, pseudofermion fields $\Phi_F$ and $\Phi_{PV}$
are generated for the gauge field configuration using a bosonic heat bath
algorithm described below.  Then, using Hamilton's equations of motion,
the extended system $(U, H)$ is evolved in the presence of the forces
due to the fixed pseudofermion background along the classical trajectory
for some finite interval of the microcanonical evolution parameter, $\tau$.
Finally, the Hamiltonian is recomputed at the end of the trajectory
and the change between endpoints $\Delta{\cal H}$ is used as input
to a Metropolis acceptance test, which makes the algorithm exact even though
the numerical integration only approximately conserves the Hamiltonian.
The result is a simultaneous update of all dynamical degrees of freedom
between each computation of the Boltzmann weight.

For lattice QCD, after formally integrating out the fermionic
(and pseudofermionic) variables, the only degrees of freedom are
the gauge fields.  The conjugate momenta give the microcanonical evolution
of the gauge fields through Hamilton's first equation of motion
\beq
\label{eq:1st_Ham_EOM}
\frac{\rm d}{{\rm d}\tau} U_\mu(x) =
i H_\mu(x) U_\mu(x) , \qquad \mbox{(no sum over $\mu$)} .
\eeq
The restriction that the gauge links $U_\mu(x)$ remain in the gauge group
${\rm SU}(N_c)$ during the evolution constrains the momenta $H_\mu(x)$
to be elements of the Lie algebra of the gauge group, hence traceless
and Hermitian.  Expectation values will be unaffected by the shift
in normalization of the partition function (\ref{eq:lat_part_fn_Ham})
due to the inclusion of the Gaussian distributed $H_\mu(x)$.

Hamilton's second equation of motion is derived from the requirement that
the Hamiltonian is a constant of the motion
\beq
\label{eq:2nd_Ham_EOM_1}
\sum_{x,\mu}{\rm tr}\left[
  H_\mu(x) \frac{{\rm d}H_\mu(x)}{{\rm d}\tau}
\right] = - \frac{{\rm d}S_{\rm eff}}{{\rm d}\tau}.
\eeq
In section \ref{sec:hmd_force}, we will explicitly compute
${\rm d}S_{\rm eff}/{\rm d}\tau$ using (\ref{eq:1st_Ham_EOM}) and show it
can be written as
\beq
\label{eq:2nd_Ham_EOM_2}
\frac{{\rm d}S_{\rm eff}}{{\rm d}\tau}
= \sum_{x,\mu}{\rm tr}\left[ i  H_\mu(x) F_\mu(x) \right]
\eeq
where $F_\mu(x)$ is the called the {\it HMD force}.  The restriction
that the momenta $H_\mu(x)$ remain in the Lie algebra of the gauge group
during the evolution means that only the traceless anti-Hermitian part
of the force may influence the momenta.  Putting (\ref{eq:2nd_Ham_EOM_1})
and (\ref{eq:2nd_Ham_EOM_2}) together gives us Hamilton's second equation
\beq
\label{eq:2nd_Ham_EOM_3}
i \frac{{\rm d}H_\mu(x)}{{\rm d}\tau}
= \left[ F_\mu(x) \right]_{\rm TA}
\eeq
where the subscript ${\rm TA}$ means to take the traceless anti-Hermitian part.
In section \ref{sec:hmd_force}, we will show that each calculation of the
fermionic force will require solving the sparse linear system
$M_F^\dagger M_F \chi_F = \Phi_F$.

The HMD update process starts with the initial gauge configuration $U_i$ and
as mentioned above, the initial momenta $H_i$ and pseudofermions are generated.
This completely defines the initial state of the dynamical variables
$\left( U_i, H_i \right)$ in the background environment of $\Phi_F$
and $\Phi_{PV}$.  The gauge and momentum contributions
to the initial Hamiltonian ${\cal H}_i$ are computed directly.
The pseudofermions are produced by a bosonic heat bath, which consists
of generating random complex fields $R_F$ and $R_{PV}$ with a Gaussian
distribution proportional to $\exp(-R^*R)$.  Defining the pseudofermion fields
\beq
\label{eq:SetPhi}
\Phi_F = M_F^\dagger R_F, \qquad
\Phi_{PV} = \left( M_{PV}^{-1} \right)^\dagger R_{PV}
\eeq
will give probability distributions proportional to
$\exp\left[-\Phi_F\left(M_F^\dagger M_F\right)^{-1}\Phi_F\right]$ and
$\exp\left(-\Phi_{PV}M_{PV}^\dagger M_{PV}\Phi_{PV}\right)$.
Using the Hermitian matrices $M^\dagger M$ allow for the use of efficient
algorithms for solving sparse linear systems.  For both Wilson and domain wall
fermions, $\det M^\dagger M = \det M^2$ so the algorithm will generate
distributions for two degenerate flavors of fermions.  Note that the
computation of $\Phi_F$ is local whereas the computation of $\Phi_{PV}$
first requires solving the sparse linear system
$\left(M_{PV}^\dagger M_{PV}\right)^{-1}\Phi_{PV}^\prime=R_{PV}$
and then computing $\Phi_{PV}=M_{PV}\Phi_{PV}^\prime$.
The fermionic and Pauli-Villars contributions to the initial Hamiltonian
are then computed using the pseudofermion fields.

The next step in the update process is to evolve $(U_i,H_i)$ by integrating
both of Hamilton's equations simultaneously over some finite interval
of $\left[\tau_i,\tau_{i+1}\right]$ to arrive deterministically
at the final state
$\left( U_{i+1}, H_{i+1} \right) \leftarrow \left( U_i, H_i \right)$,
all the while holding fixed the pseudofermion fields.  To satisfy detailed
balance, the evolution must be {\it reversible}: starting from the endpoint
of the classical trajectory with the momenta reversed,
$\left( U_{i+1}, -H_{i+1} \right)$, and integrating over the interval
$\left[ \tau_{i+1}, \tau_i \right]$ returns us to the starting point,
$\left( U_i, -H_i \right) \leftarrow \left( U_{i+1}, -H_{i+1} \right)$.
While a continuous trajectory will be trivially reversible, on the computer
care must be taken to ensure that integration using a finite difference
approximation is also reversible. 

For the simulations in chapter \ref{ch:numerical_simulations}, the reversible
finite difference approximation we used was the {\it leap frog} method.  We
choose the trajectory to start at $\tau=0$ and end at a fixed $\tau$ and divide
the interval into $N$ steps of equal length step size $\Delta\tau$.  To start
the leap frog, we evolve the momenta one half time step using
\beq
i H_\mu(x,\Delta\tau/2) = i H_\mu(x,0) + i \frac{{\rm d}H_\mu(x,0)}{{\rm d}\tau}
\frac{\Delta\tau}{2} + {\cal O}(\Delta\tau^2).
\eeq
We then alternate evolving the gauge field a single time step
\beq
U_\mu(x,\tau+\Delta\tau) = e^{i H_\mu(x,\tau+\Delta\tau/2) \Delta\tau}
U_\mu(x,\tau) + {\cal O}(\Delta\tau^3)
\eeq
and the momenta a single time step
\beq
i H_\mu(x,\tau+\Delta\tau/2) = i H_\mu(x,\tau-\Delta\tau/2) + 
i \frac{{\rm d}H_\mu(x,\tau)}{{\rm d}\tau} \Delta\tau + {\cal O}(\Delta\tau^3).
\eeq
We finish the trajectory after evolving the gauge field $N$ times
with a final half step evolution of the momenta.  We emphasize that each time
we evolve the gauge fields we must solve again the linear system
$M_F^\dagger M_F \chi_F = \Phi_F$ and recompute the fermionic force needed
to evolve the momenta (see section \ref{sec:hmd_force}).

To ensure ergodicity, the number of steps must be of order $\Delta\tau^{-1}$,
so the difference between the initial and final Hamiltonians, $\Delta{\cal H}$,
will be of order $\Delta\tau^2$.  $\Delta{\cal H}$ is then used
in a Metropolis acceptance test at the end of the HMD trajectory
to compensate for finite $\Delta\tau$ errors.  As the Hamiltonian is
an extensive quantity, the step size must be decreased as the volume is
increased to maintain a given acceptance.  After the acceptance test,
the algorithm continues by generating new momenta and pseudofermion fields
and evolving along a new classical HMD trajectory.  So, as promised, the HMC
algorithm can make simultaneous acceptable changes to all the gauge degrees
of freedom and compute changes to the Boltzmann weight each time the fermionic
linear system is solved.  However, as the lattice volume grows, more
solutions of the fermionic system are required to produce independent
configurations.

\section{The preconditioned domain wall action}
\label{sec:dwf_evenodd}

Because of the central role of algorithms for solving sparse linear systems
in the evolution of fermionic configurations, we will reformulate the domain
wall fermion matrix into a form better suited for more efficient computation.
In general, the cost of solving linear systems is proportional to the ratio
of the largest eigenvalue $\lambda_{\rm max}$ to the smallest eigenvalue
$\lambda_{\rm min}$ of the matrix, called the {\it condition number}.
A technique, called {\it preconditioning}, seeks to transform the given
matrix (usually at some cost) into a new matrix with a smaller condition
number.  If the savings in solving the new system outweigh the cost
of transforming the results back into the old system, then the preconditioner
is successful.  For a general discussion on the importance of preconditioning
when solving linear systems, see Golub and Van Loan \cite{Golub:1996}.
For fermionic evolution algorithms, if the preconditioning transformation
preserves the determinant, then the cost of the transformation is irrelevant
because we can choose to work only with the new system.  In the case
of the HMC algorithm, preconditioning will help reduce the computational cost
by at least a factor of two.

The Wilson fermion matrix (\ref{eq:frm_w_mat}) contains entries which connect
only nearest neighbor sites.  We introduce a labeling scheme where
a lattice site is labeled even if all its nearest neighbor sites are labeled
odd and vice-versa.  For example, if $\sum_{\mu=1}^4 x_\mu {\rm mod} 2 = 0$
on a given site we label it even ($e$), otherwise we label it odd ($o$).
We then rearrange the components of the fermion vectors to list odd ones first
and write the matrix in block form
\beq
\label{eq:frm_w_mat_2}
M^{(W)} = \left(
  \begin{array}{cc}
    \openone_{oo}  & -\kappa D_{oe} \\
    -\kappa D_{eo} &  \openone_{ee}
  \end{array}
\right).
\eeq
Note that we have rescaled the fermion fields by $\sqrt\kappa$ where
\beq
\kappa = \frac{1}{2 (m_0 + d r)},
\eeq
which makes the on-diagonal blocks the identity.

DeGrand \cite{DeGrand:1988vx,DeGrand:1990dk} proposed
a preconditioning technique that replaces the Wilson fermion matrix $M^{(W)}$
by a preconditioned matrix $\widetilde{M}^{(W)}$ which preserves the determinant
$\det M^{(W)} = \det \widetilde{M}^{(W)}$.  He noted that $M^{(W)}$ could be
decomposed in block form
\beq
M^{(W)} = U \widetilde{M}^{(W)} L
\eeq
by choosing $U$ and $L$ to be the upper and lower block-triangular parts
of $M^{(W)}$.  Then the block-diagonal $\widetilde{M}^{(W)}$ is
\beq
\label{eq:frm_w_mat_precond}
\widetilde{M}^{(W)} = U^{-1} M^{(W)} L^{-1} = \left(
  \begin{array}{cc}
    \openone_{oo} - \kappa^2 D_{oe} D_{eo} & 0_{oe} \\
    0_{eo} & \openone_{ee}
  \end{array}
\right) .
\eeq
We can see that $\det U = \det L = 1$ so the transformation preserves
the determinant.

To see that this odd-even technique reduces the condition number, we consider
an eigenvalue of $D$, the off-diagonal part of (\ref{eq:frm_w_mat_2}),
where $\lambda_D=1/(\kappa+\epsilon)$.  The corresponding eigenvalue
of $M^{(W)}$ is $\lambda_M = \epsilon/(\kappa+\epsilon)$.  We can choose
$\epsilon\ll\kappa$
so $\lambda_{\rm min}(M)=\lambda_M\approx\epsilon/\kappa$.  For each
$\lambda_D$, there is a corresponding eigenvalue $\lambda_D^2$ of the block
matrix $D_{oe}D_{eo}$ of (\ref{eq:frm_w_mat_precond}).  So, the smallest
eigenvalue of the preconditioned matrix is
$
  \lambda_{\rm min}(\widetilde{M}) = 1 - \kappa^2 / ( \kappa + \epsilon )^2
  \approx 2 \epsilon / \kappa
$
and the condition number is approximately reduced by half.

One additional advantage of this preconditioning choice is the decoupling
of even site fermion degrees of freedom which allows us to integrate them out
trivially, adding an irrelevant factor to the overall normalization
of the partition function.  This reduces the storage requirements when
implementing the action on a computer, since only half of the original
components contribute.

In chapter \ref{ch:QCD_with_domain_wall_fermions}, we relied on our experience
with Wilson fermions to guide us in constructing and interpreting the physics
of domain wall fermions.  Here, we can rely on that same experience to guide us
in rewriting the domain wall fermion matrix in a form better suited for solving
linear systems.  The domain wall fermion matrix $D_{x,x^\prime}^\parallel$ in
(\ref{eq:frm_d_dslash_parallel}) will have the same structure
as $M^{(W)}$ in (\ref{eq:frm_w_mat_2}) when the fermion fields are rescaled
by $\sqrt{\kappa_5}$ where
\beq
\kappa_5 = \frac{1}{2 \left( r d + 1 - m_0 \right)}.
\eeq
Before we continue, it is useful to note that the odd-even block matrices
on the $d+1$ dimensional lattice will still have a remaining Wilson-like
odd-even substructure on each $d$ dimensional plane of fixed $s$ coordinate,
which follows from (\ref{eq:frm_d_mat_1}).  Since Wilson fermions have been
studied by the lattice community for a long time, it is rather straight forward
to implement an efficient Wilson inverter and verify its implementation against
the published results of many different groups.  So, reusing the same code
to apply these Wilson $D_{oe}$ matrices to $d$ dimensional sub-vectors
in the domain wall context will require arranging those components
in a separate $d$ dimensional odd-even storage order.  As a schematic,
if the subscripts $o, e$ indicate odd and even blocks on the $d$ dimensional
lattice and $O, E$ indicate odd and even blocks on the $d+1$ dimensional
lattice then a useful arrangement of vector components is
\beq
\Psi_{x,s} = \left(
  \begin{array}{c}
    \Psi_O \\
    \Psi_E
  \end{array}
\right) = \left(
  \begin{array}{c}
    \left(
      \begin{array}{c}
        \psi_{o,s=0} \\
        \psi_{e,s=1} \\
        \vdots
      \end{array}
    \right) \\
    \left(
      \begin{array}{c}
        \psi_{e,s=0} \\
        \psi_{o,s=1} \\
        \vdots
      \end{array}
    \right)
  \end{array}
\right) .
\eeq

With vector components written in this way,
$M_{x,s;x^\prime,s^\prime}^{\it(dwf)}$ of (\ref{eq:frm_d_mat_2}) can be
written in block form
\beq
\label{eq:frm_d_mat_4}
M^{\it(dwf)} = \left(
  \begin{array}{cc}
    \openone_{OO} & -\kappa_5 D_{OE} \\
    -\kappa_5 D_{EO} & \openone_{EE}
  \end{array}
\right).
\eeq
In analogy with (\ref{eq:frm_w_mat_precond}), the preconditioned
domain wall fermion matrix, in block diagonal form, is
\beq
\label{eq:frm_d_mat_precond}
\widetilde{M}^{\rm (dwf)} = U^{-1} M^{\rm (dwf)} L^{-1} = \left(
  \begin{array}{cc}
    \openone_{OO} - \kappa_5^2 D_{OE} D_{EO} & 0_{OE} \\
    0_{EO} & \openone_{EE}
  \end{array}
\right) .
\eeq
It is straightforward to devise a scheme to test the implementation of
preconditioning.  After implementing the action of $D_{OE}$ and $D_{EO}$ on
even and odd vectors, one can simply verify the following is true:
\beq
U \widetilde{M}^{\rm (dwf)} L \Psi = M^{\it(dwf)} \Psi .
\eeq
for arbitrary $\Psi$.  The tests we have performed indicate preconditioning
was implemented within machine precision.  For the numerical simulations
we present in chapter \ref{ch:numerical_simulations}, the preconditioned
domain wall fermion linear solver typically converged in half as many
iterations as the unpreconditioned one.

\section{The hybrid molecular dynamics force term}
\label{sec:hmd_force}

In section \ref{sec:hmc_algorithm} we discussed aspects of the HMC algorithm
that did not depend on the explicit form of the action.  Here we present
an explicit derivation of the HMD force $F_\mu(x)$, as defined
in (\ref{eq:2nd_Ham_EOM_2}), for which we need to compute the derivatives
with respect to the microcanonical evolution parameter $\tau$ of all the terms
in the action.  We will consider the Wilson plaquette gauge action,
(\ref{eq:wilson_gauge_action}), the plaquette plus rectangle gauge action
(\ref{eq:rect_gauge_action}) and the preconditioned domain wall action
described in section \ref{sec:dwf_evenodd}, including the Pauli-Villars term.
The gauge action contributions to the force are well known
\cite{Gottlieb:1987mq} and are included here only for completeness.

To determine the gauge contribution to the force, we must compute
the derivative of the gauge action, which for either action can be written
\beq
\label{eq:gauge_action_deriv_1}
\frac{{\rm d}S_G}{{\rm d}\tau}
= - \frac{\beta}{2N_c} \sum_{x,\mu} {\rm tr} \left\{
  \left[ \frac{\rm d}{{\rm d}\tau} U_\mu(x) \right] V_\mu(x)
  + V_\mu^\dag(x) \left[ \frac{\rm d}{{\rm d}\tau} U_\mu^\dag(x) \right]
\right\} .
\eeq
The $V_\mu(x)$ are called staples and are made of sums of ${\rm SU}(N_c)$
matrices.  We write the staple field as the weighted sum of the plaquette
and rectangle staple fields:
$V_\mu(x) = c_0 V_\mu^{(P)}(x) + c_1 V_\mu^{(R)}(x)$.
For the plaquette field defined in (\ref{eq:plaquette}) the corresponding
plaquette staple field is 
\beqs
\label{eq:plaquette_staple}
V_\mu^{(P)}(x) = \sum_{\nu \ne \mu} & & \left[
  U_\nu(x+\hat\mu) U_\mu^\dag(x+\hat\nu) U_\nu^\dag(x)
\right. \\*
& & + \left.
  U_\nu^\dag(x+\hat\mu-\hat\nu) U_\mu^\dag(x-\hat\nu) U_\nu(x-\hat\nu)
\right] \non .
\eeqs
For the rectangle field defined in (\ref{eq:rectangle}) the corresponding
rectangle staple field is
\beqs
\label{eq:rectangle_staple}
V_\mu^{(R)}(x) = \sum_{\nu \ne \mu} & & \left[ 
  U_\mu(x+\hat\mu) U_\nu(x+2\hat\mu) U_\mu^\dag(x+\hat\mu+\hat\nu)
  U_\mu^\dag(x+\hat\nu) U_\nu^\dag(x)
\right. \\*
& & + U_\mu(x+\hat\mu) U_\nu^\dag(x+2\hat\mu-\hat\nu)
U_\mu^\dag(x+\hat\mu-\hat\nu) U_\mu^\dag(x-\hat\nu) U_\nu(x-\hat\nu) \non \\*
& & + U_\nu(x+\hat\mu) U_\nu(x+\hat\mu+\hat\nu) U_\mu^\dag(x+2\hat\nu)
U_\nu^\dag(x+\hat\nu) U_\nu^\dag(x) \non \\*
& & + U_\nu^\dag(x+\hat\mu-\hat\nu) U_\nu^\dag(x+\hat\mu-2\hat\nu)
U_\mu^\dag(x-2\hat\nu) U_\nu(x-2\hat\nu) U_\nu(x-\hat\nu) \non \\*
& & + U_\nu(x+\hat\mu) U_\mu^\dag(x+\hat\nu)
U_\mu^\dag(x-\hat\mu+\hat\nu) U_\nu^\dag(x-\hat\mu) U_\mu(x-\hat\mu) \non \\*
& & + \left.
  U_\nu^\dag(x+\hat\mu-\hat\nu) U_\mu^\dag(x-\hat\nu)
  U_\mu^\dag(x-\hat\mu-\hat\nu) U_\nu(x-\hat\mu-\hat\nu) U_\mu(x-\hat\mu)
\right] . \non
\eeqs
The expression (\ref{eq:gauge_action_deriv_1}) is valid for either gauge action
considered, as the plaquette action result is given by setting $c_1=0$.

To identify $F_\mu^{(G)}(x)$, whose traceless anti-Hermitian part will be
the pure gauge contribution to the HMD force, we substitute Hamilton's first
equation of motion (\ref{eq:1st_Ham_EOM}) into the derivative of the gauge
action (\ref{eq:gauge_action_deriv_1})
\beq
\label{eq:gauge_action_deriv_2}
\frac{{\rm d}S_G}{{\rm d}\tau} = - \frac{\beta}{2N_c} \sum_{x,\mu} {\rm tr}
\left\{
  i H_\mu(x) \left[
    U_\mu(x) V_\mu(x) - V_\mu^\dagger(x) U_\mu^\dagger(x)
  \right]
\right\} .
\eeq
By comparing this relation with Hamilton's second equation of motion
in (\ref{eq:2nd_Ham_EOM_2}) and (\ref{eq:2nd_Ham_EOM_3}), we choose
the $F_\mu^{(G)}$ to implement in our numerical simulations to be
\beq
F_\mu^{(G)} = - \frac{\beta}{N_c} U_\mu(x) V_\mu(x) ,
\eeq
which has the same traceless, anti-Hermitian part as the expression
in square brackets above.  We will make similar simplifying choices
for the numerical implementation of the other HMD forces.

It will be convenient as we compute the domain wall fermion force term
to express the result at each step as two terms related by Hermitian
conjugation.  This will be easier if we define
\beq
g(\pm\mu)_{x,s;x^\prime,s^\prime} \equiv (1 \mp \gamma_\mu)
\delta_{x\pm\hat\mu, x^\prime} \delta_{s,s^\prime}
\eeq
and
\beq
\label{eq:frm_d_matrix_G}
G_{x,s;x^\prime,s^\prime}
= \left[ 1 - ( 1 + m_f ) \delta_{0,s^\prime} \right] P_+
\delta_{x,x^\prime} \delta_{s+1,s^\prime}
+ \sum_\mu g(+\mu)_{x,s;x^\prime,s^\prime} U_\mu(x)
\eeq
then we can write the preconditioned domain wall fermion matrix which acts
only on odd sites as
\beq
\widetilde{M} = 1_{OO} - \kappa_5^2 D_{OE} D_{EO}
\eeq
where 
\beq
\label{eq:frm_d_dslash}
D_{x,s;x^\prime,s^\prime} = G_{x,s;x^\prime,s^\prime}
+ \gamma_5 {\cal R}_{s,s^{\prime\prime}}
G^\dag_{x,s^{\prime\prime};x^\prime,s^{\prime\prime\prime}}
\gamma_5 {\cal R}_{s^{\prime\prime\prime},s^\prime}
\eeq
and ${\cal R}_{s,s^\prime} = \delta_{s,L_s-1-s^\prime}$ is the reflection
operator along the extra direction.  Now, it is easy to verify the Hermiticity
relations  $D^\dagger = \gamma_5 {\cal R} D \gamma_5 {\cal R}$ and
$\widetilde{M}^\dagger = \gamma_5 {\cal R} \widetilde{M} \gamma_5 {\cal R}$.

As with the gauge part, we must compute the derivatives with respect
to microcanonical time $\tau$ of the fermion and Pauli-Villars parts
of the domain wall action, both expressed in terms of pseudofermions.
At first, we work with only the fermionic part, $S_F$, of the domain wall
action so we suppress the $F$ and $PV$ subscripts used in sections
\ref{sec:num_genl_consid} and \ref{sec:hmc_algorithm} until later
in this section when we consider the Pauli-Villars part.
An efficient iterative method is used to solve
$\widetilde{M}^\dag\widetilde{M} \chi = \Phi$, in our case based
on the conjugate gradient algorithm \cite{Hestenes:1952,Golub:1996}.
The derivative of $S_F$ can then be written in terms of the solution $\chi$ as
\beq
\label{eq:fermion_action_deriv_1}
\frac{{\rm d}S_F}{{\rm d}\tau} = - \chi^\dagger \left[
  \frac{\rm d}{{\rm d}\tau}
  \left( \widetilde{M}^\dag \widetilde{M} \right)
\right] \chi.
\eeq
Using (\ref{eq:frm_d_dslash}), the derivative of $\widetilde{M}$ can be written
\beq
\label{eq:fermion_matrix_deriv_1}
\frac{{\rm d}\widetilde{M}}{{\rm d}\tau} = -\kappa_5^2 \left[
  D_{OE} \frac{{\rm d}G_{EO}}{{\rm d}\tau}
+ \frac{{\rm d}G_{OE}}{{\rm d}\tau} D_{EO}
+ D_{OE} \gamma_5 {\cal R} \frac{{\rm d}G_{EO}^\dag}{{\rm d}\tau}
  \gamma_5 {\cal R}
+ \gamma_5 {\cal R} \frac{{\rm d}G_{OE}^\dag}{{\rm d}\tau}
  \gamma_5 {\cal R} D_{EO}
\right] .
\eeq
If we define another quantity
\beq
\label{eq:frm_d_matrix_K}
K \equiv - \kappa_5^2 \left[
  D_{OE} \frac{{\rm d}G_{EO}}{{\rm d}\tau}
+ \frac{{\rm d}G_{OE}}{{\rm d}\tau} D_{EO}
\right] ,
\eeq
then the derivative (\ref{eq:fermion_matrix_deriv_1}) can be rewritten
\beq
\label{eq:fermion_matrix_deriv_2}
\frac{{\rm d}\widetilde{M}}{{\rm d}\tau} =
K + \gamma_5 {\cal R} K^\dagger \gamma_5 {\cal R}
\eeq
after using the Hermiticity relations and noting
$\left( \gamma_5 {\cal R} \right)^2 = \openone$.  From here, it is not
difficult to show that
\beq
\frac{{\rm d}S_F}{{\rm d}\tau} = - \chi^\dagger \left[ \left(
    \gamma_5 {\cal R} K \gamma_5 {\cal R} \widetilde{M}
  + \gamma_5 {\cal R} \widetilde{M} \gamma_5 {\cal R} K
\right) + {\rm h.c.} \right] \chi,
\eeq
accomplishing our goal of expressing the derivative as two terms related
by Hermitian conjugation.

The next step is to arrange the derivative calculations
using (\ref{eq:1st_Ham_EOM}) so that $i H_\mu(x)$ appears on the left.
We start again from (\ref{eq:frm_d_matrix_G})
\beq
\frac{\rm d}{{\rm d}\tau}G_{x,s;x^\prime,s^\prime}
= \sum_\mu i H_\mu(x) g(+\mu)_{x,s;x^\prime,s^\prime} U_\mu(x) .
\eeq
While $G$ depends explicitly on the bare mass $m_f$, the derivative does not
as there are no gauge fields in the extra direction.  This result is then
substituted into (\ref{eq:frm_d_matrix_K}), recalling that $K$ is defined
only when both sets of indices are odd
\beqs
K_{x,s;x^\prime,s^\prime} & = & -\kappa_5^2 \sum_\mu \left[
  D_{x,s;x^{\prime\prime},s^{\prime\prime}}
  i H_\mu(x^{\prime\prime})
  g(+\mu)_{x^{\prime\prime},s^{\prime\prime};x^\prime,s^\prime}
  U_\mu(x^{\prime\prime})
\right. \non \\*
&& \quad \left.
  + i H_\mu(x)
  g(+\mu)_{x,s;x^{\prime\prime},s^{\prime\prime}}
  U_\mu(x)
  D_{x^{\prime\prime},s^{\prime\prime};x^\prime,s^\prime}
\right] .
\eeqs
Finally, we substitute into the derivative of the fermionic action,
and use the invariance of the trace under cyclic permutation to move
the conjugate momenta to the left
\beqs
\frac{{\rm d}S_F}{{\rm d}\tau}
&=& + \kappa_5^2 \sum_{x,\mu} {\rm~tr}_{\rm color}( i H(x,\mu) \non \\*
&& \quad \begin{array}{ll}
  \times \left\{ U_\mu(x) {\rm~tr}_{s,{\rm spin}} \right.
  & \left[ g(+\mu) \left(
    \chi \chi^\dag \widetilde{M}^\dag D
    + D \chi \chi^\dag \widetilde{M}^\dag
  \right) \right. \\*
  & + \left. g^\dag(-\mu) \left(
    \widetilde{M} \chi \chi^\dag D^\dag 
    + D^\dag \widetilde{M} \chi \chi^\dag
  \right)\right] \\*
  \quad - \left. {\rm h.c.} \right\} ). &
\end{array}
\eeqs
As with the pure gauge contribution, we compare this relation with Hamilton's
second equation in (\ref{eq:2nd_Ham_EOM_2}) and (\ref{eq:2nd_Ham_EOM_3}) and
choose $F_\mu^{(F)}(x)$ to have the same traceless, anti-Hermitian part
as the expression in the curly braces but in a form convenient for numerical
simulation.

We construct some vectors which are defined on half of the sites of the lattice
\beqs
\rho_E   = & D_{EO} \chi_O & \\
\psi_O   = & - \kappa_5^2 \widetilde{M}_{OO} \chi_O & \\
\sigma_E = & D^\dagger_{EO} \psi_O
& = - \kappa_5^2 D^\dag_{EO} \widetilde{M}_{OO} \chi_O .
\eeqs
Now we can write the fermionic contribution to the force
\beq
\label{eq:frm_d_force_1}
F_\mu^{(F)}(x) = -2 U_\mu(x) {\rm~tr}_{s,{\rm spin}} \left[
  g(+\mu) \left( \chi\sigma^\dag + \rho \psi^\dag \right)
  + g^\dag(-\mu) \left( \psi\rho^\dag + \sigma\chi^\dag \right)
\right].
\eeq
We can make one further simplification by defining two vectors which
extend over the whole lattice
\beq
v = \left( \begin{array}{c} \rho_E \\ \chi_O \end{array} \right)
\qquad
w = \left( \begin{array}{c} \sigma_E \\ \psi_O \end{array} \right).
\eeq
Finally, in terms of the vectors $v$ and $w$
\beq
\label{eq:frm_d_force_2}
F_\mu^{(F)}(x) = -2 U_\mu(x) \sum_s {\rm~tr}_{\rm spin} \left[
  (1-\gamma_\mu) v_{x+\hat\mu,s} w_{x,s}^\dag
  + (1+\gamma_\mu) w_{x+\hat\mu,s} v_{x,s}^\dag
\right] .
\eeq
With the absence of gauge fields in the $s$-direction, it is not surprising
that this fermionic force is identical in form to the force due
to $L_s$ flavors of Wilson fermions.  This implies that all of the special
properties of domain wall fermions, at this level, are encoded at the start
when finding the solution $\chi_O$ of the fermionic system.

The derivative of the Pauli-Villars part of the domain wall action is
\beq
\frac{{\rm d}S_{PV}}{{\rm d}\tau} = \Phi_{PV}^\dagger \left[
  \frac{\rm d}{{\rm d}\tau} \left(
    \widetilde{M}_{PV}^\dagger \widetilde{M}_{PV}
  \right)
\right] \Phi_{PV} .
\eeq
Note the crucial difference in sign relative to the derivative of the fermion
action (\ref{eq:fermion_action_deriv_1}).  The construction
of the Pauli-Villars contribution to the HMD force mirrors the construction
for the fermionic part provided we set the solution $\chi_{PV}$ to be
the pseudofermion field generated by the heat bath, $\chi_{PV} = \Phi_{PV}$,
set $m_f$=1 and keep track of the overall change in sign.  Using $\chi_{PV}$
in place of $\chi$ we get
\beq
\label{eq:frm_d_PV_force}
F_\mu^{(PV)}(x) = -\left. F_\mu^{(F)}(x)\right|_{m_f\!=\!1, \chi\!=\!\chi_{PV}}.
\eeq

Now that all three force contributions are known, (\ref{eq:2nd_Ham_EOM_3})
can be written
\beq
i \frac{\rm d}{{\rm d}\tau} H_\mu(x) = \left[
  F_\mu^{(G)}(x) + F_\mu^{(F)}(x) + F_\mu^{PV}(x)
\right]_{\rm TA} .
\eeq
We again emphasize that only the traceless, anti-Hermitian part of $F_\mu(x)$
contributes to the molecular dynamics evolution.

With any large coding project, it is important to devise testing strategies
to verify that the code is consistent with the analytic description.
For the dynamical domain wall action, this process is quite simple
and straight forward, provided one starts with an accurate implementation
of the Wilson fermion matrix acting on a vector, some kind of linear solver,
and the Wilson HMD force term.  Before we implemented the domain wall action
in the physics software environment of the QCDSP machines at Columbia,
we ran several dynamical Wilson fermion simulations which were in good
statistical agreement with the published results of other groups.

Since our linear solver, based on the conjugate gradient (CG) algorithm,
solves the problem $\widetilde{M}^\dagger \widetilde{M} \chi = \Phi$,
testing it simply requires choosing many different $\Phi$ sources,
solving each one for the corresponding $\chi$ solutions and then applying
$\widetilde{M}^\dagger \widetilde{M}$ to $\chi$ and comparing the result
to the original.  This defines a quantity called the {\it true residual}
\beq
\label{eq:true_residual}
r = \frac{
  \left| \Phi - \widetilde{M}^\dagger \widetilde{M} \chi \right|
}{\left| \Phi \right|}
\eeq
where we use the vector 2-norm:
$\left| \Phi \right| = \sqrt{\Phi^\dagger \Phi}$.  However, running the CG
algorithm for a sufficient number of iterations for $r$ to be smaller than
some predetermined stopping condition only verifies that some non-singular
linear system was solved.

To check that the linear system solved was actually involved the domain wall
fermion matrix $\widetilde{M}^{\dagger{\rm(dwf)}}\widetilde{M}^{\rm(dwf)}$
acting on a vector, we compared the results of the solution run on a free
lattice configuration, $U_\mu(x) \to \openone$, to the analytic solution
computed by Vranas \cite{Vranas:1997da}.  We then extended our check
to non-trivial gauge configurations by applying a random gauge transformation,
$\Lambda(x) \in {\rm SU}(N_c)$, to the free lattice configuration,
$U_\mu(x) \to \Lambda^\dagger(x) \Lambda(x+\hat\mu)$, and verified that
the true residual was invariant under gauge transformation
\beq
r = \frac{
  \left|
    \Lambda\Phi_0 - \widetilde{M}_\Lambda^\dagger \widetilde{M}_\Lambda
    \Lambda\chi_0
  \right|
}{\left|\Lambda\Phi_0\right|}
\eeq
where $\Phi_0, \chi_0$ are the free field source and solution
verified by analytic calculation and $\widetilde{M}_\Lambda$ indicates
the matrix is applied using the gauge transformed links.  If the domain wall
fermion matrix is properly implemented, then the true residual is
a gauge-invariant quantity.  When starting with a properly implemented Wilson
fermion matrix, the gauge invariance test should not fail if the free field
test passed, since the gauge fields only enter the domain wall matrix
through the Wilson portion of the code.  These tests passed within
finite precision limits of the QCDSP machine.

The only difference between the force terms of the fermionic
(\ref{eq:frm_d_force_2}) and Pauli-Villars (\ref{eq:frm_d_PV_force}) parts
of the domain wall action and the force term of $L_s$ independent flavors
of Wilson fermions is in the computation of the input solution $\chi$.
So, starting from an accurate implementation of dynamical Wilson fermions
that supports multiple independent flavors makes the implementation of
dynamical domain wall fermions straightforward.  In essence, once $\chi$
has been computed by the linear solver, $s$ is treated as a Wilson flavor
index.  As a final check, we also verified that the domain wall force terms
transform properly under random gauge transformations.

After these series of analytic tests of the numerical implementation,
it would have been nice to run an actual simulation and do a statistical
comparison with published results.  Since the simulations described
here are the first of domain wall fermions in dynamical QCD, there are
no previous published results.  However, there were still two significant
statistical tests which were run.  First, we simulated dynamical domain
wall fermions at $m_f$=1 on $8^3\!\times\!32$ volumes at $\beta$=5.7
and verified the results were consistent with quenched QCD.  This check is not
as trivial as it may seem, as the two parts of the domain wall force do not
exactly cancel trajectory by trajectory since different pseudofermion fields
are used for the fermionic and Pauli-Villars parts.  Instead, the cancellation
of forces happens stochastically, yielding quenched results in the limit
of a large number of trajectories.  Second, Vranas performed
dynamical domain wall simulations of $N_f$=2 QCD on $2^4$ volumes
\cite{Chen:2000zu} and found the results were in good statistical agreement
with dynamical simulations done directly in the overlap formalism.

Perhaps the best test of the whole implementation of the CG solver
and domain wall force working together within the context of the HMC algorithm
comes from an application of Liouville's theorem by Creutz
\cite{Creutz:1988wv} within the classical framework of molecular dynamics.
If we integrate out all of the fermion (or pseudofermion) fields,
we can write the molecular dynamics partition function as a path
integral over gauge fields and conjugate momenta
\beq
Z_{\rm HMD}
= \int \left[ {\cal D} U^\prime \right] \left[ {\cal D} H^\prime \right]
  e^{\cal -H^\prime}
= \int \left[ {\cal D} U \right] \left[ {\cal D} H \right]
  e^{\cal -H} e^{\cal H - H^\prime} .
\eeq
Since the HMD evolution is purely classical, the area preserving property
of Liouville's theorem guarantees that the integration measures are equal:
$\left[ {\cal D} U^\prime \right] \left[ {\cal D} H^\prime \right]
= \left[ {\cal D} U \right] \left[ {\cal D} H \right]$.
Dividing both sides by the partition function, $Z_{\rm HMD}$, we get
the expectation values
\beq
\label{eq:hmc_exp_delta_h}
\left\langle \exp\left( -\Delta {\cal H} \right) \right\rangle = 1
\qquad \left( \Delta {\cal H} \equiv {\cal H^\prime - H} \right)
\eeq
and, by Jensen's inequality,
\beq
\left\langle \Delta {\cal H} \right\rangle \ge 0.
\eeq
Because $\Delta {\cal H}$ is computed as part of the Metropolis acceptance
test of the HMC algorithm, it should be straightforward to calculate these
observables in any actual simulation.

In our experience, satisfying these identities seems to be the most stringent
check of the implementation of the HMC algorithm.  If the force term
is inconsistent with the Hamiltonian as implemented then Liouville's theorem
will not hold because the evolution is no longer classical and area preserving
with respect to the given Hamiltonian.  However, the code necessary to compute
the change in the Hamiltonian need not share any components with the force term
except for the linear solver.  Once the CG solver has been sufficiently
tested, checking the Hamiltonian part of the code is simple, so
these identities provide a demanding statistical test of the overall evolution,
and in particular the HMD forces.

In chapter \ref{ch:numerical_simulations}, the description of our simulations
will include the observable (\ref{eq:hmc_exp_delta_h}).  All of the runs,
taken as a whole, agree with the prediction (\ref{eq:hmc_exp_delta_h}) within
statistical error, from which we conclude that the domain wall HMC algorithm
was correctly implemented.

\clearpage


\chapter{Simulations of $N_f$=2 QCD}
\label{ch:numerical_simulations}

\thispagestyle{myheadings}
\markright{}

\section{The deconfinement phase transition}
\label{sec:phase_transition}

We are motivated to study the thermodynamics of $N_f$=2 QCD with
degenerate light quarks for two reasons.  First, it is an approximation to
$N_f$=6 QCD with the six quarks at their physical masses.
Recall from section \ref{sec:phase_transition_discussion} the discussion
of the quenched limit of QCD, where dynamical quark effects are altogether
eliminated by taking the bare quark masses to infinity.  Similarly,
the dynamics of the heavier quarks ({\it top, bottom}, $\cdots$)
will be marginal in the Euclidean path integral and hence will have
little effect on the character of the transition region.  So, $N_f$=2 QCD
is the most natural starting point for studying the role
of dynamical quark effects in QCD thermodynamics.

At first glance, this theory, corresponding to a world where the {\it up}
and {\it down} quark masses are degenerate and the rest are much heavier,
is a reasonable approximation of our world.  However, on closer inspection,
the {\it up} and {\it down} quark masses are not exactly degenerate
and the {\it strange} quark mass is only an order of magnitude heavier.
So it will be important to perform the typically more computationally demanding
$N_f$=3 QCD simulations with non-degenerate quarks to understand
the full physical theory in detail.  But, one should first study completely
(if possible) the simpler $N_f$=2 theory before looking for (perhaps)
subtle differences by turning on the dynamics of the heavier quark flavors.

A second motivation for this study is simply that $N_f$=2 QCD is
an interesting theory in its own right.  With massless quarks,
the theory minimally has ${\rm SU}_L(2)\otimes{\rm SU}_R(2)$ chiral
symmetry and the chiral condensate, $\qbq$, is an exact order parameter.
This chiral symmetry group is expected to be in the ${\rm O}(4)$ universality
class and other models in that class are known to have second order
phase transitions, so there is a possibility that the chiral transition
of $N_f$=2 massless QCD may be second order \cite{Pisarski:1984ms}.
However, if the anomalously broken ${\rm U}_A(1)$ symmetry is effectively
restored along with chiral symmetry at the transition, then the larger symmetry
is expected to be in the ${\rm O}(2)\!\otimes\!{\rm O}(4)$ universality class
and as other models in that class are known to have first order phase
transitions there is the possibility that the chiral transition could be
first order \cite{Pisarski:1984ms,Wilczek:1992sf,Rajagopal:1993qz}.

Before embarking on a large numerical project with a new fermion formulation,
we note what is already known about QCD thermodynamics
using other formulations, as all fermion formulations should give
the same answer in the continuum limit.  On $24^3\!\times\!4$ volumes
in quenched QCD, there is a first order deconfinement phase transition
at $\beta_c$=5.6925 \cite{Brown:1988qe}.  Along with other quenched simulations,
this corresponds in physical units to a critical temperature
$T_c\!\simeq\!270$ MeV \cite{Karsch:1999vy}.

For $N_f$=2 QCD with staggered fermions on $16^3\!\times\!4$ volumes,
the finite temperature phase transition occurs at $\beta_c$=5.265 for $m$=0.01
and $\beta_c$=5.291 for $m$=0.025 \cite{Vaccarino:1991jy,Vaccarino:1991kn}
which suggests that $\beta_c\!\approx\!5.25$ in the chiral limit, approximately
150 MeV.  At these quark masses, there is no evidence for a first order
phase transition.  Other $N_f$=2 dynamical simulations with different actions
have critical temperatures in physical units in the range
$T_c \simeq$ (170--190) MeV \cite{Karsch:1999vy}.

At finite lattice spacing the chiral symmetry of staggered fermions
is only a one dimensional subgroup of the continuum symmetry
and the vector flavor symmetry is explicitly broken.  Consequently, QCD
with staggered fermions at accessible lattice spacings is a theory
with one physically light pion and two unphysically heavy pions
in the range 500--600 MeV.  This may obscure the true nature of the transition
until the lattice spacing is reduced or the action is improved.

In section \ref{sec:beta_crit_8nt4}, we will locate the transition
region of $N_f$=2 QCD with moderately light degenerate quarks
in small $8^3\!\times\!4$ volumes.  Using the chiral condensate we will show
strong evidence for a chiral phase transition in the massless quark limit.
However, the small volumes used will make it difficult to establish
detailed properties of the transition region.  Thus, in section
\ref{sec:beta_crit_16nt4} we will focus in on the transition region
in larger $16^3\!\times\!4$ volumes with somewhat lighter quarks in the hope
of seeing an indication of the order of the transition.

Implicit throughout the preceding discussion is the assumption that
the domain wall fermion formulation provides us with the nearly exact
chiral symmetry needed to observe the critical phenomena related to spontaneous
chiral symmetry breaking.  In section \ref{sec:GMOR_study}, we will perform
the analysis of the GMOR relation outlined in section \ref{sec:finite_ls}
as a quantitative measure of the effectiveness of the domain wall method
for producing nearly massless fermions on the lattice.

\section{Locating the transition in $8^3\!\times\!4$ volumes}
\label{sec:beta_crit_8nt4}

In the Euclidean path integral formulation, the temporal extent of the lattice
$L_t$ is related to the inverse temperature $T^{-1}$
through the lattice spacing:  $L_t=(aT)^{-1}$.  If we hold $L_t$ fixed,
then the determination of the temperature $T_c$ of the chiral phase transition
requires, in general, searching in the space of the four parameters
$(\beta, m_0, L_s, m_f)$ that determine the lattice spacing.  The problem
of searching this space is made simpler as we are only interested
in the location of the transition in the chiral limit, as $m_f \to 0$
and $L_s\!\to\!\infty$, over an appropriate range of $m_0$.
In the massless limit, the location is specified by the critical coupling,
$\beta_c(m_0)$.

We first studied the behavior of both the Wilson line and chiral condensate
in $8^3\!\times\!4$ volumes \cite{Vranas:1998vm,Christ:1998,Vranas:1999pm,
Fleming:2000xe,Vranas:1999dg,Chen:2000zu}.  Choosing the spatial extent
of the lattice $L$ to be much larger than the temporal extent, $L\!\gg\!L_t$,
is essential to eliminating the finite volume effects that obscure
the character of the transition.  However, past experience and this work
will demonstrate that searching in small volumes with moderately heavy quarks
gives a surprisingly good estimate of $\beta_c$.  Further simulations
at larger spatial volumes and smaller masses can refine the estimate
of $\beta_c$ and reveal the character of the transition.

At very small lattice spacings, when the gauge fields are near the identity,
the condition for bound domain wall states is essentially the free condition
(\ref{eq:free_dwf_condition}).  For a large range of $m_0$, one light flavor
will be localized on the domain wall defect. Of course, to use the HMC
algorithm described in section \ref{sec:hmc_algorithm}, we must square
the determinant which gives us two degenerate flavors.  However, $m_0$
also determines the maximum four-momentum a bound surface state may have,
see (\ref{eq:free_dwf_condition}) for the condition in the free case,
and is the mass scale of the heavy Wilson doubler states that are not localized
on the defect.  So, at coarser lattice spacing, these scales become relevant
and the bare couplings and masses will have $m_0$-dependent contributions
which should cause shifts in $\beta_c$.  By locating $\beta_c$
for several $m_0$, we hope to establish that this parameter will not have to be
fine-tuned on coarse lattices for the domain wall method to work,
but some care must be taken in choosing a value to ensure the simulation
represents the correct number of light flavors
with a sufficiently large phase space for the light degrees of freedom.

Consideration of the existing numerical studies of QCD thermodynamics
with quenched and dynamical $N_f$=2 staggered fermions provides us
with a rough estimate of the location of $\beta_c$.  For quenched $L_t=4$
thermodynamics, corresponding to $m_f~=~1$ in our parameter space,
$\beta_c \approx 5.7$ and for dynamical staggered fermions in the chiral limit,
similar to $m_f=0$ and $L_s\to\infty$, $\beta_c \approx 5.25$.  Since we are
interested in the thermodynamics of domain wall fermions near the chiral limit,
if we choose $m_f$ sufficiently small and $L_s$ sufficiently large
then for some reasonable range of $m_0$ we should expect to find $\beta_c$
in the vicinity of the chiral staggered value.

To perform a fully dynamical search of the parameter space requires
starting a new simulation each time any one of the basic run parameters
$(L,L_t,\beta,m_0,L_s,m_f)$ is changed.  Although the process of analyzing
any given simulation is interesting in its own right, the sheer number
of simulations included in this work prevent us from providing that level
of detail.  So, we encourage the reader to refer to the extensive tables
and figures provided for details of any particular evolution.

Our search strategy for $8^3\!\times\!4$ volumes was to fix $m_f$=0.1, $L_s$=12
and for several values of $m_0$ from 1.15 to 2.4, we scan in $\beta$ from
weak to strong coupling over the range $\beta$ = 5.95 -- 4.65
until crossover behavior was observed in $\qbq$ and $\w3m$.  The run parameters
for the simulations are listed in tables
\ref{tab:8nt4_h1.15_beta_crit} -- \ref{tab:8nt4_h2.4_beta_crit}.
There were no difficulties with the CG algorithm failing to converge,
characteristic of the ``exceptional configuration'' problem
in dynamical Wilson fermions.  When the even-odd preconditioned domain wall
Dirac operator was used, as described in section \ref{sec:dwf_evenodd},
100--130 CG iterations were typically needed per HMD step in the confined phase
and 40--70 were needed in the deconfined phase.  Simulations took three to five
days to complete 800 HMC trajectories when run on a six Gflops portion
of the QCDSP computer at Columbia.  In some cases, when preconditioning
was not used, two and a half times the number of CG iterations were needed
per HMD step.

As the deconfined phase exists at weaker coupling than the confined phase,
the gauge configurations are more ordered, {\it i.~e.}~closer to the identity
configuration.  So, the initial configuration for each simulation
was deliberately chosen to be as far as possible from the anticipated state
after thermalization.  For simulations where a deconfined thermal state
was expected, the initial lattice was disordered, while an initial ordered
lattice was used where a confined thermal state was expected.  This way,
the thermalization time scale could be estimated by observing the system
tunneling from the wrong phase through the transition into the correct phase.

\begin{figure}[ptb]
  \centering
  \includegraphics
    [height=0.9\textheight,width=\textwidth]
    {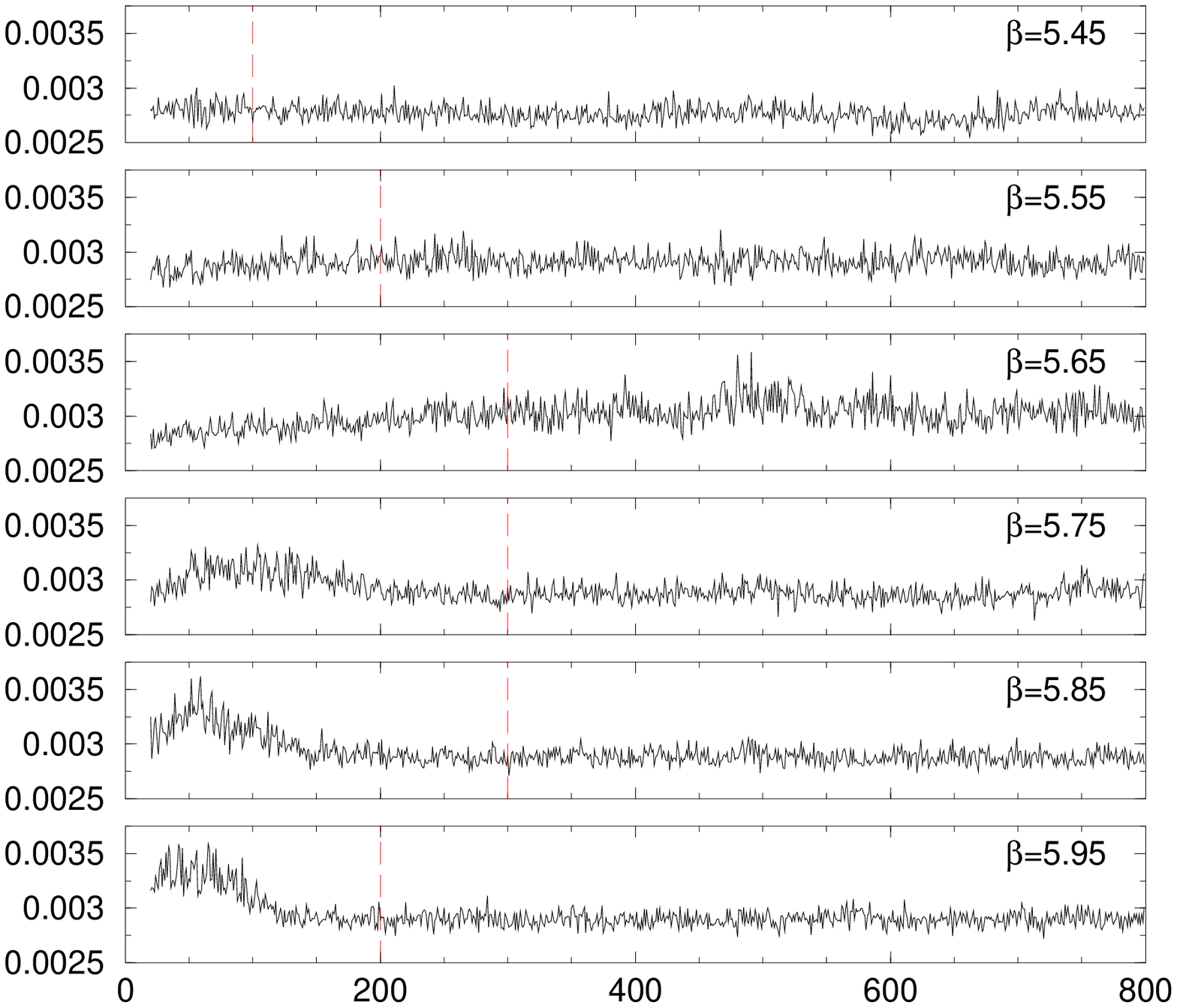}
  \caption[$\qbq$ evol: $8^3\times 4, m_0=1.15, L_s=12, m_f=0.1$]
    {$\qbq$ evol: $8^3\times 4, m_0=1.15, L_s=12, m_f=0.1$}
  \label{fig:pbp_wd_8nt4_h1.15_l12_m0.1}
\end{figure}

\begin{figure}[ptb]
  \centering
  \includegraphics
    [height=0.9\textheight,width=\textwidth]
    {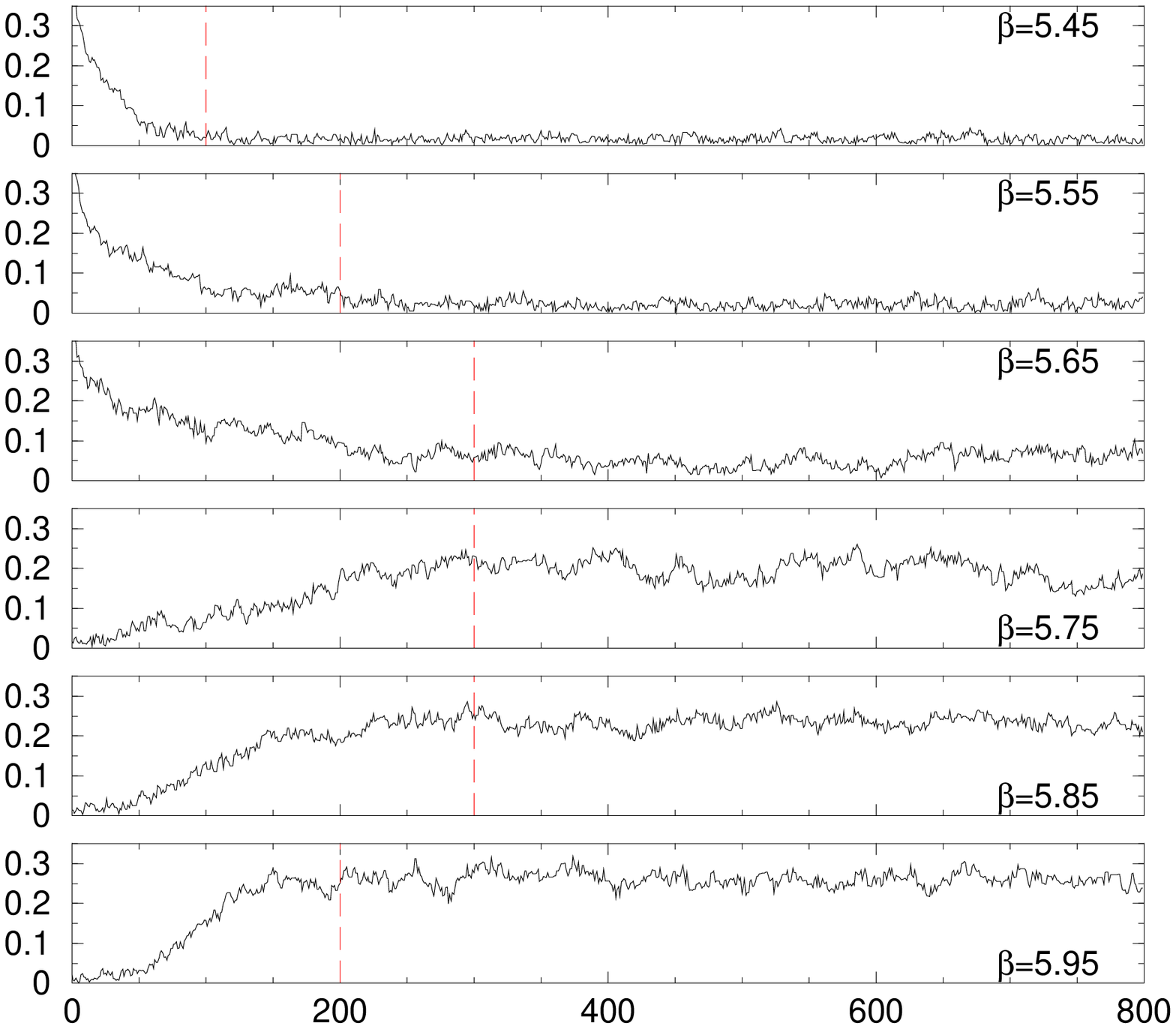}
  \caption[$\w3m$ evol: $8^3\times 4, m_0=1.15, L_s=12, m_f=0.1$]
    {$\w3m$ evol: $8^3\times 4, m_0=1.15, L_s=12, m_f=0.1$}
  \label{fig:w3m_wd_8nt4_h1.15_l12_m0.1}
\end{figure}

\begin{figure}[ptb]
  \centering
  \includegraphics
    [height=0.9\textheight,width=\textwidth]
    {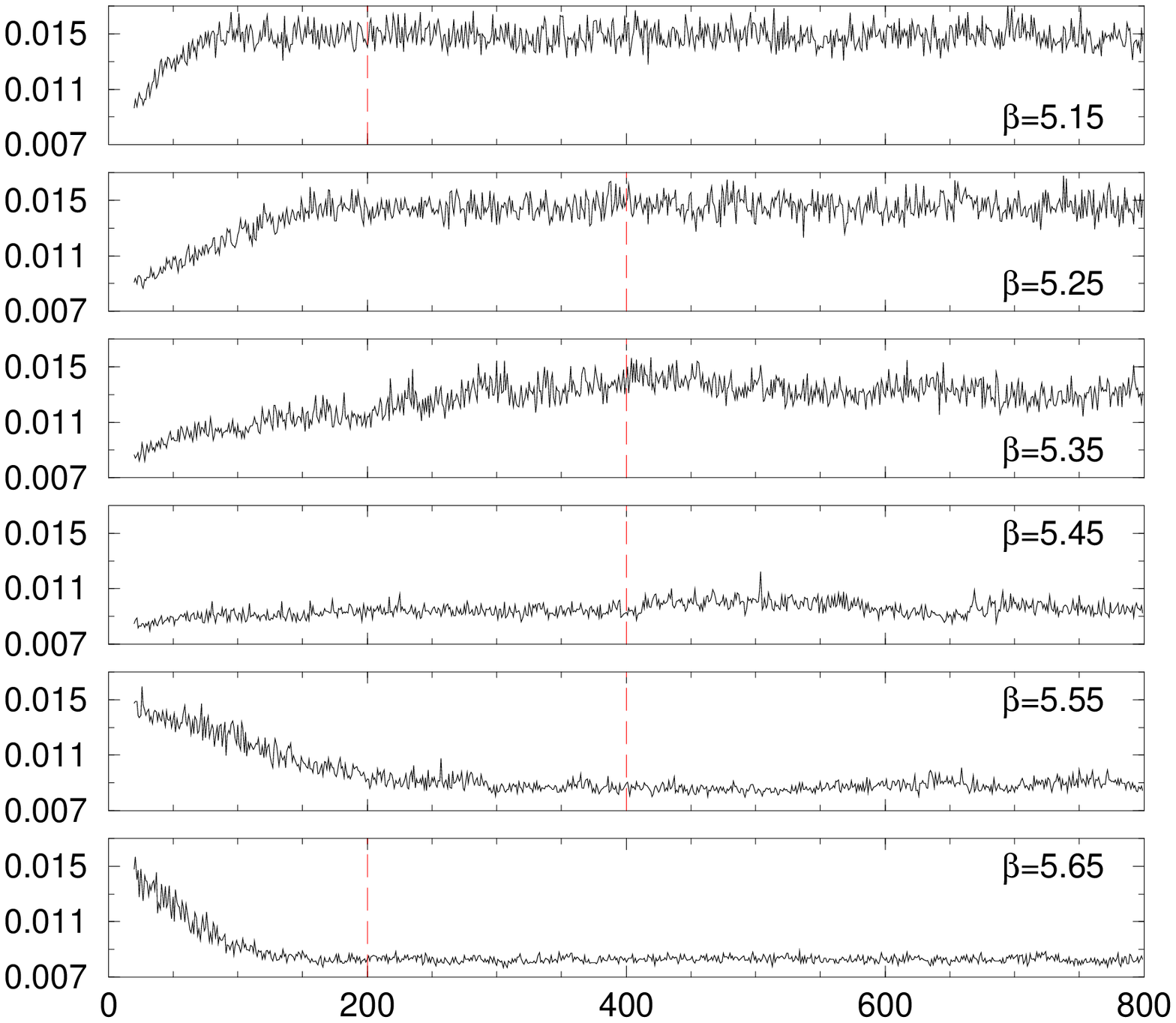}
  \caption[$\qbq$ evol: $8^3\times 4, m_0=1.4, L_s=12, m_f=0.1$]
    {$\qbq$ evol: $8^3\times 4, m_0=1.4, L_s=12, m_f=0.1$}
  \label{fig:pbp_wd_8nt4_h1.4_l12_m0.1}
\end{figure}

\begin{figure}[ptb]
  \centering
  \includegraphics
    [height=0.9\textheight,width=\textwidth]
    {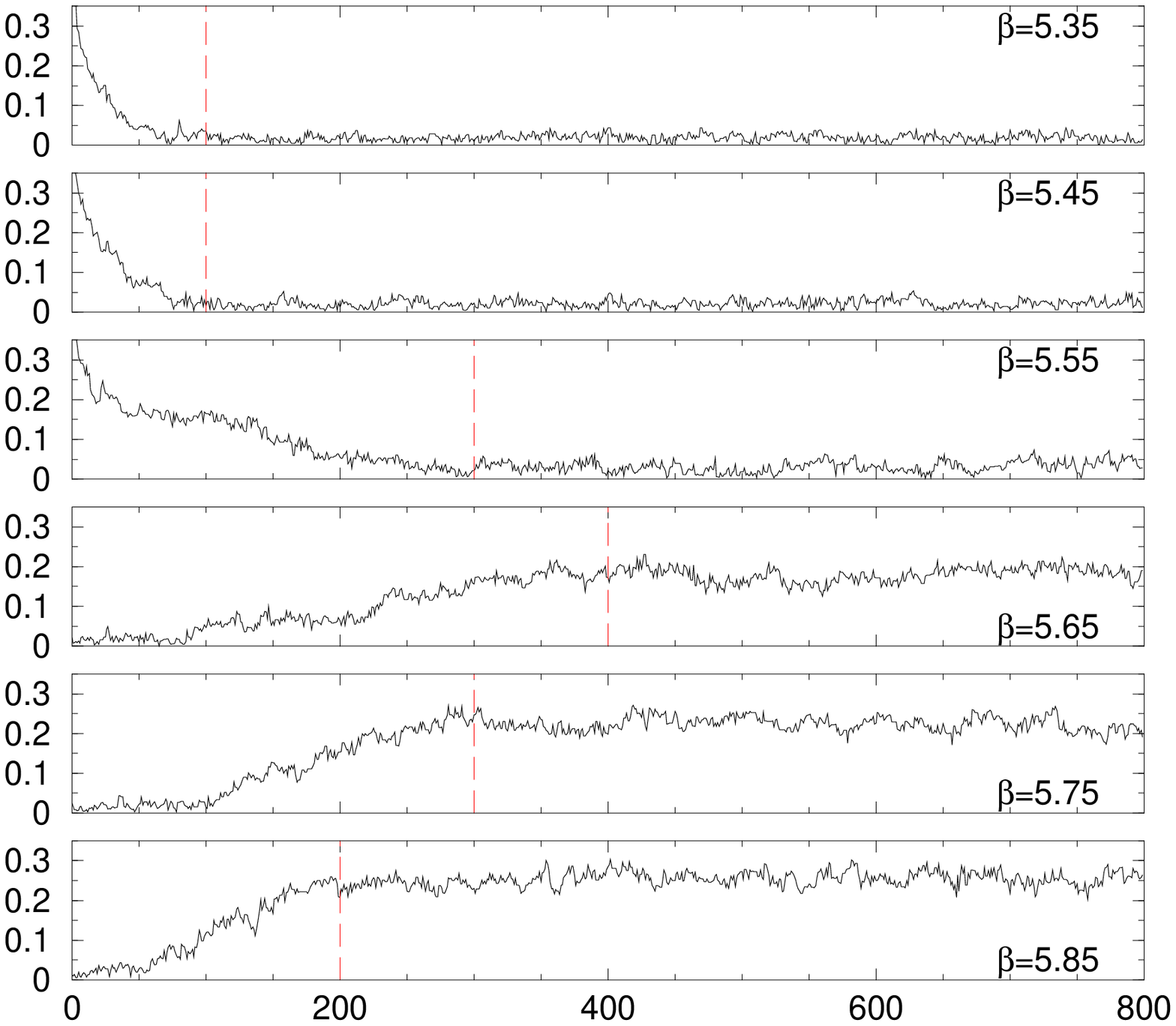}
  \caption[$\w3m$ evol: $8^3\times 4, m_0=1.4, L_s=12, m_f=0.1$]
    {$\w3m$ evol: $8^3\times 4, m_0=1.4, L_s=12, m_f=0.1$}
  \label{fig:w3m_wd_8nt4_h1.4_l12_m0.1}
\end{figure}

\begin{figure}[ptb]
  \centering
  \includegraphics
    [height=0.9\textheight,width=\textwidth]
    {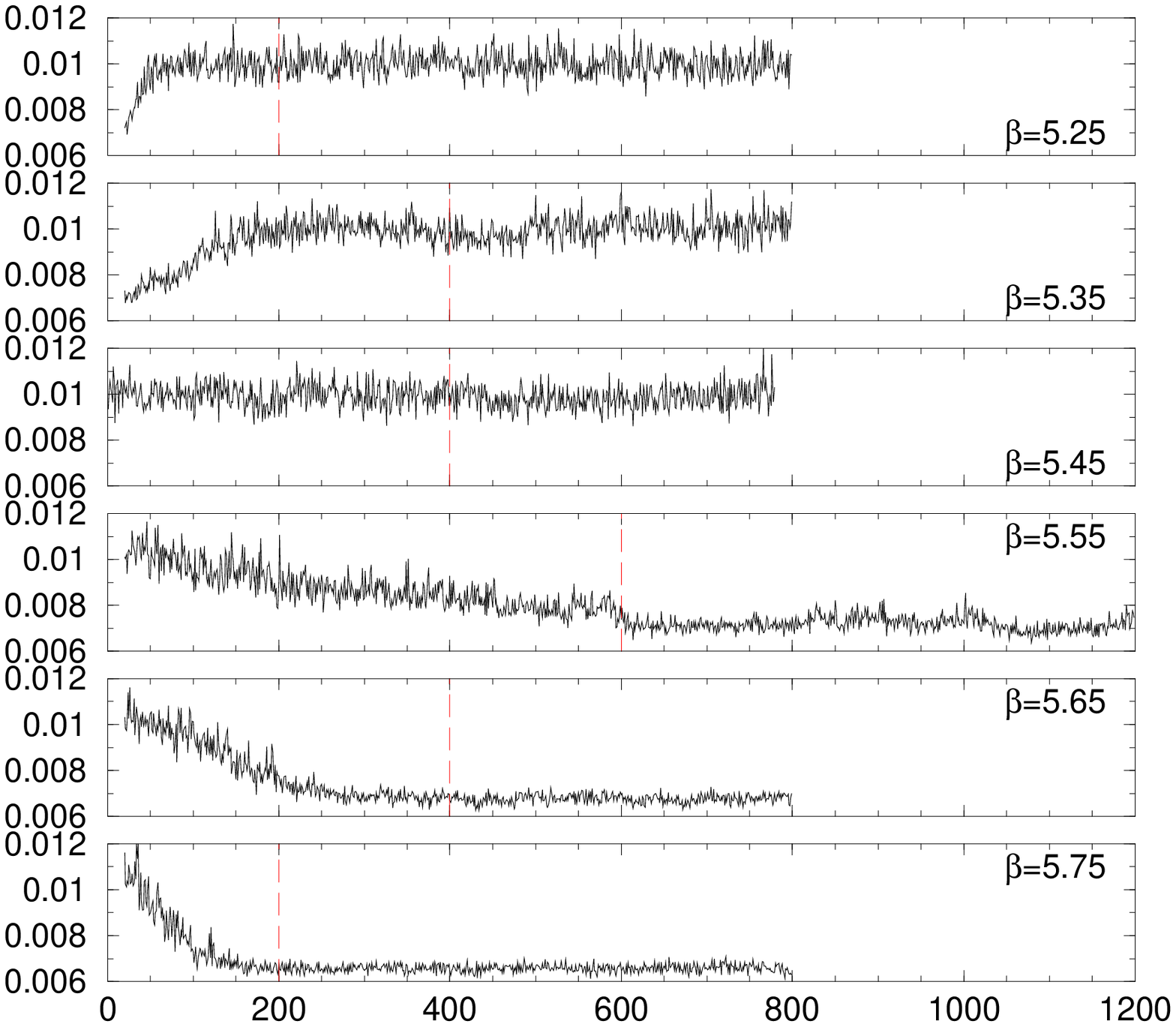}
  \caption[$\qbq$ evol: $8^3\times 4, m_0=1.65, L_s=12, m_f=0.1$]
    {$\qbq$ evol: $8^3\times 4, m_0=1.65, L_s=12, m_f=0.1$}
  \label{fig:pbp_wd_8nt4_h1.65_l12_m0.1}
\end{figure}

\begin{figure}[ptb]
  \centering
  \includegraphics
    [height=0.9\textheight,width=\textwidth]
    {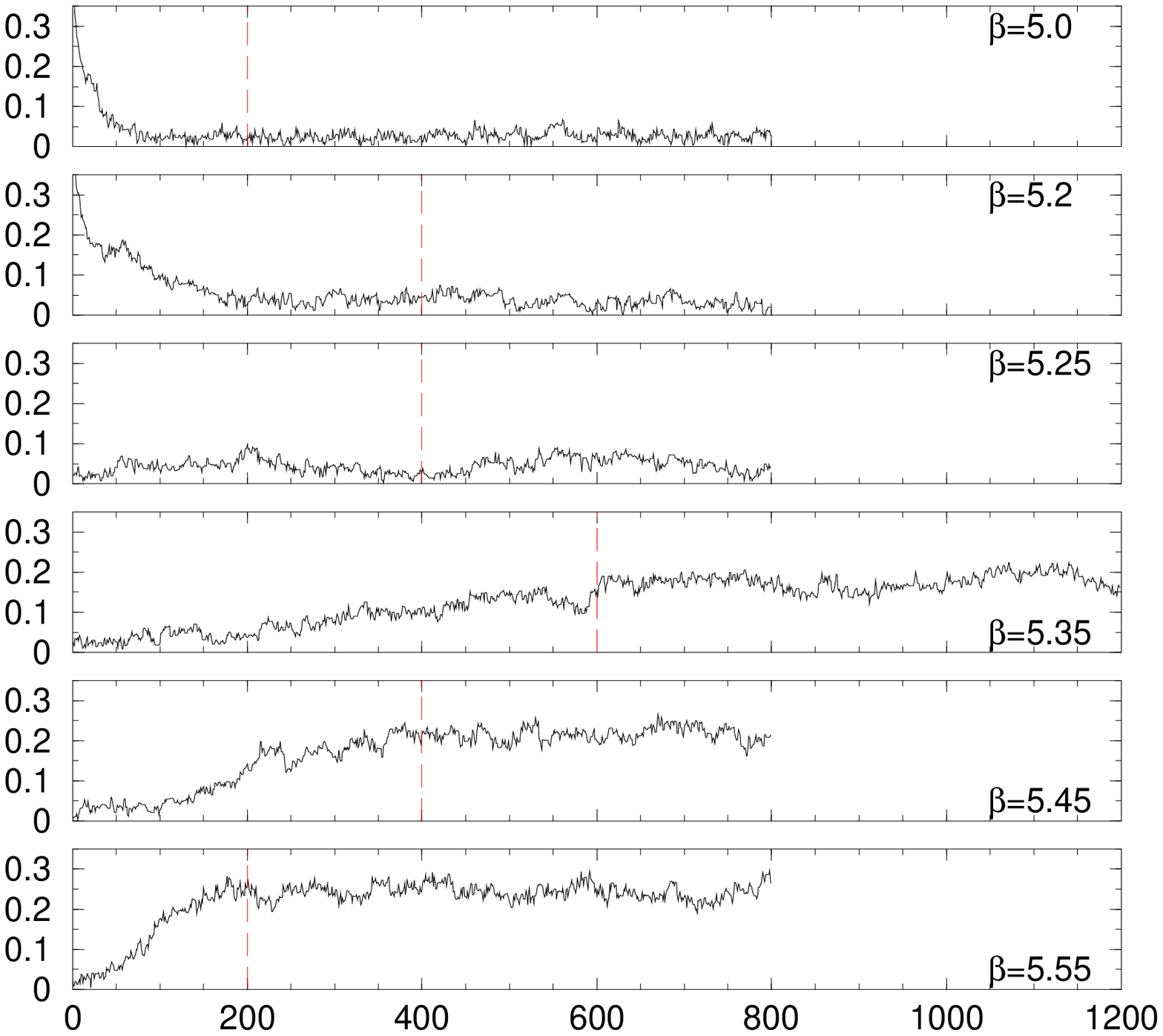}
  \caption[$\w3m$ evol: $8^3\times 4, m_0=1.65, L_s=12, m_f=0.1$]
    {$\w3m$ evol: $8^3\times 4, m_0=1.65, L_s=12, m_f=0.1$}
  \label{fig:w3m_wd_8nt4_h1.65_l12_m0.1}
\end{figure}

\begin{figure}[ptb]
  \centering
  \includegraphics
    [height=0.9\textheight,width=\textwidth]
    {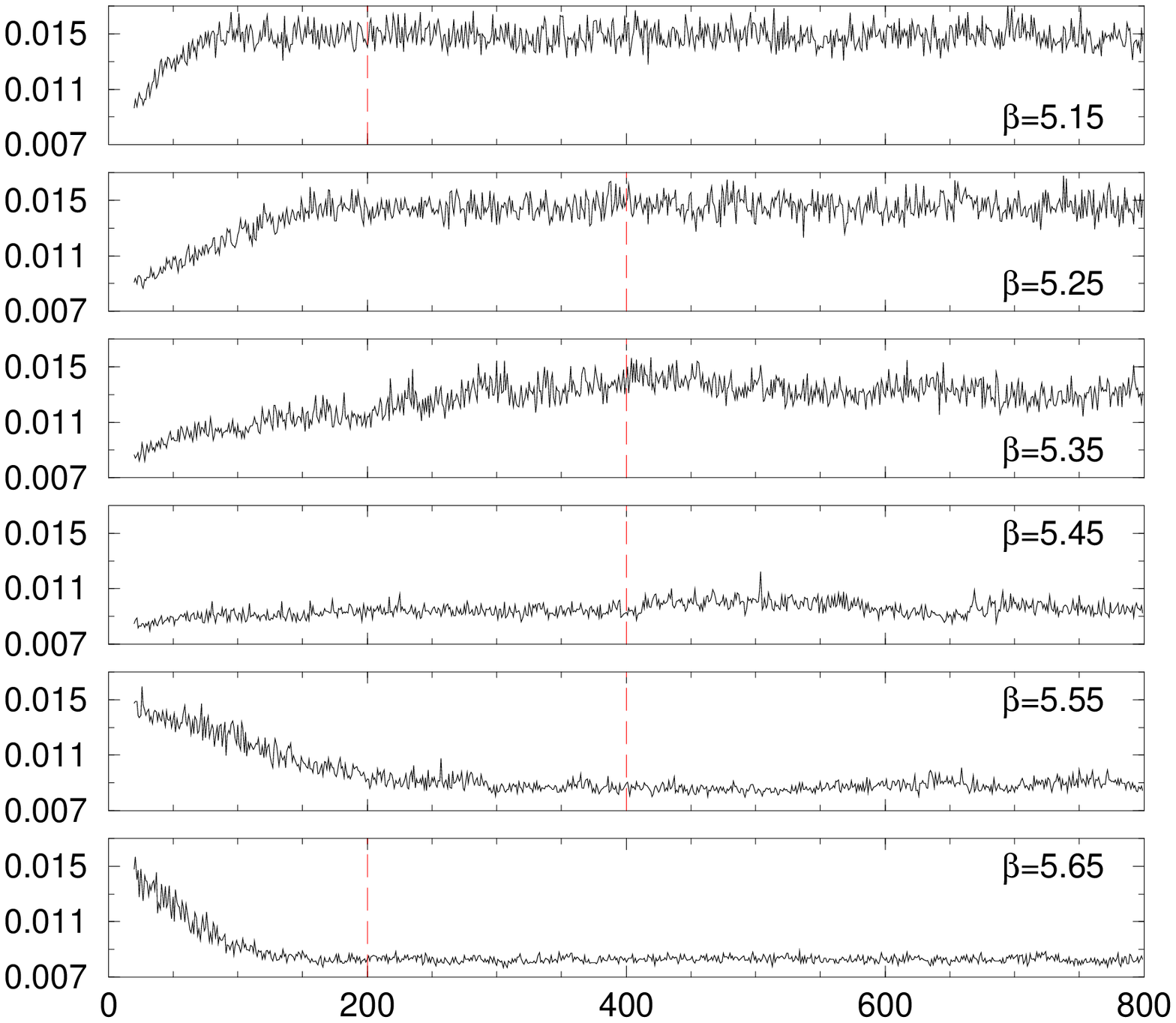}
  \caption[$\qbq$ evol: $8^3\times 4, m_0=1.8, L_s=12, m_f=0.1$]
    {$\qbq$ evol: $8^3\times 4, m_0=1.8, L_s=12, m_f=0.1$}
  \label{fig:pbp_wd_8nt4_h1.8_l12_m0.1}
\end{figure}

\begin{figure}[ptb]
  \centering
  \includegraphics
    [height=0.9\textheight,width=\textwidth]
    {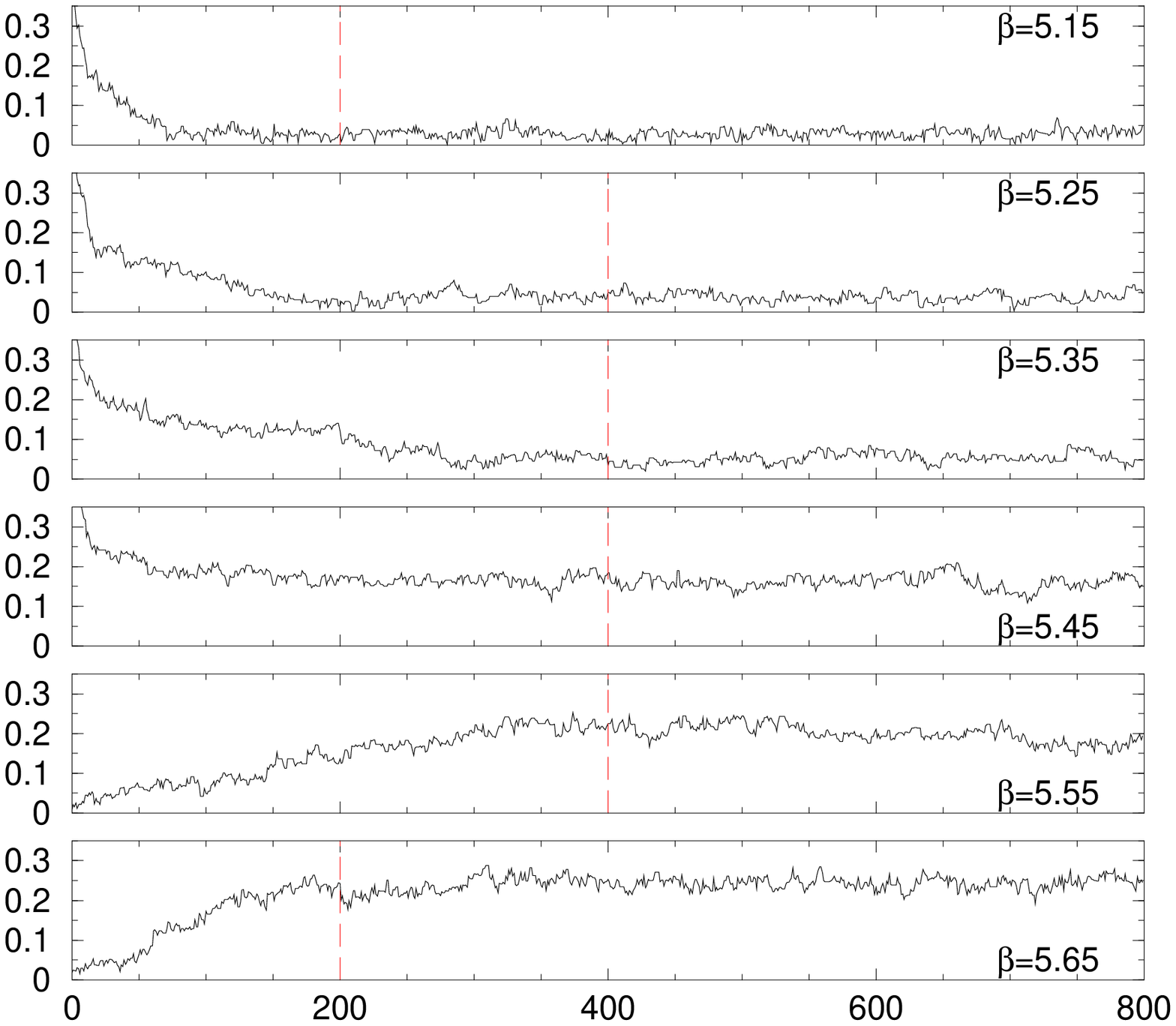}
  \caption[$\w3m$ evol: $8^3\times 4, m_0=1.8, L_s=12, m_f=0.1$]
    {$\w3m$ evol: $8^3\times 4, m_0=1.8, L_s=12, m_f=0.1$}
  \label{fig:w3m_wd_8nt4_h1.8_l12_m0.1}
\end{figure}

\begin{figure}[ptb]
  \centering
  \includegraphics
    [height=0.9\textheight,width=\textwidth]
    {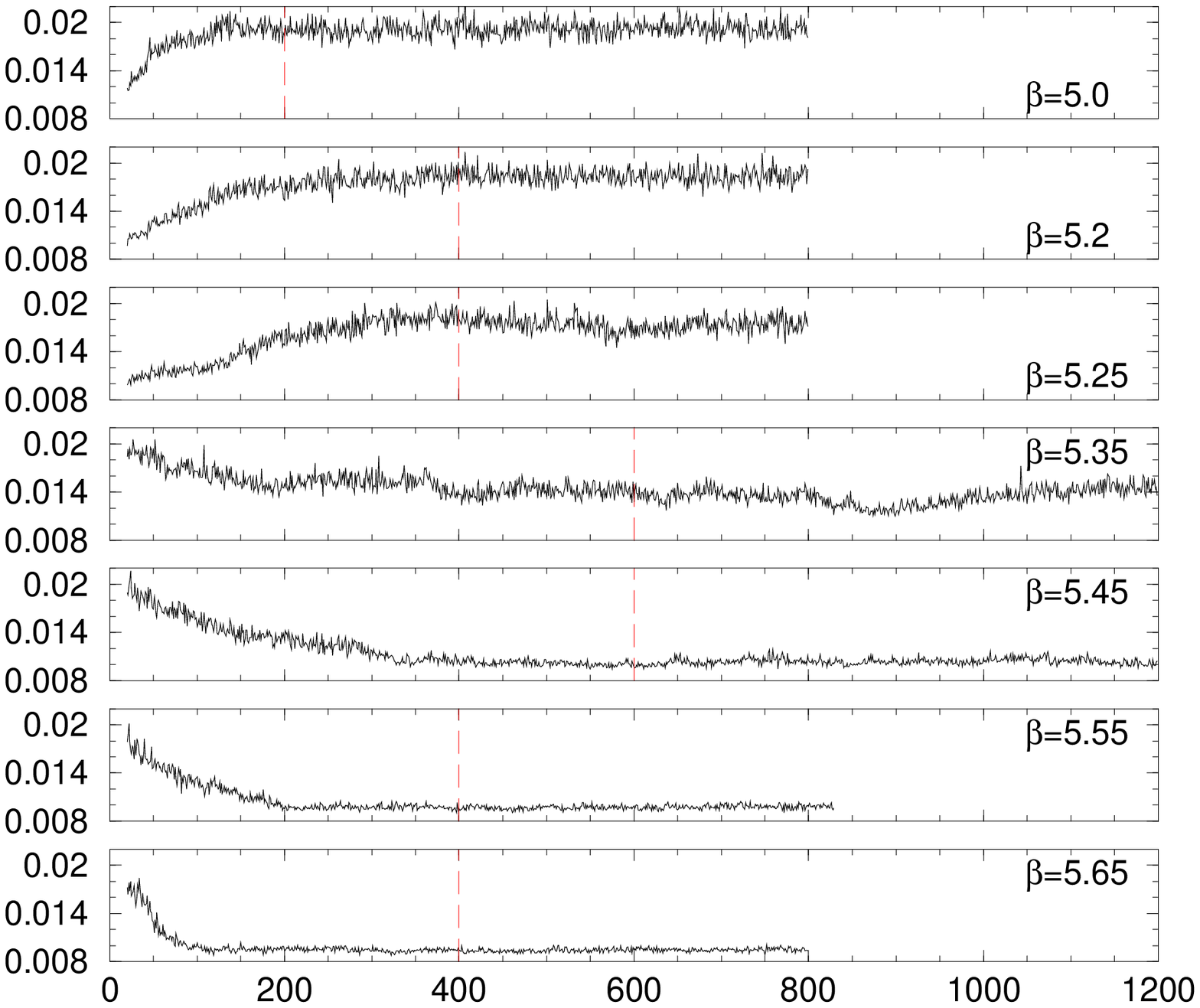}
  \caption[$\qbq$ evol: $8^3\times 4, m_0=1.9, L_s=12, m_f=0.1$]
    {$\qbq$ evol: $8^3\times 4, m_0=1.9, L_s=12, m_f=0.1$}
  \label{fig:pbp_wd_8nt4_h1.9_l12_m0.1}
\end{figure}

\begin{figure}[ptb]
  \centering
  \includegraphics
    [height=0.9\textheight,width=\textwidth]
    {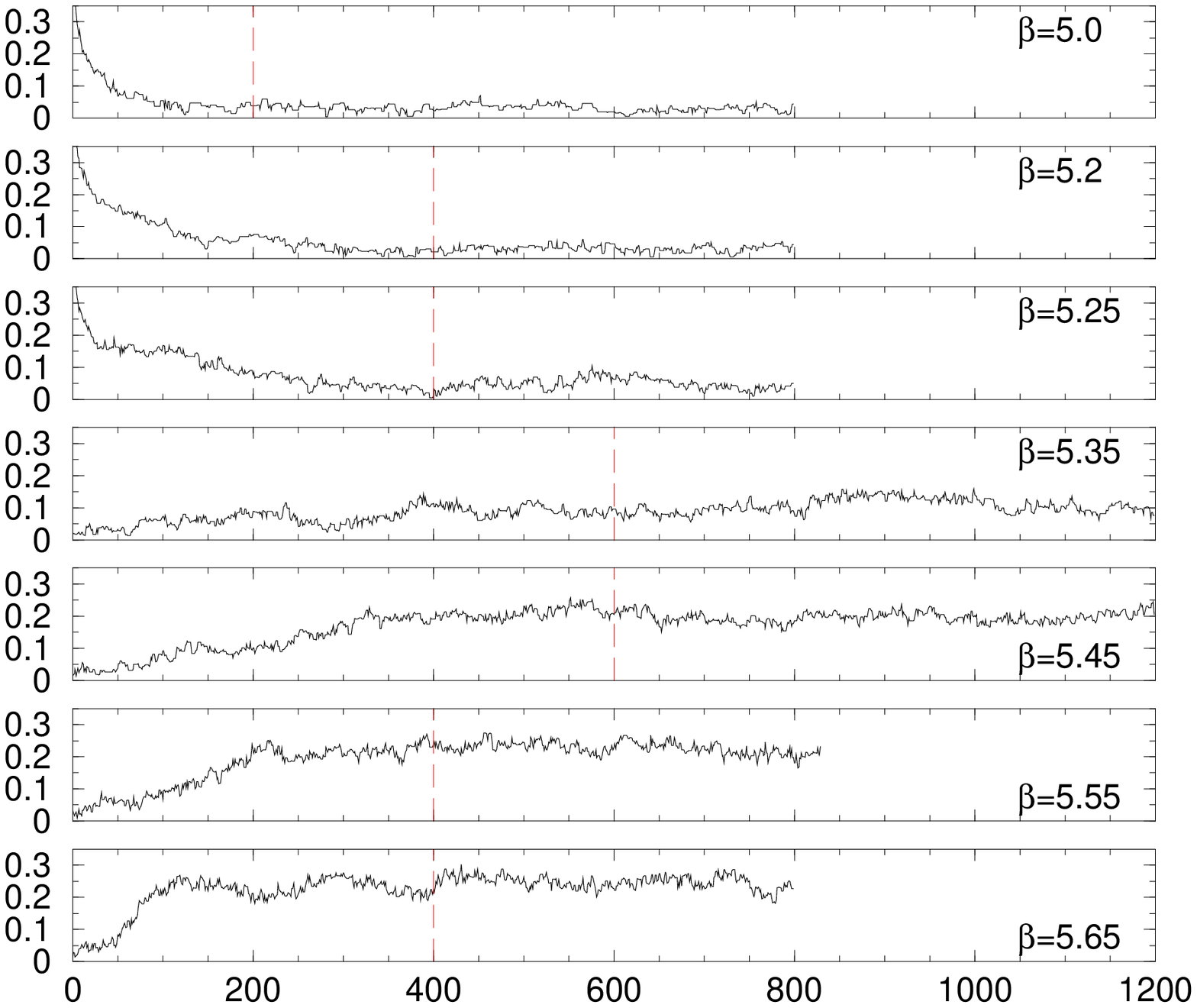}
  \caption[$\w3m$ evol: $8^3\times 4, m_0=1.9, L_s=12, m_f=0.1$]
    {$\w3m$ evol: $8^3\times 4, m_0=1.9, L_s=12, m_f=0.1$}
  \label{fig:w3m_wd_8nt4_h1.9_l12_m0.1}
\end{figure}

\begin{figure}[ptb]
  \centering
  \includegraphics
    [height=0.9\textheight,width=\textwidth]
    {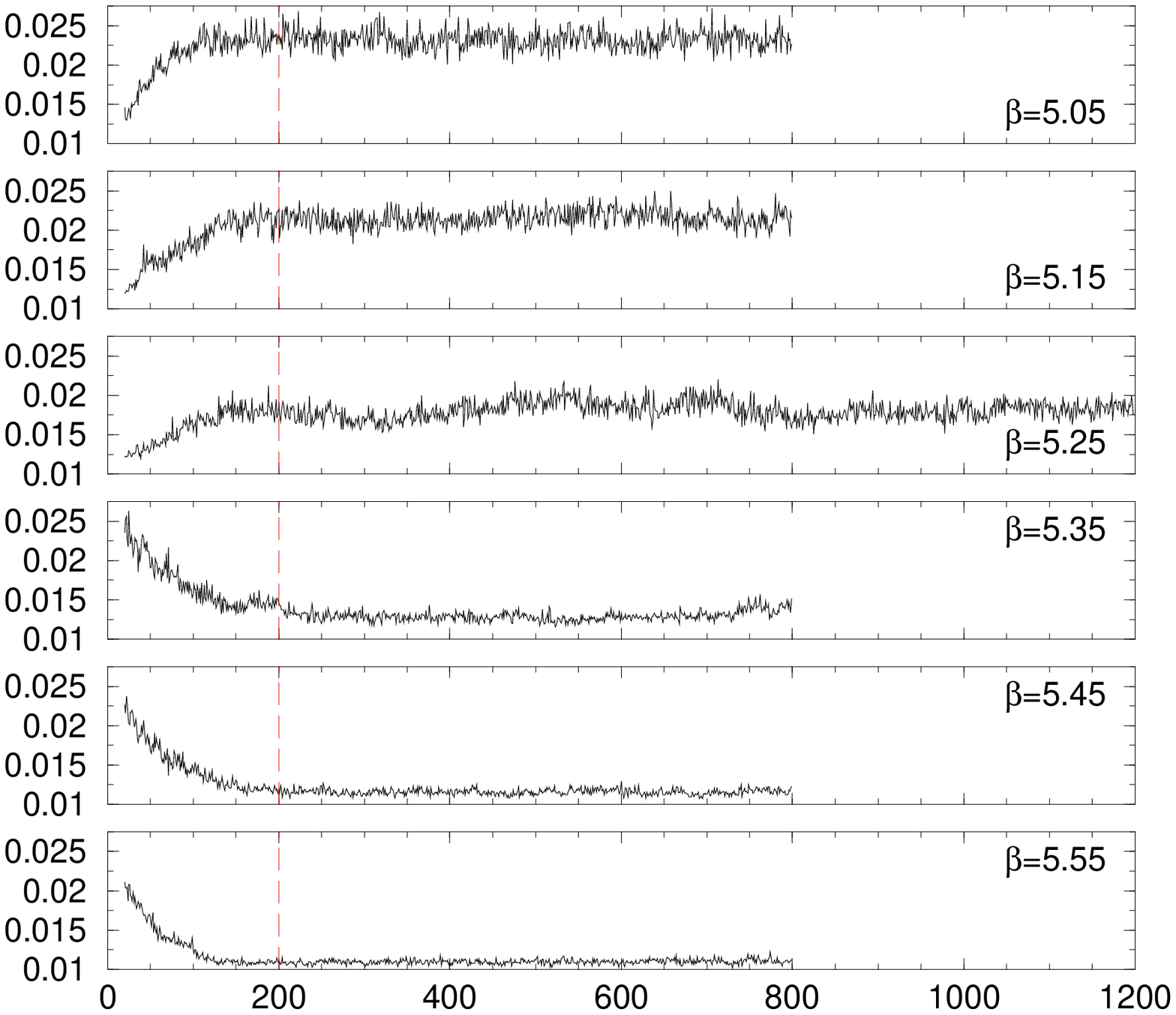}
  \caption[$\qbq$ evol: $8^3\times 4, m_0=2.0, L_s=12, m_f=0.1$]
    {$\qbq$ evol: $8^3\times 4, m_0=2.0, L_s=12, m_f=0.1$}
  \label{fig:pbp_wd_8nt4_h2.0_l12_m0.1}
\end{figure}

\begin{figure}[ptb]
  \centering
  \includegraphics
    [height=0.9\textheight,width=\textwidth]
    {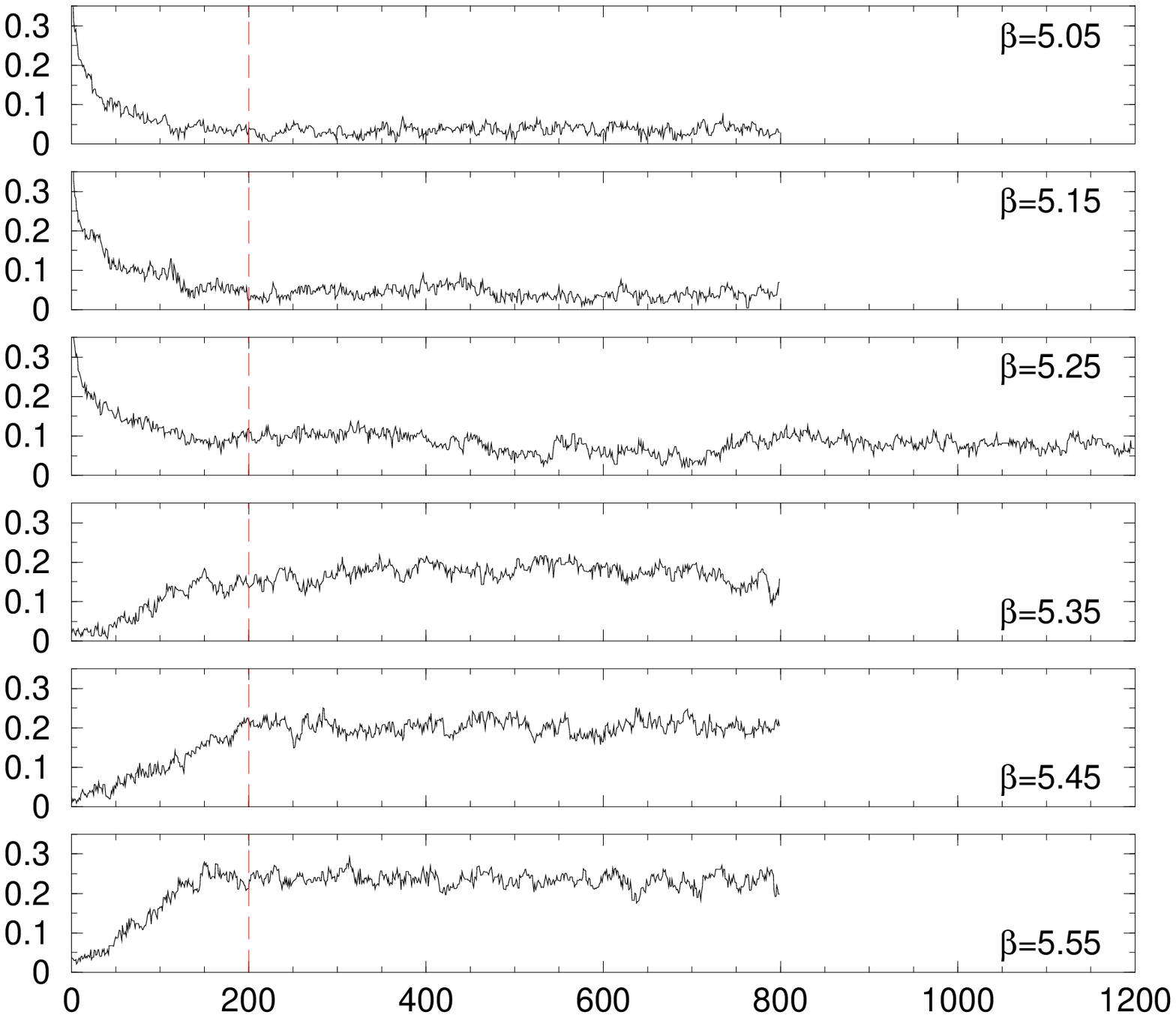}
  \caption[$\w3m$ evol: $8^3\times 4, m_0=2.0, L_s=12, m_f=0.1$]
    {$\w3m$ evol: $8^3\times 4, m_0=2.0, L_s=12, m_f=0.1$}
  \label{fig:w3m_wd_8nt4_h2.0_l12_m0.1}
\end{figure}

\begin{figure}[ptb]
  \centering
  \includegraphics
    [height=0.9\textheight,width=\textwidth]
    {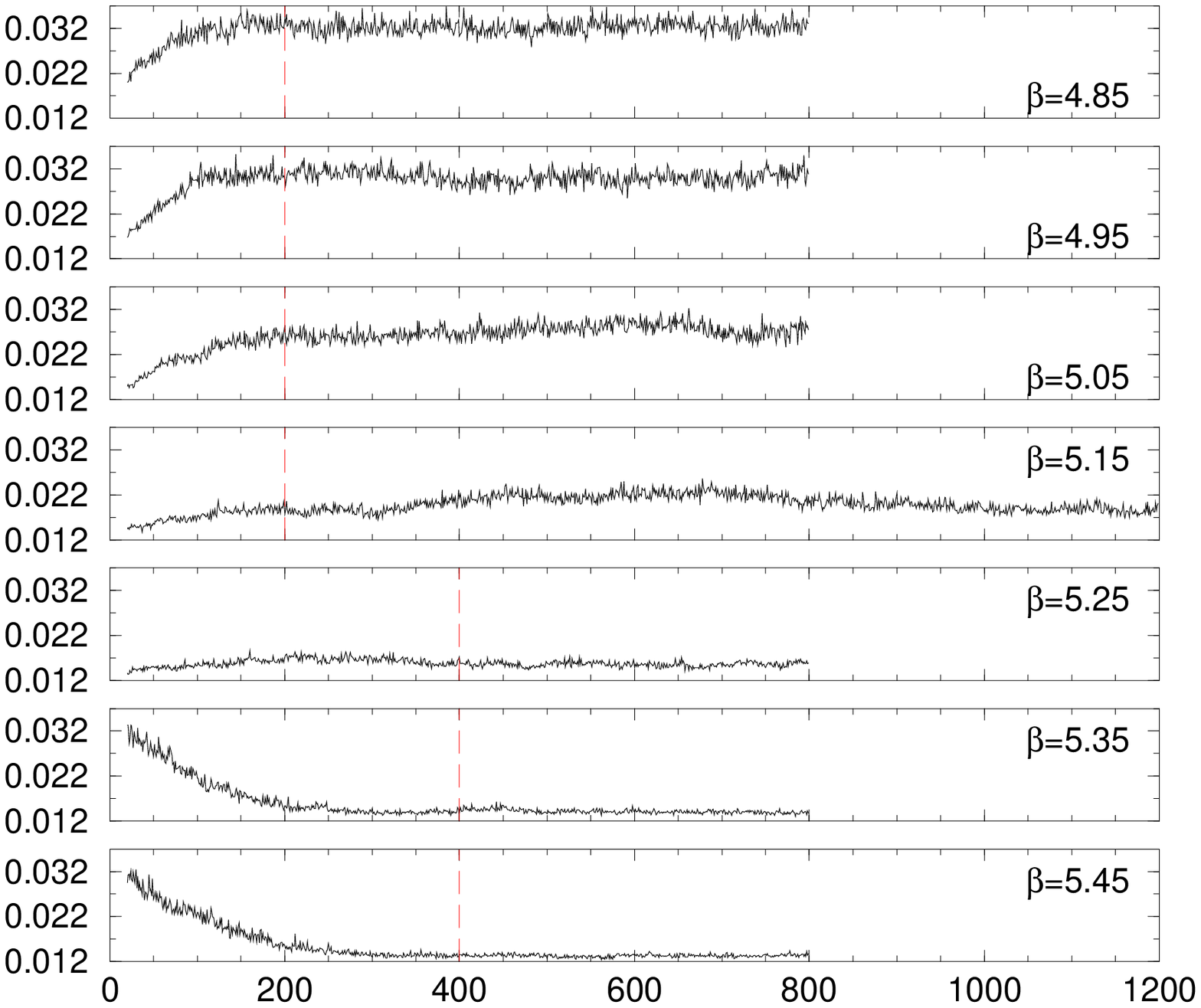}
  \caption[$\qbq$ evol: $8^3\times 4, m_0=2.15, L_s=12, m_f=0.1$]
    {$\qbq$ evol: $8^3\times 4, m_0=2.15, L_s=12, m_f=0.1$}
  \label{fig:pbp_wd_8nt4_h2.15_l12_m0.1}
\end{figure}

\begin{figure}[ptb]
  \centering
  \includegraphics
    [height=0.9\textheight,width=\textwidth]
    {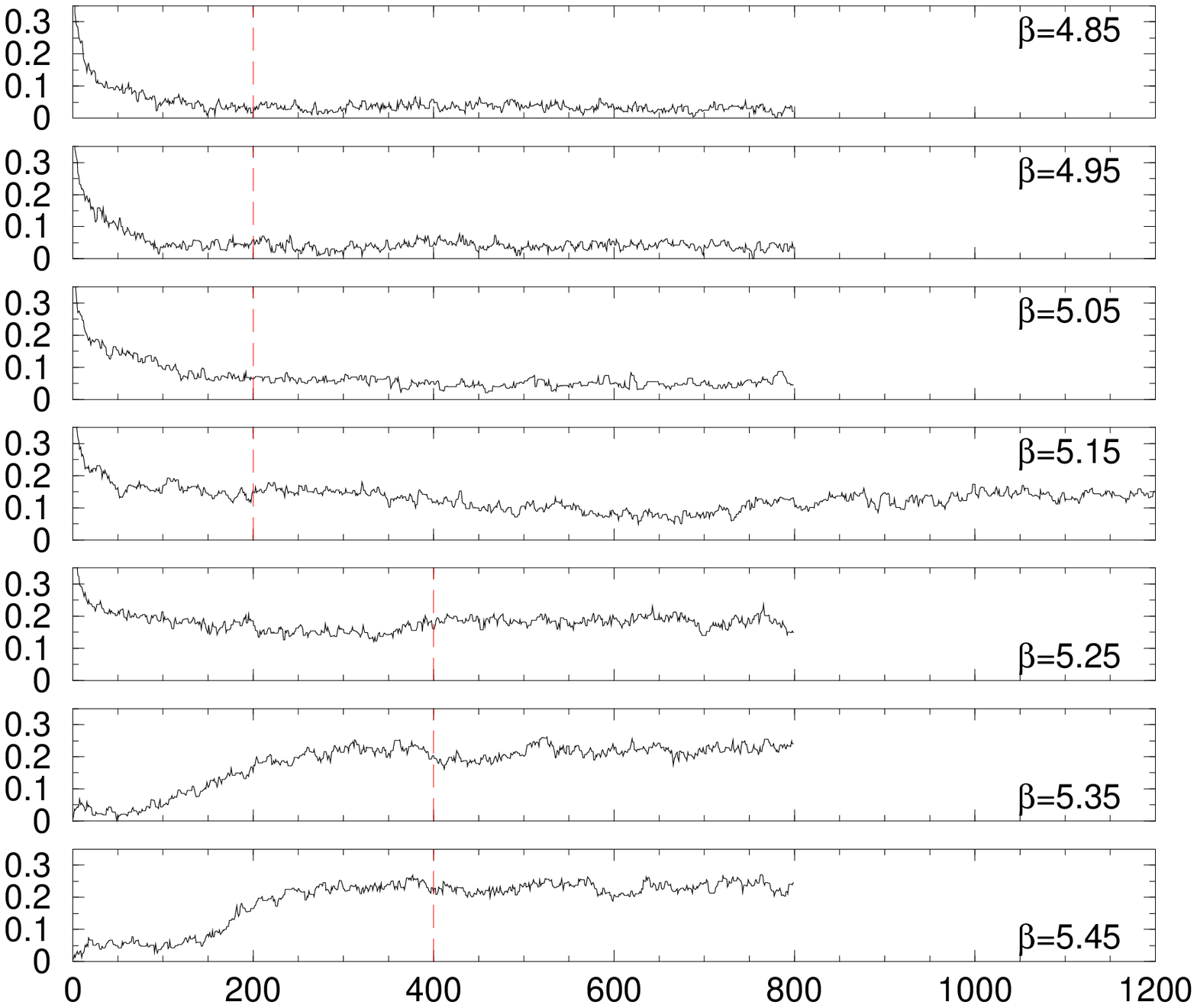}
  \caption[$\w3m$ evol: $8^3\times 4, m_0=2.15, L_s=12, m_f=0.1$]
    {$\w3m$ evol: $8^3\times 4, m_0=2.15, L_s=12, m_f=0.1$}
  \label{fig:w3m_wd_8nt4_h2.15_l12_m0.1}
\end{figure}

\begin{figure}[ptb]
  \centering
  \includegraphics
    [height=0.9\textheight,width=\textwidth]
    {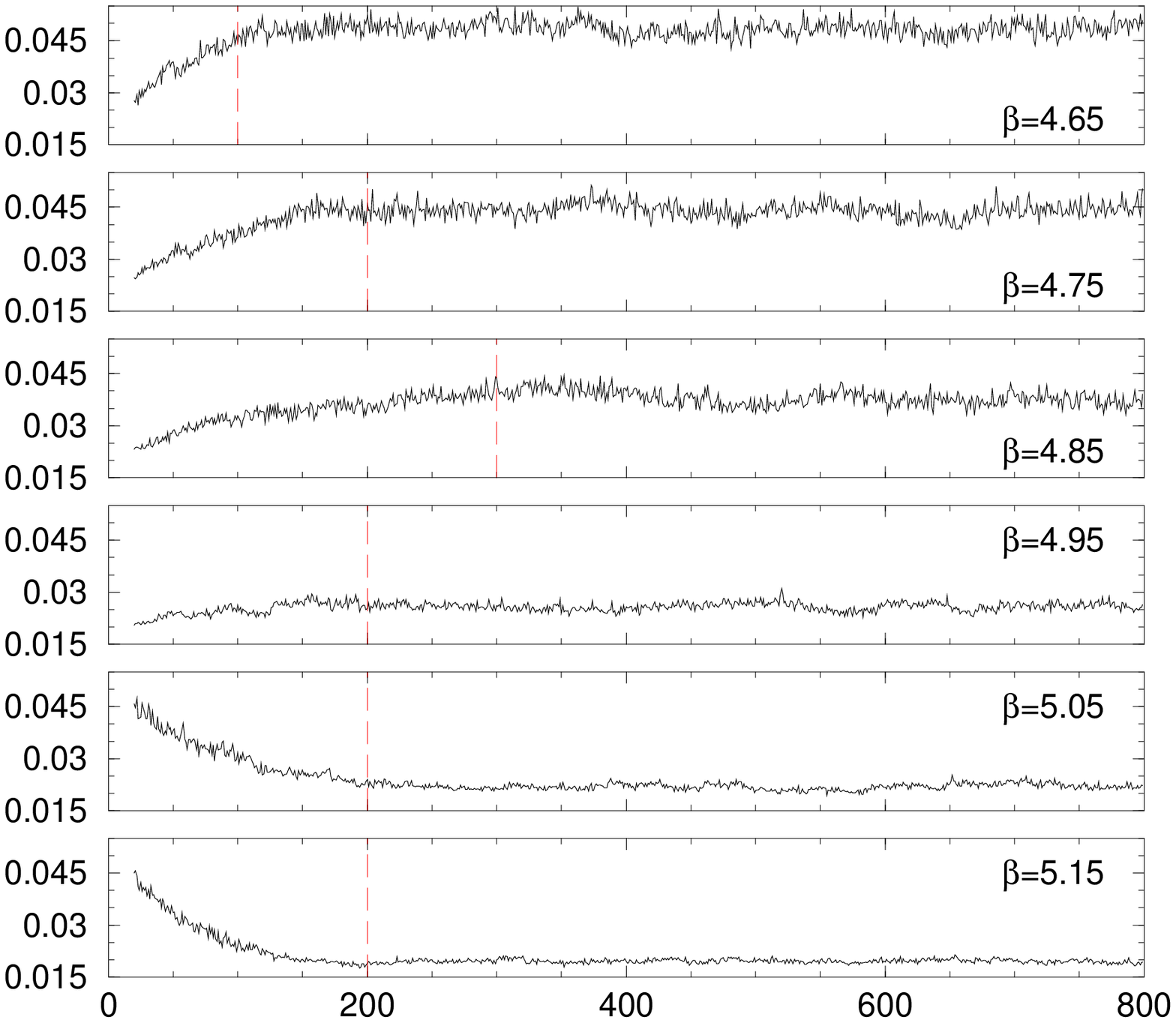}
  \caption[$\qbq$ evol: $8^3\times 4, m_0=2.4, L_s=12, m_f=0.1$]
    {$\qbq$ evol: $8^3\times 4, m_0=2.4, L_s=12, m_f=0.1$}
  \label{fig:pbp_wd_8nt4_h2.4_l12_m0.1}
\end{figure}

\begin{figure}[ptb]
  \centering
  \includegraphics
    [height=0.9\textheight,width=\textwidth]
    {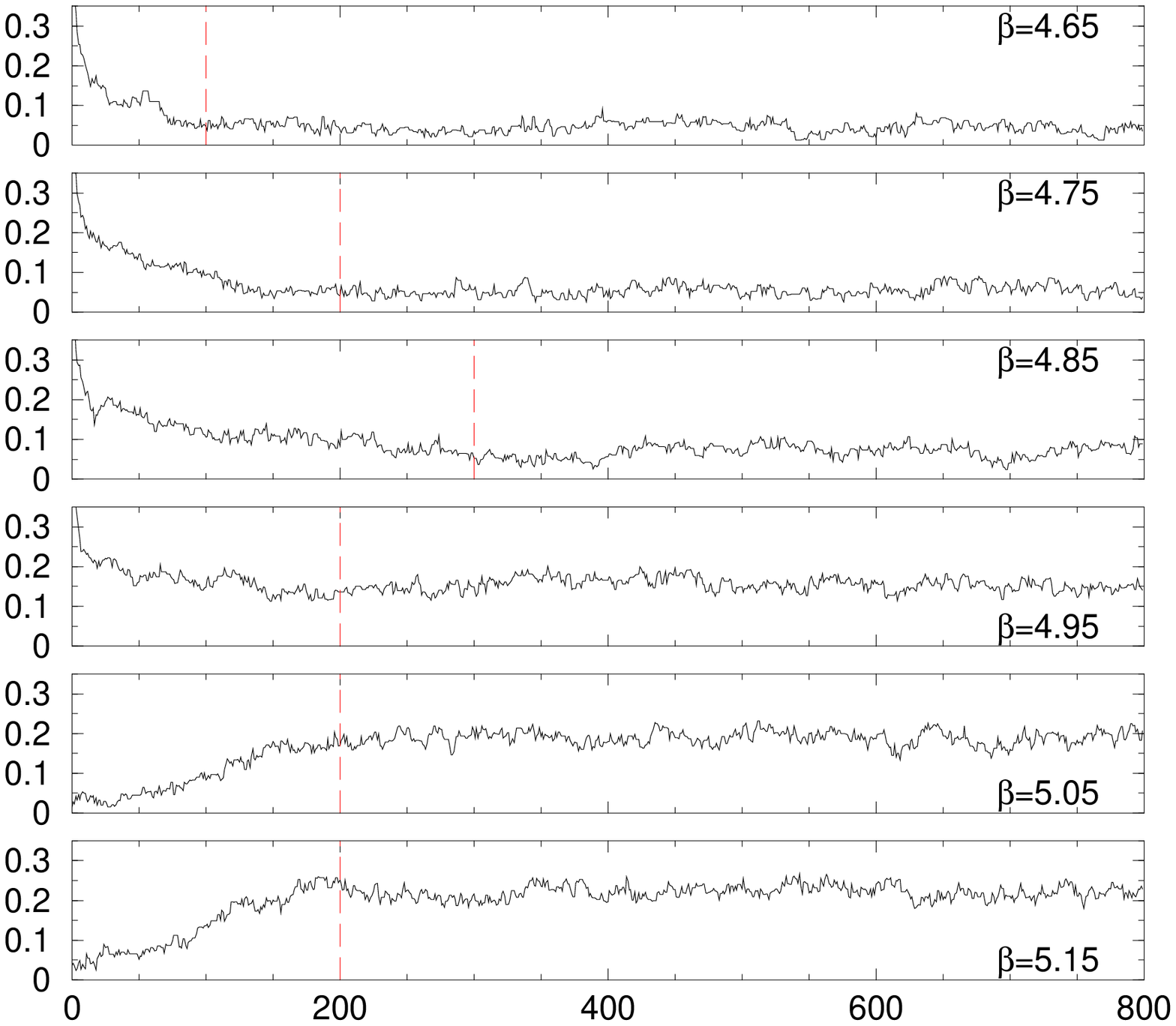}
  \caption[$\w3m$ evol: $8^3\times 4, m_0=2.4, L_s=12, m_f=0.1$]
    {$\w3m$ evol: $8^3\times 4, m_0=2.4, L_s=12, m_f=0.1$}
  \label{fig:w3m_wd_8nt4_h2.4_l12_m0.1}
\end{figure}

The evolutions of $\qbq$ and $\w3m$ are shown in figures
\ref{fig:pbp_wd_8nt4_h1.15_l12_m0.1} -- \ref{fig:w3m_wd_8nt4_h2.4_l12_m0.1}.
The vertical dashed line indicates that data points to the left are considered
part of thermalization and were not used to compute averages
and autocorrelated errors.  The figures clearly show that the short
(microcanonical) time fluctuations are much larger in the confined phase
than in the deconfined phase.  Note that the data show a tendency
for the thermalization time and the scale of large time fluctuations
to increase towards the crossover region.  This is an encouraging sign that
correlation lengths are increasing towards the center of the crossover region,
suggesting the long range order typical of phase transitions.

\begin{figure}[ptb]
  \centering
  \includegraphics
    [height=0.9\textheight,width=\textwidth]
    {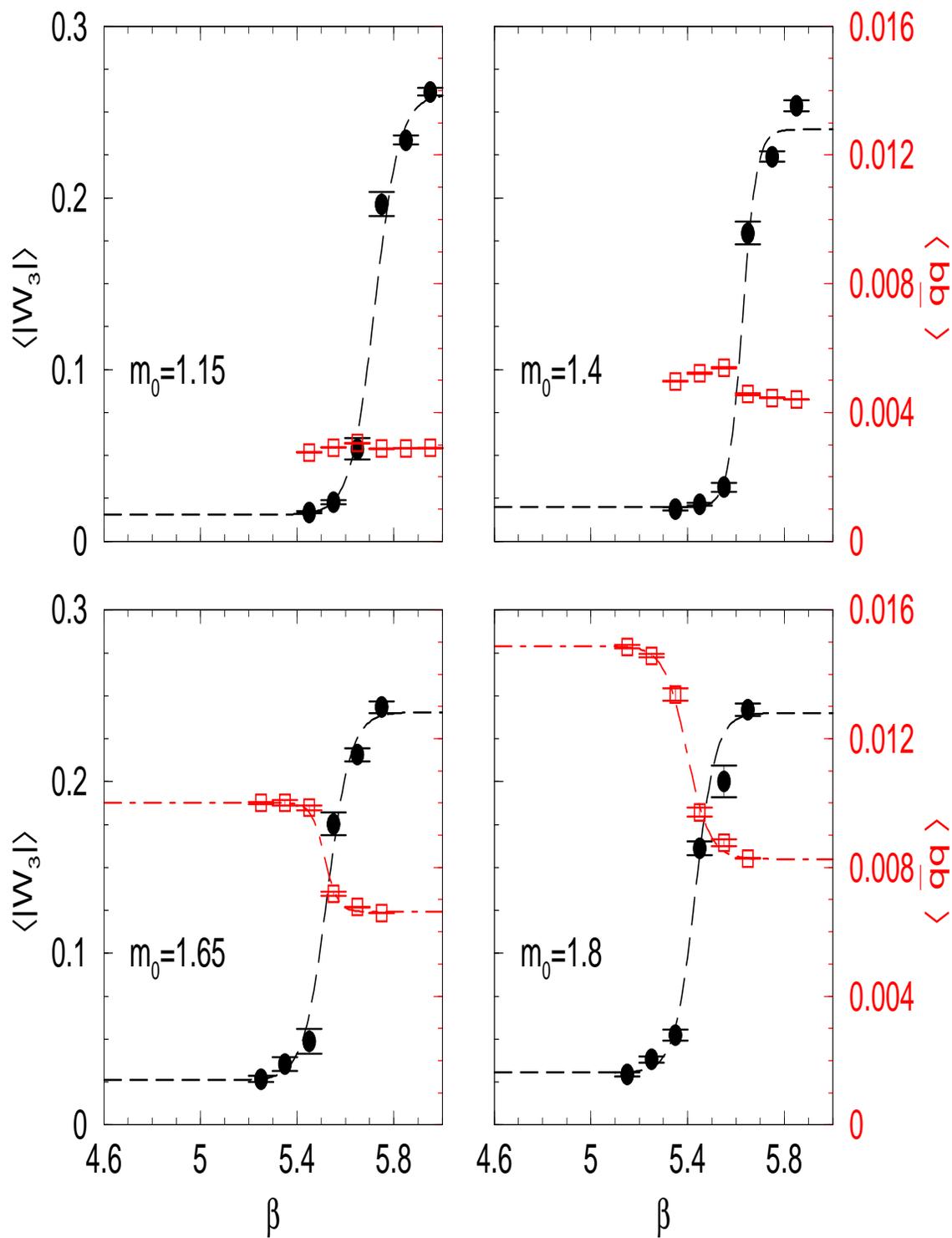}
  \caption[$\tanh$ fits of $\beta_c$: $8^3\times 4;m_0=1.15,\cdots,1.8;L_s=12;m_f=0.1$]
    {$\tanh$ fits of $\beta_c$: $8^3\times 4;m_0=1.15,\cdots,1.8;L_s=12;m_f=0.1$}
  \label{fig:pbp_w3m_wd_8nt4_h1.15-1.8_l12_m0.1}
\end{figure}

\begin{figure}[ptb]
  \centering
  \includegraphics
    [height=0.9\textheight,width=\textwidth]
    {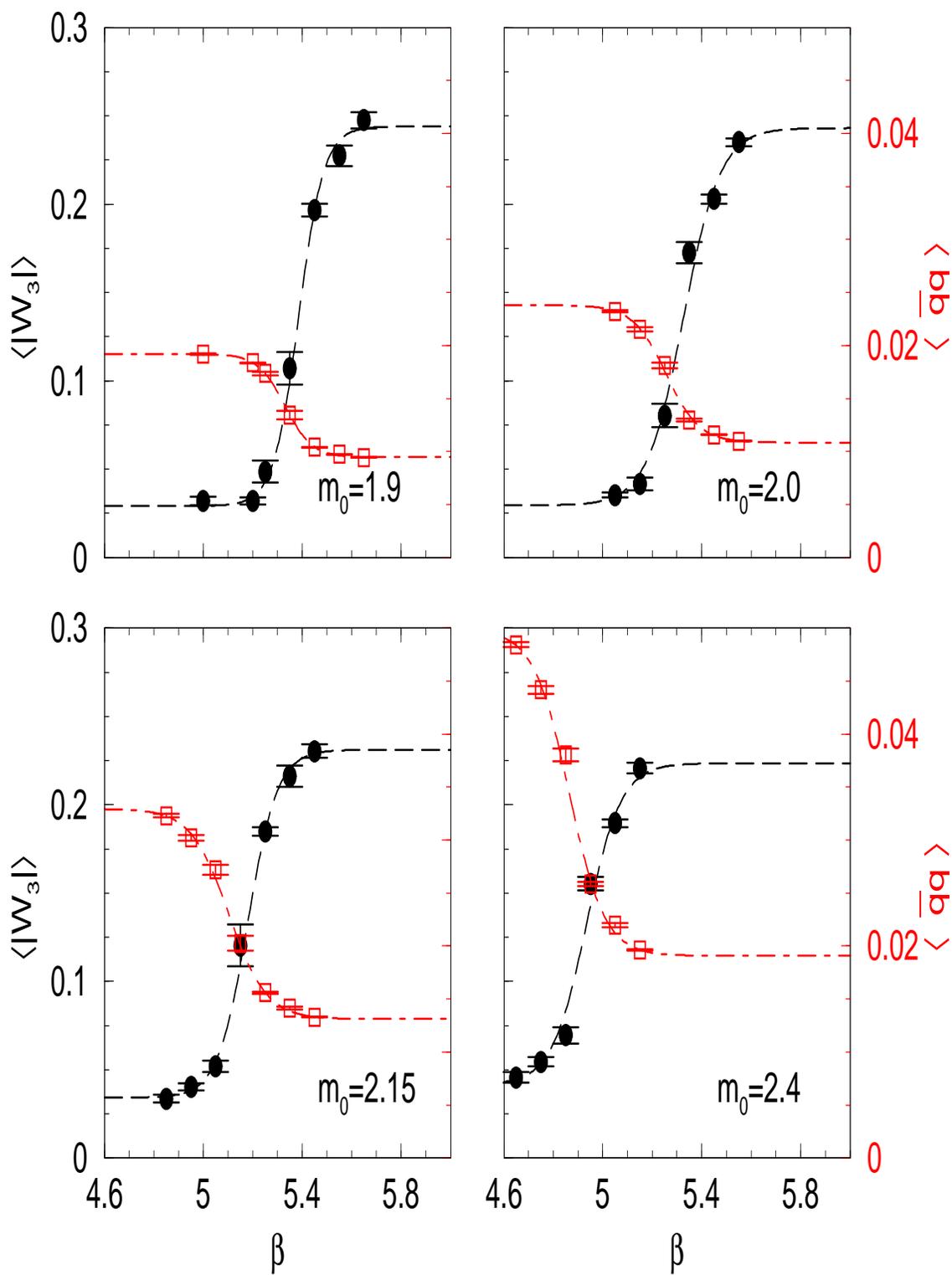}
  \caption[$\tanh$ fits of $\beta_c$: $8^3\times 4;m_0=1.9,\cdots,2.4;L_s=12;m_f=0.1$]
    {$\tanh$ fits of $\beta_c$: $8^3\times 4;m_0=1.9,\cdots,2.4;L_s=12;m_f=0.1$}
  \label{fig:pbp_w3m_wd_8nt4_h1.9-2.4_l12_m0.1}
\end{figure}

The averages and autocorrelated errors of $\qbq$ and $\w3m$ are plotted
versus $\beta$ for $m_0$ = 1.15 -- 1.8
in figure \ref{fig:pbp_w3m_wd_8nt4_h1.15-1.8_l12_m0.1} and $m_0$ = 1.9 -- 2.4
in figure \ref{fig:pbp_w3m_wd_8nt4_h1.9-2.4_l12_m0.1}.  For $m_0 \gtrsim 1.65$
there is a clear indication of simultaneous rapid change in $\qbq$ and $\w3m$
versus $\beta$, indicative of a potential phase transition smoothed out
by the moderately light quark mass and small volume.  The reader
should be careful to note the change in scale for $\qbq$
between the two figures.  For $m_0 \lesssim 1.4$, there is still a crossover
signal in $\w3m$, suspiciously close to the transition region for quenched QCD,
but any signal in $\qbq$ is nearly gone.

\begin{figure}[ptb]
  \centering
  \includegraphics
    [height=0.9\textheight,width=\textwidth]
    {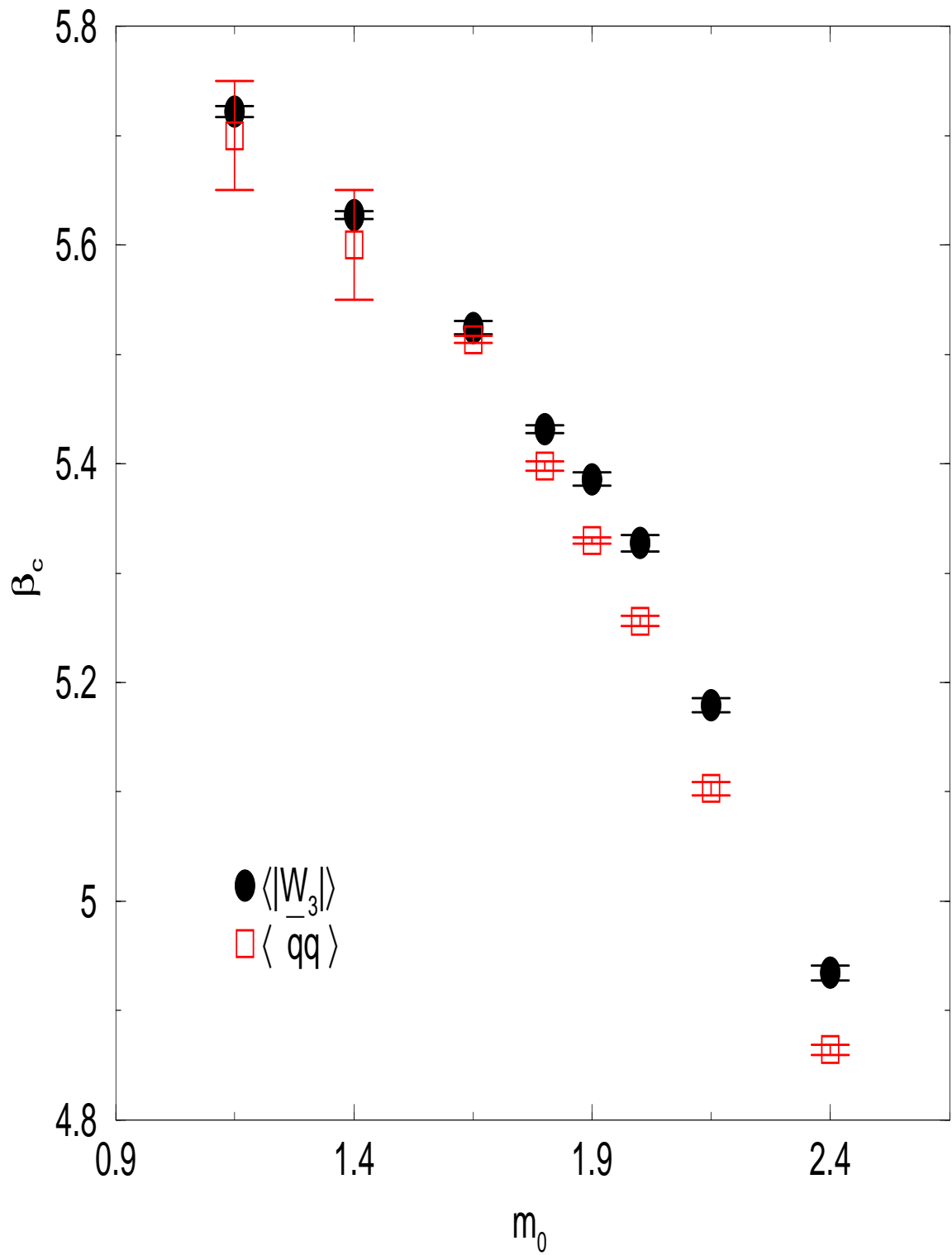}
  \caption[$\beta_c$ versus $m_0$: $8^3\times 4, L_s=12, m_f=0.1$]
    {$\beta_c$ versus $m_0$: $8^3\times 4, L_s=12, m_f=0.1$}
  \label{fig:beta_crit_wd_8nt4_l12_m0.1}
\end{figure}

Once we have identified the crossover region, we can estimate the location
of $\beta_c$ through a fit of the approximate order parameters to the ansatz
\beq
\label{eq:beta_c_tanh_ansatz}
f(\beta)=c_0\left\{c_1+\tanh\left[c_2\left(\beta-\beta_c\right)\right]\right\}.
\eeq
The results of the fits are in tables \ref{tab:w3m_tanhfit_8nt4_l12_m0.1}
and \ref{tab:pbp_tanhfit_8nt4_l12_m0.1}, plots of the fit curves versus
the data are also shown
in figures \ref{fig:pbp_w3m_wd_8nt4_h1.15-1.8_l12_m0.1} and
\ref{fig:pbp_w3m_wd_8nt4_h1.9-2.4_l12_m0.1}, and a plot of $\beta_c$
versus $m_0$ is in figure \ref{fig:beta_crit_wd_8nt4_l12_m0.1}.

Although the purpose of the ansatz (\ref{eq:beta_c_tanh_ansatz}) is to estimate
the inflection point $\beta_c$, the other parameters of the fit can be used
to quantitatively describe what our eyes have already told us by looking
at the figures.  After extracting the overall scale factor $c_0$ we can compare
$c_1$ and $c_2$ between the various fits.  When ${\rm sign}(c_0) c_1 = 1$
then the fit describes an approximate order parameter that goes exactly to zero
deep in the the symmetric phase.  $c_0 c_2$ is the slope of the fit curve
at the inflection point, so $c_2$ indicates the relative rate of crossover
between the two phases.  Examining table \ref{tab:w3m_tanhfit_8nt4_l12_m0.1},
we see from $c_1$ that the Wilson line is a good approximate order parameter
and from $c_2$ that there is a tendency for the transition to broaden
as $m_0$ increases. 

In table \ref{tab:pbp_tanhfit_8nt4_l12_m0.1} we see more interesting
behavior.  First, we note
in figure \ref{fig:pbp_w3m_wd_8nt4_h1.15-1.8_l12_m0.1}
that the $\beta$-dependence of $\qbq$ for $m_0 \lesssim 1.4$ is not well
modeled by (\ref{eq:beta_c_tanh_ansatz}).  The tendency of $\left|
c_0 \right|$ to increase with increasing $m_0$ is expected since this is
just the change in the overlap between our effective quark fields
(\ref{eq:dwf_4d_fermions})  and the actual light fermion modes suggested
by (\ref{eq:free_m_eff}).  The vanishing of this overlap as $m_0 \lesssim
1.4$ is our first indication that any light fermion modes are, on average,
no longer localized to the domain wall defect.  Next, we note that $\left|
c1 \right| \gg 1$ for $m_0 \lesssim 1.4$ which indicates that $\qbq$ is no
longer approaching zero in the symmetric phase and again suggests the
phase space for light fermions on the defect is nearly gone.  We
conclude that domain wall fermions are unreliable for $N_f$=2 and $L_t$=4
thermodynamics with $m_0 \lesssim 1.4$.  Because the fit ansatz is only
intended to give a qualitative description of the data, we expect
that $\chi^2/N_{\rm dof} \lesssim 10$ indicates a reasonable agreement
between the data and the ansatz.

Finally, for $c_2$ we see the same tendency for the transition to broaden
towards larger $m_0$.  We expect that for some $m_0 > 2.0$ the simulations
should behave like $N_f$=8 QCD but our data do not indicate any sharp
crossover from $N_f$=2 QCD to $N_f$=8 QCD for $m_0\!\leq\!2.4$.  One
interpretation is that the simulations are $N_f$=2 with the shift in
$\beta_c$ due to the change in the amount of phase space available to the
light fermions as $m_0$ varies, in analogy to the free field condition
(\ref{eq:free_dwf_condition}).  A related interpretation is that as the
phase space for the two light modes decreases while the modes themselves
become heavy for $m_0 \gtrsim 2.0$ the phase space for the eight doubler
modes is increasing while the doubler modes become light, so that $m_0$
smoothly interpolates between two flavor and eight flavor thermodynamics. 

Although the rapid crossovers seen in $8^3\!\times\!4$ volumes with $L_s$=12
and $m_f$=0.1 are indicative of a phase transition, the dynamical quarks
are likely too heavy to expect good chiral behavior.  Furthermore,
it is difficult at this point to quantify the residual chiral symmetry breaking
effects due to finite $L_s$.  We can strengthen our evidence for the existence
of the phase transition by performing a dynamical extrapolation
to the massless quark limit:  $L_s\!\to\!\infty, m_f\!\to\!0$.  Thus,
if we study the order parameter $\qbq$ outside the transition region
and demonstrate that it is zero in the symmetric phase and non-zero
in the broken phase then we are reassured that the rapid cross over
is strong evidence for a phase transition in the chiral limit.

We choose $m_0$=1.9 to study the separate phases for several reasons. 
First, $m_0$=1.9 is sufficiently far from the region $m_0 \lesssim 1.4$
where the light surface states of domain wall fermions vanishes.  Second,
choosing $m_0$<2 guarantees that at low momentum only a single light
surface mode is bound to the domain wall so ensuring the low momentum phase
space of an $N_f$=2 theory after squaring the determinant.  Finally,
although not strictly necessary, the value of $\beta_c$ is similar to the value
for $N_f$=2 staggered fermions \cite{Brown:1990ev,Vaccarino:1991kn}.

To study the confining (broken) phase in an $8^3\!\times\!4$ volume
with $m_0$=1.9, we choose $\beta$=5.2 as a good compromise between the desire
to work at as weak a coupling as possible to suppress the residual chiral
chiral symmetry breaking effects of finite $L_s$ and the need to be far enough
from the crossover region that the results will be unaffected by the presence
of a nearby phase transition.  To perform the chiral extrapolation, we varied
the bare quark mass $m_f$ over the range 0.02 to 0.18 and the extent of the
extra dimension $L_s$ over the range 4 to 40.  The run parameters and results
are summarized in table \ref{tab:8nt4_b5.2_h1.9}.

\begin{figure}
  \centering
  \includegraphics
    [height=0.9\textheight,width=\textwidth]
    {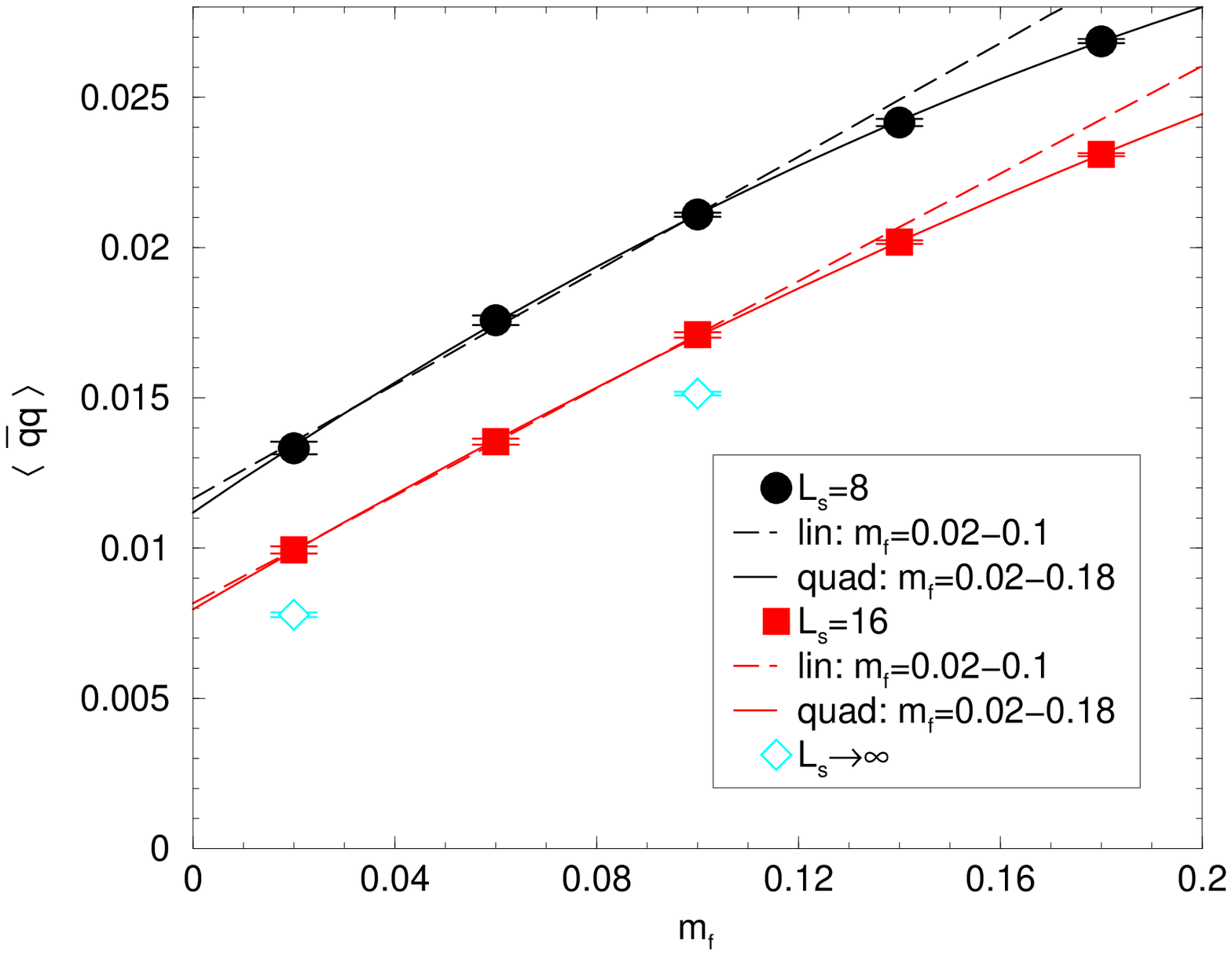}
  \caption{
    Fit of $\qbq = c_0 + c_1 m_f +c_2 m_f^2$:
    $8^3\times 4; \beta=5.2; m_0=1.9; L_s=8,16$
  }
  \label{fig:pbp_mf_fit_wd_8nt4_b5.2_h1.9}
\end{figure}

Focusing on the chiral condensate, we expect from perturbation theory that
$\qbq$ will be linear in the dynamical quark mass, $m_f$.  In figure
\ref{fig:pbp_mf_fit_wd_8nt4_b5.2_h1.9}, we show constrained fits to both
a linear ansatz and a quadratic ansatz for $L_s = 8, 16$
and $m_f$ = 0.02 -- 0.18.  By examining the $\chi^2/N_{\rm dof}$
for each constrained fit in table \ref{tab:pbp_mf_fit_wd_8nt4_b5.2_h1.9}
we can see that the data tend to be more linear in $m_f$ as $L_s$ increases. 

\begin{figure}
  \centering
  \includegraphics
    [height=0.9\textheight,width=\textwidth]
    {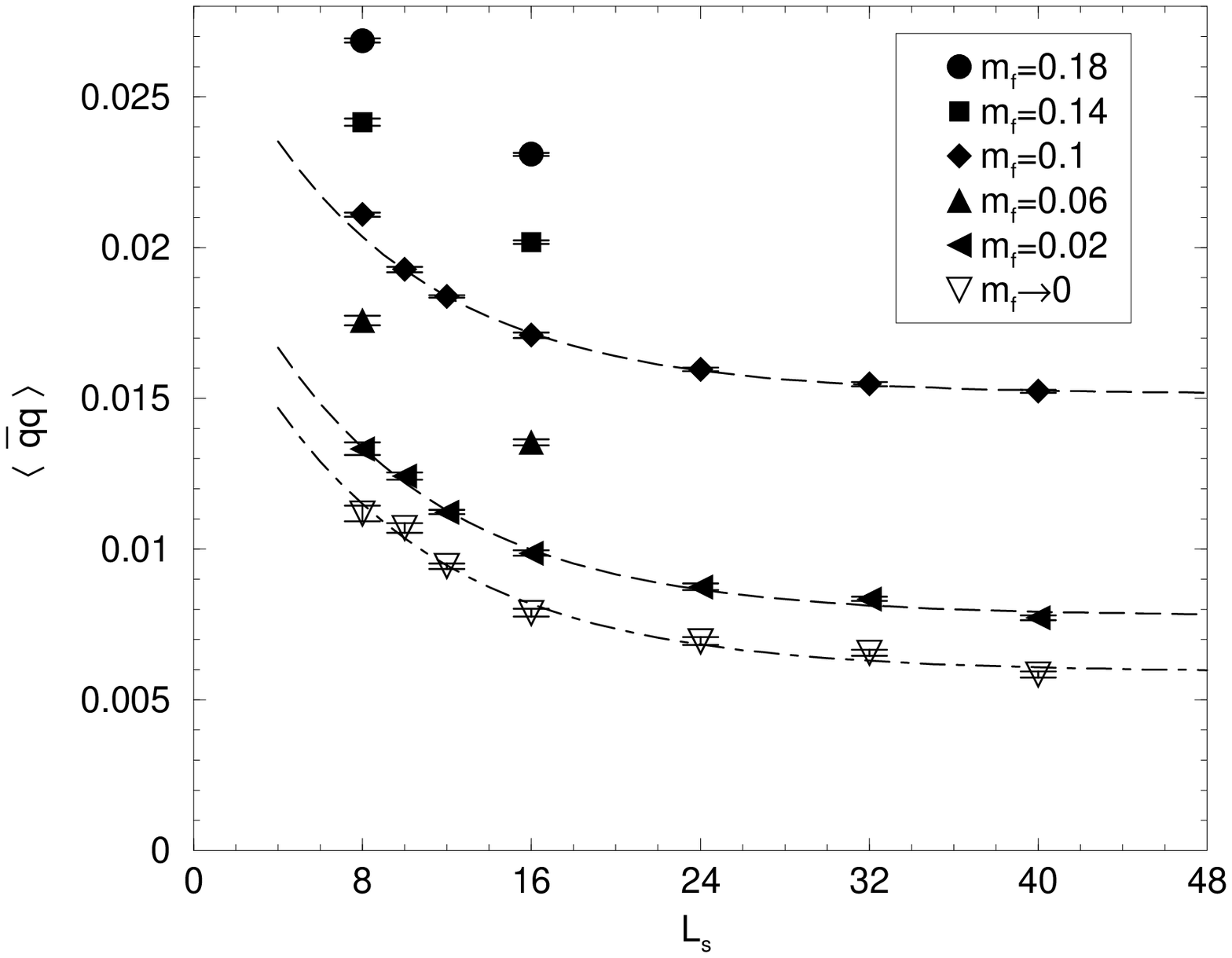}
  \caption{
    Fit of $\qbq = c_0 + c_1 \exp \left( -c_2 L_s \right)$:
    $8^3\times 4, \beta=5.2, m_0=1.9$
  }
  \label{fig:pbp_ls_fit_wd_8nt4_b5.2_h1.9}
\end{figure}

For $m_f=0.02,0.1$ there are sufficient data to extrapolate $\qbq$ to
$L_s\to\infty$ limit using an exponential ansatz.  The results of these fits
are shown in table \ref{tab:pbp_ls_fit_wd_8nt4_b5.2_h1.9} and figure
\ref{fig:pbp_ls_fit_wd_8nt4_b5.2_h1.9}.  We can see that the fit of the data
to the ansatz is reasonable.  One might worry that the $\chi^2/N_{\rm dof}$
are somewhat large but it is likely that the true autocorrelated errors
are underestimated due to the rather short length of evolutions.
Using the values of $\qbq$ as $m_f\to 0$ from the fits in table
\ref{tab:pbp_mf_fit_wd_8nt4_b5.2_h1.9}, we can also extrapolate
to $L_s\to\infty$ and we find
\beq
\lim_{L_s\to\infty} \left[ \lim_{m_f\to 0} \qbq(\beta=5.2) \right] = 0.0061(1).
\eeq
Checking to see if the order of limits changes the extrapolated value we
take the $L_s\to\infty$ limits for $m_f=0.02,0.1$ and do an unconstrained
linear extrapolation to $m_f\to 0$ and find
\beq
\lim_{m_f\to 0} \left[ \lim_{L_s\to\infty} \qbq(\beta=5.2) \right] = 0.0061(1).
\eeq
Thus, we conclude that there is a broken phase for $\beta \lesssim 5.2$
indicated by the $\qbq$ order parameter in the chiral limit.

\begin{figure}
  \centering
  \includegraphics
    [height=0.9\textheight,width=\textwidth]
    {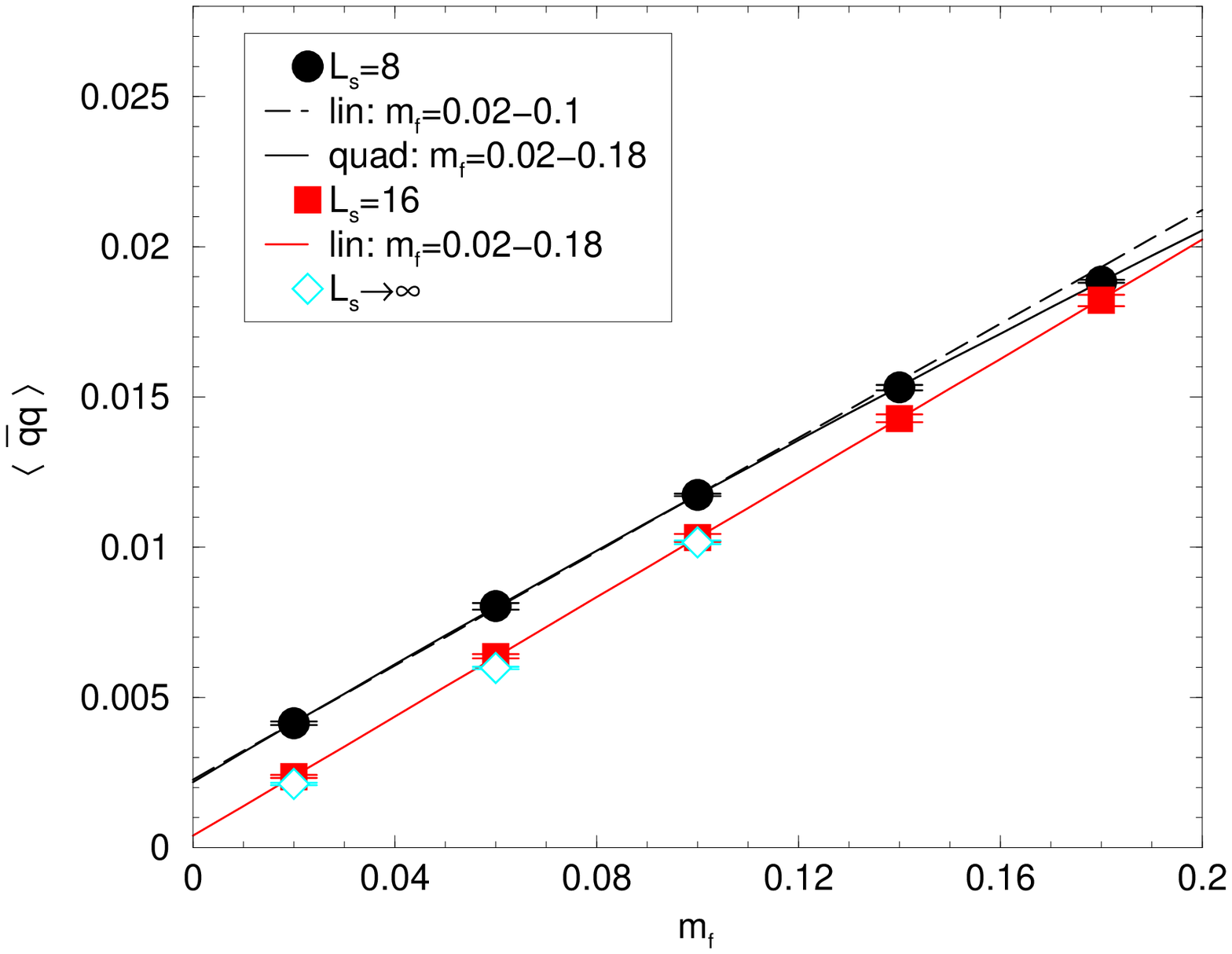}
  \caption{
    Fit of $\qbq = c_0 + c_1 m_f +c_2 m_f^2$:
    $8^3\times 4; \beta=5.45; m_0=1.9; L_s=8,16$
  }
  \label{fig:pbp_mf_fit_wd_8nt4_b5.45_h1.9}
\end{figure}

To study the deconfined (symmetric) phase in $8^3\!\times\!4$ volumes
with $m_0$=1.9, we choose to simulate with $\beta$=5.45.  As before,
we desire to run at as weak a coupling as is reasonable, but it is important
not to go too deep into the symmetric phase as our physical volume will
become too small and too hot to yield useful results.
As before we varied $m_f$ over 0.02 -- 0.18 and $L_s$ over 4 -- 32.
The run parameters and results are summarized in table
\ref{tab:8nt4_b5.45_h1.9}.  In figure \ref{fig:pbp_mf_fit_wd_8nt4_b5.45_h1.9},
we show constrained fits for a linear and quadratic ansatz
for $L_s$=8, $m_f$ = 0.02 -- 0.18 and a fit to a linear ansatz for $L_s$=16
over the same range of $m_f$.  We see immediately that the data appear
much more linear than in the broken phase.  This may occur because the rate
of exponential suppression of chiral mixing of the domain wall states
increases at weaker coupling.  Another possibility is that the rate
of exponential suppression may be larger in the symmetric phase
due to the smoother gauge fields.  This remains a question open for further
study.  Results of the fits are
in table \ref{tab:pbp_mf_fit_wd_8nt4_b5.45_h1.9}, where we have also included
fits of the $L_s$=16 data constrained be zero at $m_f$=0. 

As in the $\beta=5.2$ study, we also fit exponential extrapolations to
$L_s\to\infty$, see table \ref{tab:pbp_ls_fit_wd_8nt4_b5.45_h1.9} and
following the same double limit procedure above, if we do not constrain
the $L_s\to\infty, m_f\to 0$ limit of $\qbq$ to be zero, we get the
extrapolated values
\beqs
\lim_{L_s\to\infty} \left[ \lim_{m_f\to 0} \qbq(\beta=5.45) \right]
& = & 0.00010(5) . \\
\lim_{m_f\to 0} \left[ \lim_{L_s\to\infty} \qbq(\beta=5.45) \right]
& = & 0.00011(5) .
\eeqs
Furthermore, if we constrain the $L_s\to\infty, m_f\to 0$ limit of $\qbq$
to be zero, the quality of the fits does not change significantly.
Thus, we conclude that our observations are explained by a
deconfining chiral phase transition with $\qbq$ as the order parameter
for that transition in the chiral limit.

\section{The transition region in $16^3\!\times\!4$ volumes}
\label{sec:beta_crit_16nt4}

In section \ref{sec:beta_crit_8nt4}, we provided strong evidence
of a chiral phase transition for $N_f$=2 domain wall fermions
in the massless limit:  $m_f\!\to\!0$, $L_s\!\to\!\infty$.
After having bracketed the transition region in smaller spatial volumes,
we would like to characterize the order of the phase transition.
Working in large spatial volumes with light quarks is essential to observing
the long range correlations typical of true critical behavior.  We chose
to increase the volume from $8^3\!\times\!4$ to $16^3\!\times\!4$.
To make the quarks lighter, we chose a larger $L_s$=24 and a smaller $m_f$=0.02.

Recalling from QCD with staggered fermions that $\beta_c$ decreased slightly
as the quark mass is decreased, we first ran simulations in $8^3\!\times\!4$
volumes to map out the shift in $\beta_c$.  Further, having simulations in
the transition region in two different spatial volumes will allow us to study
systematic errors due to finite spatial volumes.  The simulations were run
at $\beta$ = 5.2, 5.275, 5.325 and 5.45 with $m_0$=1.9.
Table \ref{tab:wd_8nt4_h1.9_l24_m0.02} contains the other run parameters
for these simulations.  As before, there were no difficulties with convergence
of the CG algorithm.  300--450 CG iterations were typically needed per HMD step
in the broken phase and transition region and 100--120 CG iterations
were needed in the symmetric phase.  Simulations in the broken phase took
about thirty days to complete 800 HMC trajectories when run on a six Gflops
partition of the QCDSP computer.

The evolutions of $\qbq$ and $\w3m$ are shown in figures
\ref{fig:pbp_wd_8nt4_h1.9_l24_m0.02} and \ref{fig:w3m_wd_8nt4_h1.9_l24_m0.02}. 
The vertical dashed line indicates that data to the left are considered part
of thermalization and were not used to compute averages and autocorrelated
error estimates.  Even more dramatically than in section
\ref{sec:beta_crit_8nt4}, the short time fluctuations of $\qbq$ are much larger
in the broken phase than the symmetric phase and the longer time correlations
are clearly visible at $\beta$=5.325.  In figure
\ref{fig:pbp_w3m_wd_8nt4_h1.9_l24_m0.02} we show $\qbq$ and $\w3m$
{\it vs}.~$\beta$ and we see that $\beta$=5.325 is clearly in the middle
of the transition region.

For $16^3\!\times\!4$ volumes, we chose $\beta$ = 5.25 -- 5.35, spaced at
0.025 intervals.  Table \ref{tab:wd_16nt4_h1.9_l24_m0.02} contains
the other run parameters.  Both ordered and disordered starts were run
for each set of run parameters until observables measured on last three hundred
trajectories of each simulation were in good statistical agreement.
There were no difficulties with CG convergence.  260--300 CG iterations
were typically needed per HMD step, with more iterations required in
the middle of the transition region $(\beta \sim 5.325)$ than on the edges.
To complete 800 HMC trajectories on a 25 Gflops partition of the QCDSP computer
would take between 48 and 72 days.

The evolutions of $\qbq$ and $\w3m$ are shown
in figures \ref{fig:pbp_wd_16nt4_h1.9_l24_m0.02}
and \ref{fig:w3m_wd_16nt4_h1.9_l24_m0.02}.  Since two evolutions appear
in each panel of the figure, separate vertical lines are used to indicate
the start of the last 300 trajectories for each evolution in the panel.
The evolutions show no convincing evidence for a two-state signal
indicative of a first order phase transition.

The summary plots which display $\qbq$ and $\w3m$ versus $\beta$ are shown
in figures \ref{fig:pbp_wd_h1.9_l24_m0.02} and \ref{fig:w3m_wd_h1.9_l24_m0.02}. 
The most obvious feature of the transition for domain wall fermions is that
it does not appear particularly steep.  For comparison, we have also plotted
$N_f$=2 staggered fermion data \cite{Vaccarino:1991kn}.  Based on the different
symmetries of the two fermion formulations, the staggered data show the
transition when one pion is light and two are heavy whereas the DWF data
show the transition for three degenerate pions.

Comparing the $8^3$ and $16^3$ volumes, we also see that the larger volume
did reduce the short distance fluctuation in the signal.  However, the
point in the middle of the transition region, $\beta=5.325$, does not show
signs of the critical fluctuations indicative of a second order transition
nor does it show the two state signal of a first order phase transition.
Based on the broadness of the transition region and the absence of
critical fluctuations, we believe that the domain wall states are still
too heavy to reveal any critical behavior present in the massless theory.

To bolster our claim, we performed an $8^3\!\times\!32$ zero temperature
scale setting calculation, with $\beta$=5.325, $m_0$=1.9, $L_s$=24, $m_f$=0.02
\cite{Wu:1999cd}.  We found that $m_\rho$=1.18(3) and $m_\pi$=0.654(3)
in lattice units.  Using $m_\rho$ to set the lattice scale, $\beta_c$=5.325
corresponds to $T_c = 163(4)$ MeV and $m_\pi = 427(11)$ MeV in physical units.
By comparison, for staggered fermions near $T_c$ for
$L_t=4$, one pion has a mass of $m_{\pi} \sim 230$ MeV and the other two
are $m_{\pi_2} \sim 600$ MeV \cite{Brown:1990ev,Vaccarino:1991kn}.
This leads to several possible explanations for the difference in the widths
of the crossover region for staggered and domain wall fermions.
One possibility is that the width may depend on the difference
in the pion spectrum: one light and two heavy pions for staggered
{\it vs.}\ three moderately heavy pions for domain wall fermions.

We scanned through the domain wall crossover region by changing $\beta$
holding other parameters, particularly $m_f$ and $L_s$, fixed.  We expect
that additive renormalizations of the bare quark mass, $m_{\rm res}$
in the discussion of section \ref{sec:finite_ls}, will be smaller
at weaker coupling.  The mass of the domain wall pions suggest
$m_{\rm res} \gtrsim m_f$ so it is possible that the effective bare quark mass,
$m_{\rm eff} = m_f + m_{\rm res}$ may also be rapidly changing through
the transition region, leading to qualitatively different behavior
than staggered fermions, where the effective bare mass did not change
through the transition region.  This will be discussed further
in section \ref{sec:GMOR_study}.

Besides the relatively heavy domain wall quark masses to account
for the broadening of the crossover region relative to staggered fermions,
a final possibility is that the staggered lattice spacing evolves rapidly
as a function of bare coupling in the transition region, implying
the crossover behavior for domain wall and staggered would be similar
if shown in physical units.  The dependence of the staggered lattice spacing
on the bare parameters in the transition region was extensively studied
by Blum, {\it et al.}~\cite{Blum:1995zf}.  Summary plots which display
the staggered chiral condensate $\left\langle\overline{\chi}\chi\right\rangle$
and Wilson line $\w3m$ {\it vs.}\ temperature $T$ in MeV are shown in figures
\ref{fig:pbp_wd_vs_temp} and \ref{fig:w3m_wd_vs_temp}.

As we ran only a single zero temperature domain wall scale setting calculation
in the transition region, there are insufficient data to make
a non-perturbative determination of the functional dependence
of the lattice spacing on the bare parameters, as was done for staggered
fermions.  Close to the continuum limit, it would only be necessary
to determine the lattice spacing for one set of bare parameters
as the lattice spacings at other bare parameters are given
by the asymptotic scaling relation for two flavors of massless fermions
\begin{equation}
\label{eq:asymptotic_scaling}
a \Lambda_{\rm LAT} = \left( 8 \pi^2 \beta / 29 \right)^{345/841}
\exp\left(-4 \pi^2 \beta / 29 \right)
\end{equation}
where $\Lambda_{\rm LAT}$ is the lattice scale parameter which must be
determined by a scale setting calculation.  The relation
(\ref{eq:asymptotic_scaling}) is the result of a two loop perturbative
calculation and only contains terms which are independent of renormalization
scheme.  However, the value of the parameter $\Lambda$ will depend
upon the renormalization scheme

Solving (\ref{eq:asymptotic_scaling}) for $\Lambda_{\rm LAT}$
in our $8^3\times 32$ scale setting calculation at $\beta$=5.325 mentioned
above, we find $\Lambda_{\rm LAT}$=1.388(2) MeV\footnote{
The difference between this $\Lambda_{\rm LAT}$ and the more familiar
$\Lambda_{\overline{\rm MS}} \sim 200$ MeV is merely due to large
scheme-dependent factors.  }.  Holding $\Lambda_{\rm LAT}$ fixed,
we can estimate the temperature at other couplings holding the other bare
parameters fixed.  Although the effective bare quark mass will change
as we change the coupling, as noted above, we emphasize that $m_\rho$
is typically less sensitive to small changes in the quark mass than $m_\pi$
so we expect that this may not be a large source of systematic error.
 Using this method, we also plot the domain wall chiral condensate $\qbq$
and Wilson line $\w3m$ {\it vs.}\ temperature $T$ in MeV
in figures \ref{fig:pbp_wd_vs_temp} and \ref{fig:w3m_wd_vs_temp}.
Plotting both the staggered and domain wall data {\it vs.}\ temperature
has not resolved the difference between the widths of the crossover regions
for the two formulations, assuming that the violations of the scaling relation
(\ref{eq:asymptotic_scaling}) are not large due to the relatively strong
couplings.  We will examine this issue below where we have two scale setting
calculations at different couplings to estimate the size of the scaling
violations.

After realizing that the pion mass was too heavy to reveal critical
phenomena, we could, in principle, merely increase $L_s$ until we achieve
the desired pion mass.  Unfortunately, the $16^3\times 4, L_s=24$ calculation
is essentially the largest scale study we were able to conduct
in a reasonable amount of time.  So, we decided to pursue improvements
to our lattice action which we hoped would reduce the pion mass significantly
for roughly the same computational effort. 

Encouraged by published quenched results which indicated that replacing
the standard Wilson gauge action (\ref{eq:wilson_gauge_action})
with the improved rectangle action (\ref{eq:rect_gauge_action})
can reduce the quenched pion mass \cite{Malureanu:1999,Wu:1999cd}, we located
the $N_f$=2, $L_t$=4 transition region for this action in $8^3\times 4$ volumes
with $c_1$=-0.331, $m_0$=1.9, $L_s$=24, $m_f$=0.02
\cite{Fleming:2000xe,Vranas:1999dg}.  The choice of $c_1=-0.331$ stems
from early work by Iwasaki \cite{Iwasaki:1983ck,Iwasaki:1985we}.
Although the rationale of his approach probably has little relevance
to this work, choosing this value eliminates the need for costly study
of the effects of varying $c_1$.  Simulating with the additional term
in the gauge action did not significantly increase the computational cost
relative to the standard action because the cost is dominated
by the CG algorithm.  Table \ref{tab:rd_8nt4_h1.9_l24_m0.02}
contains the other run parameters for these simulations.

The evolutions of $\qbq$ and $\w3m$ are shown in figures
\ref{fig:pbp_rd_8nt4_h1.9_l24_m0.02} and \ref{fig:w3m_rd_8nt4_h1.9_l24_m0.02}.
As before, data to the left of the vertical dashed line in each panel
were considered thermalization and not used to compute averages and
autocorrelated errors.  Ordered and disordered starts were run
at each $\beta$ until the averages of observables agreed within errors
for at least the last 100 trajectories.  As with domain wall fermions
with the Wilson gauge action, the short distance fluctuations of $\qbq$
are diminished in the symmetric phase.

Following the $8^3\times 4$ study to locate the transition region, we
performed a detailed study of the transition region in $16^3\times 4$
volumes.  Table \ref{tab:rd_16nt4_h1.9_l24_m0.02} contains the run
parameters for these simulations.  The evolutions of $\qbq$ and $\w3m$ are
shown in figures \ref{fig:pbp_rd_16nt4_h1.9_l24_m0.02} and
\ref{fig:w3m_rd_16nt4_h1.9_l24_m0.02}. The summary plot which display
$\qbq$ versus $\beta$ is shown in figure \ref{fig:pbp_rd_h1.9_l24_m0.02}
for both $8^3\times 4$ and $16^3\times 4$ volumes.
Contrary to the observed decrease in the pion mass for quenched domain wall
fermions with the Iwasaki gauge action at zero temperature, the broadening
the transition region at finite temperature again indicates a moderately
heavy pion for dynamical domain wall fermions with Iwasaki gauge action
($c_1$=-0.331) as was the case with the Wilson gauge action ($c_1$=0).

We performed $8^3\times 32$ zero temperature scale setting
calculations in the transition region, with $\beta$=1.9 and 2.0, $c_1$=-0.331,
$m_0$=1.9, $L_s$=24 and $m_f$=0.02 \cite{Wu:1999cd}.  For $\beta$=1.9,
which appears to be in the central part of the transition region,
the dimensionless lattice meson masses are $m_\rho$=1.16(2)
and $m_\pi$=0.604(3) and, using $m_\rho$ to determine the lattice spacing
give the physical temperature $T$= 166(3) MeV and physical pion mass
$m_\pi^{\rm(phys)}$= 400(7) MeV.  This confirms our suspicion
that the improved gauge action (\ref{eq:rect_gauge_action})
does not significantly improve the chiral behavior of dynamical domain wall
fermions at strong coupling.

We used the same asymptotic scaling technique as before to extrapolate
the lattice spacing as a function of the bare coupling $\beta$
and find $\Lambda_{\rm LAT}$=98.0(1) MeV.  We can now directly compare
the transition regions of the two domain wall QCD actions.
In figures \ref{fig:pbp_rd_vs_temp} and \ref{fig:w3m_rd_vs_temp} we show
the chiral condensate $\qbq$ and the Wilson line $\w3m$ for the standard
Wilson gauge action ($c_1$=0) and the Iwasaki gauge action ($c_1$=-0.331).
While the relative offset between the two curves may be due to uncertainties
in the overall temperature scale of the scaling functions for each data set,
superficially it appears that the rectangle improvement in the gauge action
did not reduce the width of the transition region.

For the $8^3\times 32$ zero temperature scale setting calculations
with $\beta$=2.0, $c_1$=-0.331, $m_0$=1.9, $L_s$=24 and $m_f$=0.02
\cite{Wu:1999cd}, the dimensionless lattice meson masses are $m_\rho$=0.95(3)
and $m_\pi$=0.475(7) and, using $m_\rho$ to set the scale gives a physical
temperature $T$=202(6) MeV.  Note that $m_\pi / m_\rho$=0.5, within errors,
for both $\beta$=1.9 and $\beta$=2.0 simulations.
 Using (\ref{eq:asymptotic_scaling}) to extrapolate the scale from $\beta$=1.9
gives a physical temperature of $T_{\rm ext}$=186 MeV.  Violations
of asymptotic scaling are around eight percent for a five percent change
in $\beta$.  So, determining the scale from the asymptotic scaling relation
(\ref{eq:asymptotic_scaling}) tends to make the transition region appear
narrower than it would be if a non-perturbative scaling relation were used.

While there were many interesting indications that observations
of true critical behavior might be possible on $L_t=4$ lattices
with domain wall fermions, it is clear from the size of the pion mass
in physical units that $L_s=24$ is too small to suppress
the large chiral symmetry breaking effects at strong coupling.
In section \ref{sec:GMOR_study}, we will address the issue of exactly how large
an $L_s$ is needed to substantially reduce the mass of the dynamical pions
in the broken phase and transition region.  However, we can also tell
from our preceding discussion that chiral symmetry breaking effects are much
smaller in the symmetric phase.  Not only does the symmetric phase occur
at weaker coupling, but reduced average CG iterations per HMC step
and the increased rate of exponential decay as $L_s\to\infty$ for $\qbq$
seen in section \ref{sec:beta_crit_8nt4} are strong indications
that $L_s=24$ or even $L_s=16$ may be sufficient for simulations
in the chirally symmetric phase of QCD on $L_t=4$ lattices.

\section{Effects of finite $L_s$ in the confined phase}
\label{sec:GMOR_study}

One of the primary advantages of using domain wall fermions to study
two flavor QCD thermodynamics is the extra parameter $L_s$ which allows us
to control chiral symmetry breaking.  In section \ref{sec:phase_transition}
we concluded that the mass of the pion state indicated that the amount
of chiral symmetry breaking for domain wall fermions with $L_s=24$ was,
at best, only moderately better than existing calculations, which are
computationally less demanding by an order of magnitude.  See figure 2
of Karsch's LATTICE 99 review talk \cite{Karsch:1999vy}.

In defense of the domain wall method, it is important to note that it is
currently not possible to simulate in the $(a T_c)^{-1}$=4 region
of the parameter space with {\it three} light pions close to their physical
mass scale for any of the standard fermion actions.  So, the relevant questions
are why is the pion so heavy for $L_s$=24 and how large do we need to make
$L_s$ to reduce all three pions to their physical range.  After addressing
these questions, we will have established a framework to study what
improvements are possible to reduce chiral symmetry breaking without
increasing $L_s$. The framework we propose is the Gell-Mann, Oakes, and
Renner relation (\ref{eq:dwf_GMOR}) discussed in section \ref{sec:finite_ls}.

In conjunction with the low temperature phase study of the chiral condensate
in $8^3\times 4$ volumes with $\beta$=5.2 and $m_0$=1.9, described
in section \ref{sec:beta_crit_8nt4}, we also measured $\qbq$ and $\chi_\pi$
at the dynamical mass and various valence masses for several values of $L_s$.
Tables \ref{tab:pbp_spc_8nt4_b5.2_h1.9_m0.02_1}
and \ref{tab:pbp_spc_8nt4_b5.2_h1.9_m0.02_2} summarize these observables.

In figure \ref{fig:dj5_wd_8nt4_b5.2_h1.9_m0.02}, we show
the mid-plane contribution to the axial current,
$\Delta J_5 = \qbq - m_f \chi_\pi$, from (\ref{eq:dwf_GMOR}).
Since the GMOR relation (\ref{eq:dwf_GMOR}) holds configuration
by configuration, we could define a $\Delta J_5$ for each configuration.
However, in our simulations we only measured the chiral condensate
averaged over the spacetime volume $\qbq$ and not the contribution
from a single point $\qbqo$ required in (\ref{eq:dwf_GMOR}).
However, the expectation values of two condensates will agree
in the ensemble average, $\qbq \approx \qbqo$, our definition
of $\Delta J_5$ is valid as the difference of two expectation values.
We used the jackknife method to estimate the error of $\Delta J_5$.
The long dashed line in the figure is the fit
$\Delta J_5 = 0.0096(2) \exp\left[-0.0190(8) L_s\right]$ over the range
$L_s \in [16,40]$ with $\chi^2/{\rm dof}$=6.5/2.  Although the $\chi^2$
of the fit is a little high, we see that the points appear to scatter
along the straight line on the log scale.  So, our data are consistent
with chiral symmetry restoration in the $L_s\to\infty$ limit.

We can decompose the effects of finite $\Delta J_5$ by extracting
the additive mass renormalization $m_{\rm res}$ and
the chiral condensate subtraction $b_0$ using the valence data
in tables \ref{tab:pbp_spc_8nt4_b5.2_h1.9_m0.02_1}
and \ref{tab:pbp_spc_8nt4_b5.2_h1.9_m0.02_1} and the ansatz
\beq
\qbq = \left( m_f + m_{\rm res} \right) \chi_\pi + b_0
\eeq
from (\ref{eq:dwf_GMOR}) and (\ref{eq:qbq_o_spc_ansatz}).
We simultaneously fit $\qbq$ and $\chi_\pi$
over $m_f^{\rm(val)} \in [0.02,0.14]$ for each $L_s$.  To perform this fit,
we must further assume $\chi_\pi$ can be parameterized
as in (\ref{eq:chi_pi_param}).  We then fit $m_{\rm res}$ and $b_0$
to a decreasing exponential in $L_s$.  The whole fitting procedure is run
under a jackknife loop.  The fit did not account for correlations
between data points because there were insufficient data
to accurately determine the correlation matrix.  Our previous experience
with this fit ansatz using quenched data \cite{Fleming:1999eq} suggests
that including the effects of correlations does not change the answer
significantly.

The results are in table \ref{tab:GMOR_wd_8nt4_b5.2_h1.9_m0.02},
all errors quoted are jackknife errors.  In figure
\ref{fig:GMOR_wd_8nt4_b5.2_h1.9_m0.02}, we show the $L_s$-dependence
of $m_{\rm res}$ and $-b_0$ using ten jackknife blocks.  The exponential fits
shown, with jackknife errors, are
\beqs
& -b_0 = 0.0106(8) \exp\left[-0.016(3) L_s\right], &
\chi^2/{\rm dof} = 0.19(11) \\
& m_{\rm res} = 0.18(1) \exp\left[-0.026(3) L_s\right], &
\chi^2/{\rm dof} = 0.08(9)
\eeqs
The fitting range is $L_s \in [10,40]$ and there are four degrees of freedom.
Although the rate of exponential decay is slow, again we note that our data
are consistent with chiral symmetry restoration in the $L_s\to\infty$ limit.
If we use the asymptotic scaling relation (\ref{eq:asymptotic_scaling})
to extrapolate the lattice spacing from the $8^3\times 32$, $\beta$=5.325
scale setting calculation described in section \ref{sec:beta_crit_16nt4},
we get $a^{-1} \approx 550$ MeV with an unknown systematic error
due to violations of asymptotic scaling.  So, in physical units,
the bare lattice quark mass is $m_f^{\rm(phys)} \approx 10$ MeV
and the bare residual quark mass at $L_s$=24
is $m_{\rm res}^{\rm(phys)} \approx 50$ MeV.

In table \ref{tab:GMOR_val_scale_setting}, we show the valence GMOR parameters
$m_{\rm res}$ and $b_0$ for the scale setting calculations \cite{Wu:1999cd}
which supported the transition region study in section
\ref{sec:phase_transition}.  In the first column, ``W'' implies
the Wilson gauge action (\ref{eq:wilson_gauge_action}) and ``R'' implies
the rectangle action (\ref{eq:rect_gauge_action}) with the choice $c_1=-0.331$.
As before, there were insufficient data to accurately estimate the correlation
matrix, so it was not included in the fit.  From the $8^3\times 32$
scale setting calculations described in section \ref{sec:beta_crit_16nt4},
we know that the lattice spacing for domain wall fermions
with the Wilson gauge action at $\beta=5.325$ is $a^{-1}$=652(17) MeV.
Using the lattice spacing, we express the dynamical bare lattice quark mass as
$m_f^{\rm(phys)}$=13.0(3) MeV, in physical units, and the bare residual quark
mass as $m_{\rm res}^{\rm(phys)}$=37(2) MeV.  When compared
with the $\beta$=5.2 simulations above, we find clear evidence supporting
the perturbative expectation that chiral symmetry breaking effects are smaller
at weaker coupling, for a fixed $L_s$=24.

Now, we can estimate whether a fourfold increase in $L_s$ to $L_s$=96
would be sufficient to reduce the pion masses to a physically interesting
scale.  First, we will imagine that a zero temperature calculation would
be done on $8^3\times 32$ with $\beta$=5.325, $m_0$=1.9 and $L_s$=96.
A conservative estimate of the exponential decay of the residual mass
at $\beta=5.325$ is $m_{\rm res} \propto \exp(-0.03 L_s)$ since this
agrees with the observed rate at $\beta$=5.2 within error.
With this decay rate, the bare lattice would be $m_{\rm res} \approx 0.007$
at $L_s$=96.  Choosing the bare lattice quark mass of $m_f$=0.01 gives
$m_f \gtrsim m_{\rm res}$ and will reduce the possibility of large changes
in the effective bare mass while scanning through the transition region.
In \cite{Wu:1999cd}, the dynamical bare quark mass dependence of the pion mass 
was determined to be $m_\pi^2 \propto 5.38 m_f$ when $L_s$=24.
Assuming the same slope will be found when $L_s$=96 and that the pion mass
vanishes (in an infinite volume) at $m_f \approx -0.007$, then
the dimensionless pion mass will be $m_\pi \approx 0.3$.  Using the lattice
spacing found at $L_s$=24 and $m_f$=0.02, the physical pion mass should be
$m_\pi^{\rm(phys)} \approx 200$ MeV at $L_s$=96, a physically interesting
scale.

The GMOR relation does provide a useful framework to study the chiral symmetry
breaking effects of finite domain wall separation.  However, one must be
careful when choosing a parameterization of $\Delta J_5$ that
the physical assumptions remain valid throughout the region of coupling space
of interest.  Since our assumption was based on the existence of a pion pole
due to spontaneous chiral symmetry breaking, the method described here is only
applicable in the confined phase.

\clearpage


\chapter*{Concluding remarks}
\addcontentsline{toc}{chapter}{Concluding remarks}

\thispagestyle{myheadings}
\markright{}

We have established that domain wall fermions preserve the full chiral
and flavor symmetries of QCD.  However, these features come at a substantial
cost:  one and sometimes two orders of magnitude increase in computational
effort.  So, it is important to carefully choose problems where the role
of chiral symmetry is crucial to determining the correct answer to the problem
and similar quality results could not be obtained by simply redirecting
the increased computational effort into working at smaller lattice spacings
using less demanding formulations, like staggered or Wilson fermions.

The problem we chose to study is the finite temperature phase transition
and quark-gluon plasma phase of $N_f$=2 QCD.  This is a particularly
challenging problem because the finite temperature calculations
are usually performed at stronger couplings than comparable calculations
at zero temperature.  This is also a particularly challenging problem
for the domain wall fermion formulation as the suppression of chiral symmetry
breaking effects at stronger couplings requires larger values of $L_s$.

From this work it seems clear that domain wall fermions work as advertised.
The hybrid Monte Carlo (HMC) and conjugate gradient (CG) algorithms worked
without any difficulty beyond the linear increase in computational cost
proportional to $L_s$.  The only negative aspect of this work was that
the computational resources were not four times larger, the amount necessary
to simulate on $16^\times 4$ lattices with dynamical pions close
to their physical masses.

Despite this limitation, several important physical results were reported
here.  On $8^3\times 4$ volumes, extrapolations to the chiral limit,
$m_f\to 0$ and $L_s\to\infty$, on either side of the crossover region
give a clear evidence of the existence of a finite temperature phase transition
that restores chiral symmetry in the chiral limit.

For the first time ever \cite{Karsch:1999vy}, we performed simulations
on $16^3\times 4$ volumes with three dynamical pions whose masses were
on the order of 400 MeV.  Unfortunately, simulations with 400 MeV pions
are not sufficiently close to the chiral limit to exhibit convincing evidence
of the order of the chiral transition.  Of course, even reducing the mass
of the dynamical pions to 200 MeV may not be sufficient to uncover evidence
of the chiral transition.  However, pions are not massless in nature,
so the width of the crossover region with dynamical pions near their physical
value is certainly of phenomenological interest.

Finally, from our study of the GMOR relation, we can determine the residual
chiral symmetry breaking effects of domain wall fermions at finite $L_s$.
Using the GMOR relation, we found these effects were characterized in terms
of a residual mass $m_{\rm res}$ which decreases exponentially in $L_s$.
In the transition region, for $L_s$=24 and $m_f$=0.02 the effective bare quark
mass was dominated by finite $L_s$ effects that were too large to allow
dynamical pions near the physical pion mass.  However, we argued
quite conservatively that a fourfold increase in $L_s$ with $m_f$=0.01
should be sufficient for the effective bare quark mass to be dominated
by $m_f$ in the transition region and allow for the masses of the three
dynamical pions should be near their physical values.

We believe the full characterization of the $L_t=4$ transition region will be
completely accessible to the next generation of supercomputers, even without
substantial improvement of the domain wall action or algorithms.  However,
many exciting questions about the high temperature phase,
like the equation of state \cite{Fleming:2000bk}, can still be addressed
using the current generation of supercomputers.

\clearpage


\thispagestyle{myheadings}
\markright{}

\clearpage


\chapter*{Tables}
\addcontentsline{toc}{chapter}{Tables}
\thispagestyle{myheadings}
\markright{}

\clearpage


\begin{table}
  \centering
  \caption[Parameters for $\beta_c(m_0=1.15)$ search]
    {Run parameters for $\beta_c(m_0=1.15)$ search}
  \label{tab:8nt4_h1.15_beta_crit}
  vol: $8^3\times 4$, \qquad $m_0=1.15$, \qquad $L_s=12$, \qquad $m_f=0.1$ \\
  HMC traj.~len: $\frac{1}{50}\times 25$, \qquad CG stop cond: $10^{-6}$
  \begin{tabular}{cccccccc}
    \hline\hline
    $\beta$ & start & \# traj & acc. &
    $\left\langle e^{-\Delta H}\right\rangle$ & plaq. & $\w3m$ & $\qbq$ \\
    \hline\hline
    5.45 & O & 100-800 & 0.87 & 0.99(1)  & 0.470(1)  & 0.0168(6) &
      0.00276(1)  \\
    5.55 & O & 200-800 & 0.87 & 0.98(1)  & 0.4933(6) & 0.023(1)  &
      0.002916(6) \\
    5.65 & O & 300-800 & 0.87 & 1.00(2)  & 0.5218(9) & 0.054(6)  &
      0.00305(2)  \\
    5.75 & D & 300-800 & 0.86 & 0.986(9) & 0.5571(7) & 0.196(7)  &
      0.002875(7) \\
    5.85 & D & 300-800 & 0.85 & 1.00(1)  & 0.5719(7) & 0.234(3)  &
      0.002881(3) \\
    5.95 & D & 200-800 & 0.87 & 0.99(1)  & 0.5857(5) & 0.262(2)  &
      0.002898(3)
  \end{tabular}
\end{table}

\begin{table}
  \centering
  \caption[Parameters for $\beta_c(m_0=1.4)$ search]
    {Run parameters for $\beta_c(m_0=1.4)$ search}
  \label{tab:8nt4_h1.4_beta_crit}
  vol: $8^3\times 4$, \qquad $m_0=1.4$, \qquad $L_s=12$, \qquad $m_f=0.1$ \\
  HMC traj.~len: $\frac{1}{50}\times 25$, \qquad CG stop cond: $10^{-6}$ \\
  \begin{tabular}{cccccccc}
    \hline\hline
    $\beta$ & start & \# traj & acc. &
    $\left\langle e^{-\Delta H}\right\rangle$ & plaq. & $\w3m$ & $\qbq$ \\
    \hline\hline
    5.35 & O & 100-800 & 0.87 & 1.01(1)  & 0.4435(7) & 0.0189(8) &
      0.00497(1)  \\
    5.45 & O & 100-800 & 0.86 & 0.99(1)  & 0.4630(7) & 0.0215(8) &
      0.00522(1)  \\
    5.55 & O & 300-800 & 0.86 & 1.00(1)  & 0.487(1)  & 0.032(3)  &
      0.00539(3)  \\
    5.65 & D & 400-800 & 0.86 & 1.00(2)  & 0.540(2)  & 0.180(6)  &
      0.00457(4)  \\
    5.75 & D & 300-800 & 0.85 & 0.99(1)  & 0.5598(8) & 0.224(3)  &
      0.00445(1)  \\
    5.85 & D & 200-800 & 0.89 & 1.011(8) & 0.5744(3) & 0.254(3)  &
      0.004409(5)
  \end{tabular}
\end{table}

\begin{table}
  \centering
  \caption[Parameters for $\beta_c(m_0=1.65)$ search]
    {Run parameters for $\beta_c(m_0=1.65)$ search}
  \label{tab:8nt4_h1.65_beta_crit}
  vol: $8^3\times 4$, \qquad $m_0=1.65$, \qquad $L_s=12$, \qquad $m_f=0.1$ \\
  HMC traj.~len: $\frac{1}{50}\times 25$, \qquad CG stop cond: $10^{-6}$
  \begin{tabular}{cccccccc}
    \hline\hline
    $\beta$ & start & \# traj & acc. &
    $\left\langle e^{-\Delta H}\right\rangle$ & plaq. & $\w3m$ & $\qbq$ \\
    \hline\hline
    5.25 & O & 200-800  & 0.82 & 0.97(2)  & 0.4289(5) & 0.027(2) & 0.01000(2) \\
    5.35 & O & 400-800  & 0.68 & 0.98(4)  & 0.451(3)  & 0.035(4) & 0.01000(9) \\
    5.45 & D & 400-800  & 0.74 & 1.09(5)  & 0.4769(8) & 0.049(7) & 0.00985(7) \\
    5.55 & D & 600-1200 & 0.80 & 1.00(3)  & 0.531(1)  & 0.175(7) & 0.00718(7) \\
    5.65 & D & 400-800  & 0.79 & 0.96(2)  & 0.5507(9) & 0.216(4) & 0.00677(2) \\
    5.75 & D & 200-800  & 0.88 & 0.989(7) & 0.5663(4) & 0.243(3) & 0.00658(1)
  \end{tabular}
\end{table}

\begin{table}
  \centering
  \caption[Parameters for $\beta_c(m_0=1.8)$ search]
    {Run parameters for $\beta_c(m_0=1.8)$ search}
  \label{tab:8nt4_h1.8_beta_crit}
  vol: $8^3\times 4$, \qquad $m_0=1.8$, \qquad $L_s=12$, \qquad $m_f=0.1$ \\
  HMC traj.~len: $\frac{1}{50}\times 25$, \qquad CG stop cond: $10^{-6}$
  \begin{tabular}{cccccccc}
    \hline\hline
    $\beta$ & start & \# traj & acc. &
    $\left\langle e^{-\Delta H}\right\rangle$ & plaq. & $\w3m$ & $\qbq$ \\
    \hline\hline
    5.15 & O & 200-800 & 0.83 & 1.02(2) & 0.4191(8) & 0.029(1) & 0.01485(5) \\
    5.25 & O & 400-800 & 0.66 & 0.97(5) & 0.4381(6) & 0.038(2) & 0.01458(5) \\
    5.35 & O & 400-800 & 0.63 & 0.97(5) & 0.471(2)  & 0.052(3) & 0.0134(2)  \\
    5.45 & O & 400-800 & 0.76 & 1.01(3) & 0.515(2)  & 0.161(4) & 0.0097(1)  \\
    5.55 & D & 400-800 & 0.79 & 1.05(5) & 0.540(1)  & 0.200(9) & 0.0088(1)  \\
    5.65 & D & 200-800 & 0.89 & 1.01(2) & 0.5570(5) & 0.242(4) & 0.00828(2)
  \end{tabular}
\end{table}

\begin{table}
  \centering
  \caption[Parameters for $\beta_c(m_0=1.9)$ search]
    {Run parameters for $\beta_c(m_0=1.9)$ search}
  \label{tab:8nt4_h1.9_beta_crit}
  vol: $8^3\times 4$, \qquad $m_0=1.9$, \qquad $L_s=12$, \qquad $m_f=0.1$ \\
  CG stop cond: $10^{-6}$
  \begin{tabular}{ccccccccc}
    \hline\hline
    $\beta$ & start & traj.~len. & \# traj & acc. &
    $\left\langle e^{-\Delta H}\right\rangle$ & plaq. & $\w3m$ & $\qbq$ \\
    \hline\hline
    5.0  & O & $\frac{1}{40}\times 20$ & 200-800  & 0.37 & 0.8(1)  & 0.4002(8) &
      0.032(2) & 0.01919(5) \\
    5.2  & O & $\frac{1}{40}\times 20$ & 400-800  & 0.43 & 1.2(3)  & 0.437(1)  &
      0.032(2) & 0.01838(4) \\
    5.25 & O & $\frac{1}{50}\times 25$ & 400-800  & 0.65 & 1.10(9) & 0.452(1)  &
      0.049(6) & 0.0174(2)  \\
    5.35 & D & $\frac{1}{50}\times 25$ & 600-1200 & 0.69 & 0.95(5) & 0.493(2)  &
      0.107(9) & 0.0135(4)  \\
    5.45 & D & $\frac{1}{50}\times 25$ & 600-1200 & 0.74 & 1.01(4) & 0.528(1)  &
      0.197(4) & 0.01039(7) \\
    5.55 & D & $\frac{1}{50}\times 25$ & 400-830  & 0.82 & 1.00(1) & 0.5463(5) &
      0.227(6) & 0.00974(4) \\
    5.65 & D & $\frac{1}{50}\times 25$ & 400-800  & 0.88 & 1.03(1) & 0.5613(8) &
      0.248(5) & 0.00943(4)
  \end{tabular}
\end{table}

\begin{table}
  \centering
  \caption[Parameters for $\beta_c(m_0=2.0)$ search]
    {Run parameters for $\beta_c(m_0=2.0)$ search}
  \label{tab:8nt4_h2.0_beta_crit}
  vol: $8^3\times 4$, \qquad $m_0=2.0$, \qquad $L_s=12$, \qquad $m_f=0.1$ \\
  HMC traj.~len: $\frac{1}{50}\times 25$, \qquad CG stop cond: $10^{-6}$
  \begin{tabular}{cccccccc}
    \hline\hline
    $\beta$ & start & \# traj & acc. &
    $\left\langle e^{-\Delta H}\right\rangle$ & plaq. & $\w3m$ & $\qbq$ \\
    \hline\hline
    5.05 & O & 200-800  & 0.77 & 0.99(3) & 0.4192(8) & 0.035(1) & 0.02324(8) \\
    5.15 & O & 200-800  & 0.75 & 0.98(3) & 0.442(1)  & 0.042(3) & 0.0215(2)  \\
    5.25 & O & 200-1200 & 0.79 & 1.03(1) & 0.474(1)  & 0.080(7) & 0.0181(3)  \\
    5.35 & D & 200-800  & 0.83 & 1.00(2) & 0.5130(7) & 0.173(6) & 0.0130(2)  \\
    5.45 & D & 200-800  & 0.87 & 1.02(2) & 0.5349(5) & 0.203(3) & 0.01157(4) \\
    5.55 & D & 200-800  & 0.85 & 1.01(1) & 0.5503(4) & 0.235(3) & 0.01099(2)
  \end{tabular}
\end{table}

\begin{table}
  \centering
  \caption[Parameters for $\beta_c(m_0=2.15)$ search]
    {Run parameters for $\beta_c(m_0=2.15)$ search}
  \label{tab:8nt4_h2.15_beta_crit}
  vol: $8^3\times 4$, \qquad $m_0=2.15$, \qquad $L_s=12$, \qquad $m_f=0.1$ \\
  HMC traj.~len: $\frac{1}{50}\times 25$, \qquad CG stop cond: $10^{-6}$
  \begin{tabular}{cccccccc}
    \hline\hline
    $\beta$ & start & \# traj & acc. &
    $\left\langle e^{-\Delta H}\right\rangle$ & plaq. & $\w3m$ & $\qbq$ \\
    \hline\hline
    4.85 & O & 200-800  & 0.69 & 0.95(3) & 0.4004(8) & 0.034(2) & 0.0323(2)  \\
    4.95 & O & 200-800  & 0.72 & 0.97(3) & 0.419(2)  & 0.040(2) & 0.0302(3)  \\
    5.05 & O & 200-800  & 0.48 & 0.92(4) & 0.443(2)  & 0.052(3) & 0.0272(5)  \\
    5.15 & O & 200-1200 & 0.62 & 1.01(3) & 0.480(3)  & 0.12(1)  & 0.0203(7)  \\
    5.25 & O & 400-800  & 0.70 & 0.97(5) & 0.5105(4) & 0.185(2) & 0.01559(8) \\
    5.35 & D & 400-800  & 0.69 & 1.01(4) & 0.529(1)  & 0.216(6) & 0.0141(1)  \\
    5.45 & D & 400-800  & 0.71 & 1.00(3) & 0.5453(7) & 0.230(4) & 0.01330(5)
  \end{tabular}
\end{table}

\begin{table}
  \centering
  \caption[Parameters for $\beta_c(m_0=2.4)$ search]
    {Run parameters for $\beta_c(m_0=2.4)$ search}
  \label{tab:8nt4_h2.4_beta_crit}
  vol: $8^3\times 4$, \qquad $m_0=2.4$, \qquad $L_s=12$, \qquad $m_f=0.1$ \\
  HMC traj.~len: $\frac{1}{50}\times 25$, \qquad CG stop cond: $10^{-6}$
  \begin{tabular}{cccccccc}
    \hline\hline
    $\beta$ & start & \# traj & acc. &
    $\left\langle e^{-\Delta H}\right\rangle$ & plaq. & $\w3m$ & $\qbq$ \\
    \hline\hline
    4.65 & O & 100-800 & 0.63 & 1.02(5) & 0.3953(6) & 0.046(3) & 0.0484(3)  \\
    4.75 & O & 200-800 & 0.68 & 0.99(5) & 0.4156(9) & 0.054(3) & 0.0442(3)  \\
    4.85 & O & 300-800 & 0.70 & 0.94(4) & 0.439(2)  & 0.069(5) & 0.0380(6)  \\
    4.95 & O & 200-800 & 0.77 & 1.02(4) & 0.4779(6) & 0.155(4) & 0.0257(2)  \\
    5.05 & D & 200-800 & 0.80 & 1.01(2) & 0.4987(9) & 0.190(2) & 0.0220(2)  \\
    5.15 & D & 200-800 & 0.84 & 1.01(3) & 0.5170(5) & 0.221(3) & 0.01962(7)
  \end{tabular}
\end{table}


\begin{table}
  \centering
  \caption{
    Fit of $\w3m=c_0\left\{c_1+\tanh\left[c_2(\beta-\beta_c)\right]\right\}$:
    $8^3\times 4, L_s=12, m_f=0.1$
  }
  \label{tab:w3m_tanhfit_8nt4_l12_m0.1}
  \begin{tabular}{ccccccc}
    \hline\hline
    $m_0$ & $\beta_c$ & $c_o$ & $c_1$ & $c_2$ & $N_{\rm dof}$ &
      $\chi^2/N_{\rm dof}$ \\
    \hline\hline
    1.15 & 5.722(5) & 0.122(1) & 1.126(7) & 9.8(5)  & 2 & 14.50 \\
    1.4  & 5.627(4) & 0.110(1) & 1.180(7) & 17(1)   & 2 & 21.39 \\
    1.65 & 5.524(6) & 0.107(2) & 1.24(2)  & 10(1)   & 2 &  9.60 \\
    1.8  & 5.431(4) & 0.105(2) & 1.29(1)  & 11.7(8) & 2 & 11.07 \\
    1.9  & 5.386(6) & 0.107(3) & 1.27(2)  & 9.9(8)  & 3 &  2.61 \\
    2.0  & 5.327(7) & 0.107(3) & 1.27(3)  & 6.7(6)  & 2 & 10.33 \\
    2.15 & 5.179(6) & 0.098(2) & 1.35(3)  & 8.3(6)  & 3 &  1.13 \\
    2.4  & 4.934(7) & 0.091(4) & 1.46(5)  & 7.2(6)  & 2 & 14.02
  \end{tabular}
\end{table}

\begin{table}
  \centering
  \caption{
    Fit of $\qbq=c_0\left\{c_1+\tanh\left[c_2(\beta-\beta_c)\right]\right\}$:
    $8^3\times 4, L_s=12, m_f=0.1$
  }
  \label{tab:pbp_tanhfit_8nt4_l12_m0.1}
  \begin{tabular}{ccccccc}
    \hline\hline
    $m_0$ & $\beta_c$ & $c_o$ & $c_1$ & $c_2$ & $N_{\rm dof}$ &
      $\chi^2/N_{\rm dof}$ \\
    \hline\hline
    1.65 & 5.524(6) & -0.00170(1)  & -4.89(3) & 18(2)   & 2 &  38.28 \\
    1.8  & 5.398(4) & -0.00331(5)  & -3.50(3) & 10.6(7) & 2 &   4.90 \\
    1.9  & 5.330(3) & -0.00488(3)  & -2.94(2) & 9.1(2)  & 3 &   2.85 \\
    2.0  & 5.256(5) & -0.00648(9)  & -2.67(3) & 7.5(3)  & 2 &   7.12 \\
    2.15 & 5.103(6) & -0.0099(1)   & -2.32(2) & 6.5(3)  & 3 &   3.36 \\
    2.4  & 4.864(4) & -0.0154(3)   & -2.24(3) & 6.9(3)  & 2 &  14.96
  \end{tabular}
\end{table}

\clearpage


\begin{table}
  \centering
  \caption[Parameters for $8^3\times4, \beta=5.2, m_0=1.9$ confined phase study]
    {Run parameters for $8^3\times4, \beta=5.2, m_0=1.9$ confined phase study}
  \label{tab:8nt4_b5.2_h1.9}
  \renewcommand{\arraystretch}{0.9}
  \begin{tabular}{ccccccccc}
    \hline\hline
    $m_f$ & $L_s$ & traj.~len. & \# traj. & acc. &
    $\left\langle e^{-\Delta H}\right\rangle$ & plaq. & $\w3m$ & $\qbq$ \\
    \hline\hline
    0.02 &  4 & $\frac{1}{64}\times 32$   & 200-800  & 0.94 & 1.002(8) &
      0.500(3)  & 0.16(1)  & 0.0250(7)  \\
         &  6 & $\frac{1}{64}\times 32$   & 200-800  & 0.92 & 1.00(1)  &
      0.468(2)  & 0.080(4) & 0.0167(3)  \\
         &  8 & $\frac{1}{64}\times 32$   & 400-800  & 0.89 & 0.98(1)  &
      0.456(2)  & 0.061(2) & 0.0133(2)  \\
         & 10 & $\frac{1}{64}\times 32$   & 200-2000 & 0.86 & 0.995(8) &
      0.4460(9) & 0.048(2) & 0.0124(1)  \\
         & 12 & $\frac{1}{64}\times 32$   & 200-2000 & 0.84 & 1.01(1)  &
      0.4428(6) & 0.048(2) & 0.01123(7) \\
         & 16 & $\frac{1}{64}\times 32$   & 550-2000 & 0.75 & 0.98(2)  &
      0.4388(9) & 0.049(2) & 0.00987(9) \\
         & 24 & $\frac{1}{100}\times 50$  & 350-2000 & 0.73 & 0.95(2)  &
      0.4359(7) & 0.047(3) & 0.0088(1)  \\
         & 32 & $\frac{1}{100}\times 50$  & 300-2000 & 0.72 & 1.03(2)  &
      0.4317(7) & 0.045(2) & 0.00835(7) \\
         & 40 & $\frac{1}{128}\times 64$  & 300-1350 & 0.73 & 1.02(3)  &
      0.4342(6) & 0.044(2) & 0.00772(8) \\
    \hline
    0.06 &  8 & $\frac{1}{50}\times 25$   & 200-950  & 0.83 & 0.98(1)  &
      0.450(1)  & 0.046(3) & 0.0176(2)  \\
         & 16 & $\frac{1}{64}\times 32$   & 200-820  & 0.84 & 0.99(2)  &
      0.4361(8) & 0.045(3) & 0.0135(1)  \\
    \hline
    0.1  &  4 & $\frac{1}{40}\times 20$   & 300-800  & 0.82 & 1.00(2)  &
      0.463(2)  & 0.058(3) & 0.0375(4)  \\
         &  6 & $\frac{1}{50}\times 25$   & 300-800  & 0.89 & 1.00(2)  &
      0.454(3)  & 0.048(5) & 0.0246(4)  \\
         &  8 & $\frac{1}{40}\times 20$   & 300-800  & 0.57 & 0.89(5)  &
      0.4437(8) & 0.040(3) & 0.02109(7) \\
         & 10 & $\frac{1}{50}\times 25$   & 200-800  & 0.83 & 1.00(3)  &
      0.4405(6) & 0.036(2) & 0.01927(9) \\
         & 12 & $\frac{1}{40}\times 20$   & 400-800  & 0.43 & 1.2(3)   &
      0.437(1)  & 0.032(2) & 0.01838(4) \\
         & 16 & $\frac{1}{64}\times 32$   & 200-800  & 0.80 & 0.98(2)  &
      0.435(2)  & 0.035(2) & 0.01709(9) \\
         & 24 & $\frac{1}{64}\times 32$   & 200-800  & 0.72 & 0.94(2)  &
      0.433(1)  & 0.033(2) & 0.01596(7) \\
         & 32 & $\frac{1}{100}\times 50$  & 200-800  & 0.82 & 0.99(2)  &
      0.4305(5) & 0.037(2) & 0.01547(7) \\
         & 40 & $\frac{1}{100}\times 50$  & 200-800  & 0.78 & 1.00(5)  &
      0.432(1)  & 0.035(2) & 0.01524(5) \\
    \hline
    0.14 &  8 & $\frac{1}{40}\times 20$   & 200-860  & 0.63 & 1.03(6)  &
      0.4433(7) & 0.033(5) & 0.0241(1)  \\
         & 16 & $\frac{1}{64}\times 32$   & 200-800  & 0.85 & 1.03(2)  &
      0.433(1)  & 0.030(1) & 0.02017(6) \\
    \hline
    0.18 &  8 & $\frac{1}{40}\times 20$   & 200-1200 & 0.70 & 1.02(3)  &
      0.4410(7) & 0.030(1) & 0.02686(7) \\
         & 16 & $\frac{1}{64}\times 32$   & 200-800  & 0.84 & 0.98(2)  &
      0.432(1)  & 0.033(1) & 0.02309(5) \\
  \end{tabular}
  \renewcommand{\arraystretch}{1.0}
\end{table}


\begin{table}
  \centering
  \caption{
    Fit of $\qbq = c_0 + c_1 m_f +c_2 m_f^2$,
    $8^3\times 4, \beta=5.2, m_0=1.9$
  }
  \label{tab:pbp_mf_fit_wd_8nt4_b5.2_h1.9}
  \begin{minipage}[t]{5in}
    \begin{tabular}{rr@{--}lllccr@{.}l}
      \hline\hline
      \multicolumn{1}{c}{$L_s$} &
      \multicolumn{2}{c}{fit range} &
      \multicolumn{1}{c}{$c_0$} &
      \multicolumn{1}{c}{$c_1$} &
      $c_2$ &
      $N_{\rm dof}$ &
      \multicolumn{2}{c}{$\chi^2/N_{\rm dof}$} \\
      \hline\hline
       8 & 0.02 & 0.1  & 0.0117(2)  & 0.095(2)  &   ---    & 1 &  3&67 \\
         & 0.02 & 0.14 & 0.0122(2)  & 0.088(2)  &   ---    & 2 &  9&89 \\
         & 0.02 & 0.18 & 0.0130(1)  & 0.0781(9) &   ---    & 3 & 20&48 \\
         & 0.02 & 0.14 & 0.0111(3)  & 0.118(7)  & -0.18(4) & 1 &  0&23 \\
         & 0.02 & 0.18 & 0.0112(3)\footnote{
        Shown in figure \ref{fig:pbp_ls_fit_wd_8nt4_b5.2_h1.9} and used
        as input to fits in table \ref{tab:pbp_ls_fit_wd_8nt4_b5.2_h1.9}.
      }                             & 0.114(5)  & -0.15(2) & 2 &  0&37 \\
      \hline
      10 & 0.02 & 0.1  & 0.0107(2)$^\alph{mpfootnote}$
                                    & 0.086(2)  &   ---    & 0 &
        \multicolumn{2}{c}{---} \\
      12 & 0.02 & 0.1  & 0.00944(9)$^\alph{mpfootnote}$
                                    & 0.089(1)  &   ---    & 0 &
        \multicolumn{2}{c}{---} \\
      \hline
      16 & 0.02 & 0.1  & 0.0081(1)  & 0.090(2)  &   ---    & 1 &  0&24 \\
         & 0.02 & 0.14 & 0.00830(9) & 0.0855(9) &   ---    & 2 &  6&68 \\
         & 0.02 & 0.18 & 0.00857(8) & 0.0815(6) &   ---    & 3 & 16&38 \\
         & 0.02 & 0.14 & 0.0079(2)  & 0.101(5)  & -0.10(3) & 1 &  0&73 \\
         & 0.02 & 0.18 & 0.0079(1)$^\alph{mpfootnote}$
                                    & 0.100(3)  & -0.09(1) & 2 &  0&43 \\
      \hline
      24 & 0.02 & 0.1  & 0.0069(1)$^\alph{mpfootnote}$
                                    & 0.090(2)  &   ---    & 0 &
        \multicolumn{2}{c}{---} \\
      32 & 0.02 & 0.1  & 0.00656(9)$^\alph{mpfootnote}$
                                    & 0.089(1)  &   ---    & 0 &
        \multicolumn{2}{c}{---} \\
      40 & 0.02 & 0.1  & 0.00584(9)$^\alph{mpfootnote}$
                                    & 0.094(1)  &   ---    & 0 &
        \multicolumn{2}{c}{---} \\
      $\infty$\footnote{
        Used fitted values from table \ref{tab:pbp_ls_fit_wd_8nt4_b5.2_h1.9}
        as input.
      }  & 0.02 & 0.1  & 0.0060(1)  & 0.092(1)  &   ---    & 0 &
        \multicolumn{2}{c}{---}
    \end{tabular}
  \end{minipage}
\end{table}


\begin{table}
  \centering
  \caption{
    Fit of $\qbq = c_0 + c_1 \exp \left( -c_2 L_s \right)$,
    $8^3\times 4, \beta=5.2, m_0=1.9$
  }
  \label{tab:pbp_ls_fit_wd_8nt4_b5.2_h1.9}
  \begin{minipage}[t]{5in}
    \begin{tabular}{lr@{--}llllcr@{.}l}
      \hline\hline
      \multicolumn{1}{c}{$m_f$} &
      \multicolumn{2}{c}{fit range} &
      \multicolumn{1}{c}{$c_0$} &
      \multicolumn{1}{c}{$c_1$} &
      \multicolumn{1}{c}{$c_2$} &
      $N_{\rm dof}$ &
      \multicolumn{2}{c}{$\chi^2/N_{\rm dof}$} \\
      \hline\hline
      0.0\footnote{
        Used fitted values from table \ref{tab:pbp_mf_fit_wd_8nt4_b5.2_h1.9}
        as input.
      }    &  8 & 40 & 0.0059(1)  & 0.014(1)  & 0.11(1)  & 4 & 6&62 \\
           & 10 & 40 & 0.0060(1)  & 0.016(2)  & 0.13(1)  & 3 & 7&90 \\
           & 12 & 40 & 0.0058(2)  & 0.012(2)  & 0.10(1)  & 2 & 9&41 \\
           & 16 & 40 & 0.003(4)   & 0.007(3)  & 0.02(2)  & 1 & 3&16 \\
      \hline
      0.02 &  8 & 40 & 0.00779(8)\footnote{
        Used as input to fits in table \ref{tab:pbp_mf_fit_wd_8nt4_b5.2_h1.9}.
      }                           & 0.014(1)  & 0.116(8) & 4 & 5&55 \\
           & 10 & 40 & 0.00780(9) & 0.014(1)  & 0.118(9) & 3 & 7&37 \\
           & 12 & 40 & 0.0076(1)  & 0.011(1)  & 0.10(1)  & 2 & 7&74 \\
           & 16 & 40 & 0.0065(9)  & 0.0064(4) & 0.04(2)  & 1 & 4&67 \\
      \hline
      0.1  &  8 & 40 & 0.01527(4) & 0.0188(8) & 0.149(5) & 4 & 5&14 \\
           & 10 & 40 & 0.01514(6)$^\alph{mpfootnote}$
                                  & 0.013(1)  & 0.119(8) & 3 & 0&41 \\
           & 12 & 40 & 0.01513(7) & 0.013(1)  & 0.117(9) & 2 & 0&56 \\
           & 16 & 40 & 0.0151(1)  & 0.010(3)  & 0.101(2) & 1 & 0&03 \\
    \end{tabular}
  \end{minipage}
\end{table}

\clearpage


\begin{table}
  \centering
  \caption[
    Parameters for $8^3\times4, \beta=5.45, m_0=1.9$ deconfined phase study
  ]{
    Run parameters for $8^3\times4, \beta=5.45, m_0=1.9$ deconfined phase study
  }
  \label{tab:8nt4_b5.45_h1.9}
  \begin{tabular}{ccccccccc}
    \hline\hline
    $m_f$ & $L_s$ & traj.~len. & \# traj. & acc. &
    $\left\langle e^{-\Delta H}\right\rangle$ & plaq. & $\w3m$ & $\qbq$ \\
    \hline\hline
    0.02 &  8 & $\frac{1}{64}\times 32$  & 200-800  & 0.91 & 1.005(7) &
      0.5376(7) & 0.226(4) & 0.00415(6) \\
         & 10 & $\frac{1}{64}\times 32$  & 200-1000 & 0.91 & 0.992(8) &
      0.5328(6) & 0.207(4) & 0.00319(5) \\
         & 12 & $\frac{1}{64}\times 32$  & 200-800  & 0.95 & 1.009(9) &
      0.5300(4) & 0.202(5) & 0.00270(3) \\
         & 16 & $\frac{1}{64}\times 32$  & 200-800  & 0.90 & 1.02(1)  &
      0.5266(8) & 0.199(4) & 0.00237(6) \\
         & 24 & $\frac{1}{64}\times 32$  & 400-1200 & 0.86 & 0.98(2)  &
      0.5257(7) & 0.187(3) & 0.00216(6) \\
         & 32 & $\frac{1}{100}\times 50$ & 400-800  & 0.94 & 1.00(2)  &
      0.524(2)  & 0.180(5) & 0.00209(5) \\
    \hline
    0.06 &  8 & $\frac{1}{50}\times 25$  & 200-1000 & 0.86 & 0.99(3)  &
      0.536(1)  & 0.217(3) & 0.0080(1)  \\
         & 10 & $\frac{1}{64}\times 32$  & 200-1000 & 0.92 & 0.994(7) &
      0.5313(6) & 0.203(4) & 0.00704(5) \\
         & 12 & $\frac{1}{64}\times 32$  & 200-1000 & 0.89 & 1.013(8) &
      0.5286(8) & 0.195(4) & 0.00666(5) \\
         & 16 & $\frac{1}{64}\times 32$  & 400-800  & 0.76 & 1.02(4)  &
      0.525(2)  & 0.192(4) & 0.00637(7) \\
         & 24 & $\frac{1}{64}\times 32$  & 300-1000 & 0.84 & 1.00(1)  &
      0.521(2)  & 0.174(6) & 0.00617(9) \\
         & 32 & $\frac{1}{64}\times 32$  & 500-1000 & 0.80 & 1.00(2)  &
      0.525(2)  & 0.189(3) & 0.00592(4) \\
    \hline
    0.1  &  8 & $\frac{1}{50}\times 25$  & 300-800  & 0.83 & 0.98(2)  &
      0.5336(6) & 0.211(4) & 0.01174(4) \\
         & 10 & $\frac{1}{50}\times 25$  & 300-990  & 0.88 & 0.99(1)  &
      0.5310(9) & 0.200(2) & 0.01075(5) \\
         & 12 & $\frac{1}{50}\times 25$  & 600-1200 & 0.74 & 1.01(4)  &
      0.528(1)  & 0.197(4) & 0.01838(4) \\
         & 16 & $\frac{1}{64}\times 32$  & 400-800  & 0.79 & 1.01(3)  &
      0.523(1)  & 0.170(5) & 0.0103(1)  \\
         & 24 & $\frac{1}{64}\times 32$  & 400-2000 & 0.86 & 0.991(8) &
      0.512(1)  & 0.170(8) & 0.0102(1)  \\
         & 32 & $\frac{1}{64}\times 32$  & 300-1000 & 0.81 & 0.98(2)  &
      0.519(1)  & 0.159(5) & 0.01011(9) \\
    \hline
    0.14 &  8 & $\frac{1}{50}\times 25$  & 200-800  & 0.83 & 1.01(1)  &
      0.533(1)  & 0.210(3) & 0.01531(9) \\
         & 16 & $\frac{1}{64}\times 32$  & 600-1200 & 0.76 & 0.98(2)  &
      0.520(1)  & 0.159(9) & 0.0143(1)  \\
    \hline
    0.18 &  8 & $\frac{1}{50}\times 25$  & 400-800  & 0.81 & 1.03(2)  &
      0.5314(6) & 0.202(4) & 0.01884(5) \\
         & 16 & $\frac{1}{64}\times 32$  & 600-1200 & 0.78 & 0.94(2)  &
      0.515(1)  & 0.141(8) & 0.0182(2)  \\
  \end{tabular}
\end{table}


\begin{table}
  \centering
  \caption{
    Fit of $\qbq = c_0 + c_1 m_f +c_2 m_f^2$,
    $8^3\times 4, \beta=5.45, m_0=1.9$
  }
  \label{tab:pbp_mf_fit_wd_8nt4_b5.45_h1.9}
  \begin{minipage}[t]{5in}
    \begin{tabular}{|r|r@{--}l|r@{.}l|r@{.}l|r@{.}l|c|r@{.}l|}
      \hline
      \multicolumn{1}{|c|}{$L_s$} &
      \multicolumn{2}{c|}{fit range} &
      \multicolumn{2}{c|}{$c_0$} &
      \multicolumn{2}{c|}{$c_1$} &
      \multicolumn{2}{c|}{$c_2$} &
      $N_{\rm dof}$ &
      \multicolumn{2}{c|}{$\chi^2/N_{\rm dof}$} \\
      \hline\hline
       8 & 0.02 & 0.1  & 0&00227(7) & 0&0948(9) & \multicolumn{2}{c|}{---}
                                                            & 1 &  0&62 \\
      \cline{2-12}
         & 0.02 & 0.18 & 0&00219(9)\footnote{
        Shown in figure \ref{fig:pbp_ls_fit_wd_8nt4_b5.45_h1.9} and used
        as input to fits in table \ref{tab:pbp_ls_fit_wd_8nt4_b5.45_h1.9}.
      }
                                    & 0&099(2)  & -0&037(9) & 2 &  0&10 \\
      \cline{4-12}
         & \multicolumn{2}{c|}{}
                       & 0&00246(6) & 0&0915(5) & \multicolumn{2}{c|}{---}
                                                            & 3 &  5&80 \\
      \hline
      10 & 0.02 & 0.1  & 0&00131(6)$^\alph{mpfootnote}$
                                    & 0&0947(9) & \multicolumn{2}{c|}{---}
                                                            & 1 &  1&23 \\
      \hline
      12 & 0.02 & 0.1  & 0&00077(4)$^\alph{mpfootnote}$
                                    & 0&0968(8) & \multicolumn{2}{c|}{---}
                                                            & 1 &  3&85 \\
      \hline
      16 & 0.02 & 0.1  & 0&00039(8) & 0&100(2)  & \multicolumn{2}{c|}{---}
                                                            & 1 &  0&09 \\
      \cline{4-12}
         & \multicolumn{2}{c|}{}
                       & \multicolumn{2}{c|}{---}
                                    & 0&1060(8) & \multicolumn{2}{c|}{---}
                                                            & 2 & 12&86 \\
      \cline{2-12}
         & 0.02 & 0.18 & 0&0004(1)  & 0&100(3)  & -0&01(2)  & 2 &  0&02 \\
      \cline{4-12}
         & \multicolumn{2}{c|}{}
                       & 0&00040(6)$^\alph{mpfootnote}$
                                    & 0&0992(8) & \multicolumn{2}{c|}{---}
                                                            & 3 &  0&07 \\
      \cline{4-12}
         & \multicolumn{2}{c|}{}
                       & \multicolumn{2}{c|}{---}
                                    & 0&111(1)  & -0&06(1)  & 3 &  4&81 \\
      \hline
      24 & 0.02 & 0.1  & 0&00016(8)$^\alph{mpfootnote}$
                                    & 0&100(2)  & \multicolumn{2}{c|}{---}
                                                            & 1 &  0&01 \\
      \cline{4-12}
         & \multicolumn{2}{c|}{}
                       & \multicolumn{2}{c|}{---}
                                    & 0&103(1)  & -0&07(3)  & 1 &  0&83 \\
      \cline{4-12}
         & \multicolumn{2}{c|}{}
                       & \multicolumn{2}{c|}{---}
                                    & 0&103(1)  & \multicolumn{2}{c|}{---}
                                                            & 2 &  2&20 \\
      \hline
      32 & 0.02 & 0.1  & \multicolumn{2}{c|}{---}
                                    & 0&100(5)  & \multicolumn{2}{c|}{---}
                                                            & 2 &  3&99 \\
      \cline{4-12}
         & \multicolumn{2}{c|}{}
                       & 0&00006(6)$^\alph{mpfootnote}$
                                    & 0&099(1)  & \multicolumn{2}{c|}{---}
                                                            & 1 &  7&11 \\
      \cline{4-12}
         & \multicolumn{2}{c|}{}
                       & \multicolumn{2}{c|}{---}
                                    & 0&099(2)  &  0&02(2)  & 1 &  7&40 \\
      \hline
      $\infty$\footnote{
        Used fitted values from table \ref{tab:pbp_ls_fit_wd_8nt4_b5.45_h1.9}
        as input.
      }  & 0.02 & 0.1  & \multicolumn{2}{c|}{---}
                                    & 0&1012(4) & \multicolumn{2}{c|}{---}
                                                            & 2 & 6&90 \\
      \cline{4-12}
         & \multicolumn{2}{c|}{}
                       & 0&00011(5) & 0&0995(8) & \multicolumn{2}{c|}{---}
                                                            & 1 &  8&45 \\
      \hline
    \end{tabular}
  \end{minipage}
\end{table}


\begin{table}
  \centering
  \caption{
    Fit of $\qbq = c_0 + c_1 \exp \left( -c_2 L_s \right)$,
    $8^3\times 4, \beta=5.45, m_0=1.9$
  }
  \label{tab:pbp_ls_fit_wd_8nt4_b5.45_h1.9}
  \begin{minipage}[t]{5in}
    \begin{tabular}{|r@{.}l|r@{--}l|r@{.}l|r@{.}l|r@{.}l|c|r@{.}l|}
      \hline
      \multicolumn{2}{|c|}{$m_f$} &
      \multicolumn{2}{c|}{fit range} &
      \multicolumn{2}{c|}{$c_0$} &
      \multicolumn{2}{c|}{$c_1$} &
      \multicolumn{2}{c|}{$c_2$} &
      $N_{\rm dof}$ &
      \multicolumn{2}{c|}{$\chi^2/N_{\rm dof}$} \\
      \hline
      0&0\footnote{
        Used fitted values from table \ref{tab:pbp_mf_fit_wd_8nt4_b5.45_h1.9}
        as input.
      }    &  8 & 32 & 0&00010(5) & 0&019(3) & 0&27(2)  & 3 &  0&76 \\
      \cline{5-13}
      \multicolumn{2}{|c|}{} 
           & \multicolumn{2}{c|}{}
                     & \multicolumn{2}{c|}{---}
                                  & 0&015(2) & 0&24(1)  & 4 &  1&64 \\
      \cline{3-13}
      \multicolumn{2}{|c|}{} 
           & 10 & 32 & 0&00013(5) & 0&02(1)  & 0&30(4)  & 2 &  1&45 \\
      \cline{5-13}
      \multicolumn{2}{|c|}{} 
           & \multicolumn{2}{c|}{}
                     & \multicolumn{2}{c|}{---}
                                  & 0&011(3) & 0&22(2)  & 3 &  1&74 \\
      \cline{3-13}
      \multicolumn{2}{|c|}{} 
           & 12 & 32 & \multicolumn{2}{c|}{---}
                                  & 0&005(2) & 0&15(3)  & 2 &  0&22 \\
      \cline{5-13}
      \multicolumn{2}{|c|}{} 
           & \multicolumn{2}{c|}{}
                     & 0&00014(5) & 0&03(5)  & 0&3(1)   & 1 &  4&32 \\
      \hline
      0&02 &  8 & 32 & 0&00213(4)\footnote{
        Used as input to fits
        in table \ref{tab:pbp_mf_fit_wd_8nt4_b5.45_h1.9}.
      }
                                  & 0&025(4) & 0&31(2)  & 3 &  1&07 \\
      \cline{3-13}
      \multicolumn{2}{|c|}{} 
           & 10 & 32 & 0&00212(4) & 0&018(6) & 0&28(4)  & 2 &  1&24 \\
      \cline{3-13}
      \multicolumn{2}{|c|}{} 
           & 12 & 32 & 0&00208(6) & 0&005(3) & 0&18(6)  & 1 &  0&10 \\
      \hline
      0&06 &  8 & 32 & 0&00599(4)$^\alph{mpfootnote}$
                                  & 0&015(3) & 0&26(2)  & 3 &  4&84 \\
      \cline{3-13}
      \multicolumn{2}{|c|}{} 
           & 10 & 32 & 0&00591(6) & 0&005(1) & 0&15(3)  & 2 &  2&05 \\
      \cline{3-13}
      \multicolumn{2}{|c|}{} 
           & 12 & 32 & 0&00598(5) & 0&01(1)  & 0&24(7)  & 1 &  8&04 \\

      \hline
      0&1  &  8 & 32 & 0&01016(6)$^\alph{mpfootnote}$
                                  & 0&08(3)  & 0&48(6)  & 3 &  0&42 \\
      \cline{3-13}
      \multicolumn{2}{|c|}{} 
           & 10 & 32 & 0&01015(7) & 0&04(7)  & 0&4(2)   & 2 &  0&58 \\
      \cline{3-13}
      \multicolumn{2}{|c|}{} 
           & 12 & 32 & 0&0101(1)  & 0&002(6) & 0&2(3)   & 1 &  0&30 \\
      \hline
    \end{tabular}
  \end{minipage}
\end{table}

\clearpage


\begin{table}
  \centering
  \caption[
    Parameters for transition study: $8^3\times 4, m_0=1.9, L_s=24, m_f=0.02$
  ]{
    Run parameters for transition study:
    $8^3\times 4, m_0=1.9, L_s=24, m_f=0.02$
  }
  \label{tab:wd_8nt4_h1.9_l24_m0.02}
  \begin{tabular}{lcr@{$\times$}lr@{--}llllll}
    \multicolumn{1}{c}{$\beta$} &
      \multicolumn{1}{c}{start} &
      \multicolumn{2}{c}{traj.~len.} &
      \multicolumn{2}{c}{\# traj.} &
      \multicolumn{1}{c}{acc.} &
      \multicolumn{1}{c}{$\left\langle e^{-\Delta H}\right\rangle$} &
      \multicolumn{1}{c}{plaq.} &
      \multicolumn{1}{c}{$\w3m$} &
      \multicolumn{1}{c}{$\qbq$} \\
    \hline\hline
    5.2   & O & $\frac{1}{100}$ & 50 & 300 & 2000 & 0.72 & 0.95(2) &
      0.4359(6) & 0.046(2) & 0.0087(1)  \\
    5.275 & O & $\frac{1}{100}$ & 50 & 200 & 1000 & 0.83 & 1.01(2) &
      0.459(1)  & 0.068(4) & 0.0069(2)  \\
    5.325 & D & $\frac{1}{100}$ & 50 & 200 & 1150 & 0.89 & 1.01(1) &
      0.480(3)  & 0.10(1)  & 0.0050(4)  \\
    5.45  & D & $\frac{1}{64}$  & 32 & 400 & 1200 & 0.86 & 0.98(2) &
      0.5257(7) & 0.187(3) & 0.00216(6) \\
  \end{tabular}
\end{table}


\begin{table}
  \centering
  \caption[
    Parameters for transition study: $16^3\times 4, m_0=1.9, L_s=24, m_f=0.02$
  ]{
    Run parameters for transition study:
    $16^3\times 4, m_0=1.9, L_s=24, m_f=0.02$
  }
  \label{tab:wd_16nt4_h1.9_l24_m0.02}
  HMC traj.~len: $\frac{1}{128}\times 64$, \qquad
  CG stop cond: $10^{-6}$
  \begin{tabular}{lcr@{--}llllll}
    \multicolumn{1}{c}{$\beta$} &
      \multicolumn{1}{c}{start} &
      \multicolumn{2}{c}{\# traj.} &
      \multicolumn{1}{c}{acc.} &
      \multicolumn{1}{c}{$\left\langle e^{-\Delta H}\right\rangle$} &
      \multicolumn{1}{c}{plaq.} &
      \multicolumn{1}{c}{$\w3m$} &
      \multicolumn{1}{c}{$\qbq$} \\
    \hline\hline
    5.25  & D &  140 &  440 & 0.69 & 1.01(6) & 0.4465(2) & 0.046(2) &
      0.0087(1)  \\
          & O &  350 &  650 & 0.74 & 0.99(5) & 0.4475(3) & 0.051(2) &
      0.00814(4) \\
    \hline
    5.275 & D &  360 &  660 & 0.70 & 0.92(5) & 0.4565(5) & 0.062(2) &
      0.00737(4) \\
          & O &  470 &  770 & 0.71 & 0.93(5) & 0.4561(3) & 0.061(2) &
      0.00747(4) \\
    \hline
    5.3   & D & 1100 & 1400 & 0.72 & 1.05(4) & 0.4656(8) & 0.074(2) &
      0.0066(1)  \\
          & O &  720 & 1020 & 0.75 & 1.06(4) & 0.4683(6) & 0.077(2) &
      0.00629(7) \\
    \hline
    5.325 & D &  490 &  790 & 0.82 & 1.00(3) & 0.4800(6) & 0.096(2) &
      0.00506(6) \\
          & O &  670 &  970 & 0.77 & 0.96(3) & 0.480(1)  & 0.102(4) &
      0.0049(3)  \\
    \hline
    5.35  & D &  650 &  950 & 0.86 & 1.04(3) & 0.4924(9) & 0.124(3) &
      0.0037(1)  \\
          & O &  200 &  500 & 0.84 & 1.03(3) & 0.4928(4) & 0.127(2) &
      0.00366(5) \\
  \end{tabular}
\end{table}


\begin{table}
  \centering
  \caption[
    Parameters for $8^3\times 4$ improved gauge action transition study
  ]{
    Run parameters for $8^3\times 4$ improved gauge action transition study
  }
  \label{tab:rd_8nt4_h1.9_l24_m0.02}
  vol: $8^3\times 4$, \qquad
  $c_1 = -0.331, \qquad m_0=1.9, \qquad L_s=24, \qquad m_f=0.02$ \\
  HMC traj.~len: $\frac{1}{100}\times 50$, \qquad CG stop cond: $10^{-6}$
  \begin{tabular}{lcr@{--}llllll}
    \multicolumn{1}{c}{$\beta$} &
      \multicolumn{1}{c}{start} &
      \multicolumn{2}{c}{\# traj.} &
      \multicolumn{1}{c}{acc.} &
      \multicolumn{1}{c}{$\left\langle e^{-\Delta H}\right\rangle$} &
      \multicolumn{1}{c}{plaq.} &
      \multicolumn{1}{c}{$\w3m$} &
      \multicolumn{1}{c}{$\qbq$} \\
    \hline\hline
    1.7 & D &  70 & 170 & 0.77 & 1.00(5) & 0.424(1)  & 0.032(2) &
      0.01067(9) \\
        & O &  70 & 170 & 0.76 & 0.96(4) & 0.423(1)  & 0.034(2) &
      0.0104(2)  \\
    \hline
    1.8 & D & 100 & 200 & 0.86 & 1.07(5) & 0.4620(9) & 0.037(2) &
      0.0097(2)  \\
        & O & 100 & 200 & 0.78 & 0.93(4) & 0.462(1)  & 0.033(4) &
      0.0096(2)  \\
    \hline
    1.9 & D & 130 & 230 & 0.88 & 0.98(5) & 0.507(2)  & 0.057(3) &
      0.0066(2)  \\
        & O & 130 & 230 & 0.89 & 1.11(6) & 0.5075(9) & 0.064(5) &
      0.0064(2)  \\
    \hline
    2.0 & D & 160 & 260 & 0.93 & 0.98(3) & 0.5500(8) & 0.123(6) &
      0.0028(1)  \\
        & O & 160 & 260 & 0.96 & 0.99(3) & 0.5502(9) & 0.128(3) &
      0.00245(5) \\
    \hline
    2.1 & D & 200 & 400 & 0.94 & 0.98(1) & 0.5824(6) & 0.17(1)  &
      0.0021(2)  \\
        & O & 200 & 400 & 0.94 & 0.99(3) & 0.5806(7) & 0.17(1)  &
      0.0023(2)  \\
    \hline
    2.2 & D & 140 & 240 & 0.95 & 1.01(1) & 0.6054(4) & 0.20(2)  &
      0.00191(3) \\
        & O & 140 & 240 & 0.93 & 0.99(1) & 0.6064(5) & 0.212(9) &
      0.00187(1) \\
  \end{tabular}
\end{table}


\begin{table}
  \centering
  \caption[
    Parameters for $16^3\times 4$ improved gauge action transition study
  ]{
    Run parameters for $16^3\times 4$ improved gauge action transition study
  }
  \label{tab:rd_16nt4_h1.9_l24_m0.02}
  vol: $16^3\times 4$, \qquad
  $c_1 = -0.331, \qquad m_0=1.9, \qquad L_s=24, \qquad m_f=0.02$ \\
  HMC traj.~len: $\frac{1}{128}\times 64$, \qquad CG stop cond: $10^{-6}$
  \begin{tabular}{lcr@{--}llllll}
    \multicolumn{1}{c}{$\beta$} &
      \multicolumn{1}{c}{start} &
      \multicolumn{2}{c}{\# traj.} &
      \multicolumn{1}{c}{acc.} &
      \multicolumn{1}{c}{$\left\langle e^{-\Delta H}\right\rangle$} &
      \multicolumn{1}{c}{plaq.} &
      \multicolumn{1}{c}{$\w3m$} &
      \multicolumn{1}{c}{$\qbq$} \\
    \hline\hline
    1.85 & D & 150 & 270 & 0.75 & 1.0(1)  & 0.4848(4) & 0.044(1)  &
      0.00843(6) \\
         & O & 160 & 300 & 0.73 & 1.09(9) & 0.4861(4) & 0.044(2)  &
      0.00823(7) \\
    \hline
    1.9  & D & 200 & 300 & 0.71 & 0.96(6) & 0.5079(7) & 0.063(2)  &
      0.0064(1)  \\
         & O & 200 & 300 & 0.78 & 0.99(3) & 0.5095(3) & 0.0658(8) &
      0.00598(7) \\
    \hline
    1.95 & D & 220 & 520 & 0.76 & 1.00(2) & 0.5313(3) & 0.098(2)  &
      0.00378(5) \\
         & O & 150 & 450 & 0.83 & 1.04(4) & 0.5321(3) & 0.107(2)  &
      0.00362(4) \\
    \hline
    2.0  & D & 200 & 300 & 0.93 & 1.08(3) & 0.5506(3) & 0.1242(6) &
      0.00260(7) \\
         & O & 200 & 300 & 0.90 & 1.05(3) & 0.5505(3) & 0.127(2)  &
      0.00260(5) \\
  \end{tabular}
\end{table}

\clearpage


\begin{table}
  \centering
  \caption{
    Valence $\qbq$ and $\chi_\pi$:
    $8^3\times 4, \beta=5.2, m_0=1.9, L_s=10\mbox{--}16, m_f=0.02$
  }
  \label{tab:pbp_spc_8nt4_b5.2_h1.9_m0.02_1}
  \renewcommand{\arraystretch}{0.9}
  \begin{tabular}{|r|r@{.}l|r@{--}l|r|r@{.}l|r@{--}l|r|r@{.}l|}
    \hline
    &
      \multicolumn{2}{c|}{} &
      \multicolumn{5}{c|}{$\qbq$} &
      \multicolumn{5}{c|}{$\chi_\pi$} \\
    \cline{4-13}
    \multicolumn{1}{|c|}{$L_s^{\rm (dyn)}$} &
      \multicolumn{2}{c|}{$m_f^{\rm (val)}$} &
      \multicolumn{2}{c|}{\# traj} &
      \multicolumn{1}{c|}{\# meas} &
      \multicolumn{2}{c|}{avg} &
      \multicolumn{2}{c|}{\# traj} &
      \multicolumn{1}{c|}{\# meas} &
      \multicolumn{2}{c|}{avg} \\
    \hline\hline
    10 & 0 & 02 & 200 & 2000 & 1800 & 0 & 0124(1)  &
      800 & 2000 & 240 & 0 & 130(1)  \\
    \cline{6-8}
       & 0 & 06 & \multicolumn{2}{c|}{}
                             &  360 & 0 & 0156(1)  &
      \multicolumn{2}{c|}{}
                 &     & 0 & 1202(7) \\
       & 0 & 1  & \multicolumn{2}{c|}{}
                             &      & 0 & 0186(1)  &
      \multicolumn{2}{c|}{}
                 &     & 0 & 1130(6) \\
       & 0 & 14 & \multicolumn{2}{c|}{}
                             &      & 0 & 0215(1)  &
      \multicolumn{2}{c|}{}
                 &     & 0 & 1071(5) \\
       & 0 & 18 & \multicolumn{2}{c|}{}
                             &      & 0 & 02432(9) &
      \multicolumn{2}{c|}{}
                 &     & 0 & 1022(4) \\
       & 0 & 22 & \multicolumn{2}{c|}{}
                             &      & 0 & 02695(8) &
      \multicolumn{2}{c|}{}
                 &     & 0 & 0979(4) \\
    \hline
    12 & 0 & 02 & 200 & 2000 & 1800 & 0 & 01123(7) &
      800 & 2000 & 240 & 0 & 129(1)  \\
    \cline{6-8}
       & 0 & 06 & \multicolumn{2}{c|}{}
                             &  360 & 0 & 01448(8) &
      \multicolumn{2}{c|}{}
                 &     & 0 & 1193(7) \\
       & 0 & 1  & \multicolumn{2}{c|}{}
                             &      & 0 & 01759(7) &
      \multicolumn{2}{c|}{}
                 &     & 0 & 1120(6) \\
       & 0 & 14 & \multicolumn{2}{c|}{}
                             &      & 0 & 02056(6) &
      \multicolumn{2}{c|}{}
                 &     & 0 & 1061(4) \\
       & 0 & 18 & \multicolumn{2}{c|}{}
                             &      & 0 & 02338(5) &
      \multicolumn{2}{c|}{}
                 &     & 0 & 1012(4) \\
       & 0 & 22 & \multicolumn{2}{c|}{}
                             &      & 0 & 02608(5) &
      \multicolumn{2}{c|}{}
                 &     & 0 & 0970(3) \\
    \hline
    16 & 0 & 02 & 550 & 2000 & 1450 & 0 & 00987(9) &
      800 & 2000 & 240 & 0 & 134(1)  \\
    \cline{6-8}
       & 0 & 06 & \multicolumn{2}{c|}{}
                             &  290 & 0 & 01318(8) &
      \multicolumn{2}{c|}{}
                 &     & 0 & 123(1)  \\
       & 0 & 1  & \multicolumn{2}{c|}{}
                             &      & 0 & 01639(7) &
      \multicolumn{2}{c|}{}
                 &     & 0 & 1146(8) \\
       & 0 & 14 & \multicolumn{2}{c|}{}
                             &      & 0 & 01945(6) &
      \multicolumn{2}{c|}{}
                 &     & 0 & 1082(7) \\
       & 0 & 18 & \multicolumn{2}{c|}{}
                             &      & 0 & 02236(6) &
      \multicolumn{2}{c|}{}
                 &     & 0 & 1029(6) \\
       & 0 & 22 & \multicolumn{2}{c|}{}
                             &      & 0 & 02513(5) &
      \multicolumn{2}{c|}{}
                 &     & 0 & 0985(5) \\
    \hline
  \end{tabular}
\end{table}

\begin{table}
  \centering
  \caption{
    Valence $\qbq$ and $\chi_\pi$:
    $8^3\times 4, \beta=5.2, m_0=1.9, L_s=24\mbox{--}40, m_f=0.02$
  }
  \label{tab:pbp_spc_8nt4_b5.2_h1.9_m0.02_2}
  \renewcommand{\arraystretch}{1.0}
  \begin{tabular}{|r|r@{.}l|r@{--}l|r|r@{.}l|r@{--}l|r|r@{.}l|}
    \hline
    &
      \multicolumn{2}{c|}{} &
      \multicolumn{5}{c|}{$\qbq$} &
      \multicolumn{5}{c|}{$\chi_\pi$} \\
    \cline{4-13}
    \multicolumn{1}{|c|}{$L_s^{\rm (dyn)}$} &
      \multicolumn{2}{c|}{$m_f^{\rm (val)}$} &
      \multicolumn{2}{c|}{\# traj} &
      \multicolumn{1}{c|}{\# meas} &
      \multicolumn{2}{c|}{avg} &
      \multicolumn{2}{c|}{\# traj} &
      \multicolumn{1}{c|}{\# meas} &
      \multicolumn{2}{c|}{avg} \\
    \hline\hline
    24 & 0 & 02 & 350 & 2000 & 1650 & 0 & 0088(1)  &
      800 & 2000 & 240 & 0 & 141(2)  \\
    \cline{6-8}
       & 0 & 06 & \multicolumn{2}{c|}{}
                             &  330 & 0 & 0122(1)  &
      \multicolumn{2}{c|}{}
                 &     & 0 & 127(1)  \\
       & 0 & 1  & \multicolumn{2}{c|}{}
                             &      & 0 & 01550(9) &
      \multicolumn{2}{c|}{}
                 &     & 0 & 1177(9) \\
       & 0 & 14 & \multicolumn{2}{c|}{}
                             &      & 0 & 01863(8) &
      \multicolumn{2}{c|}{}
                 &     & 0 & 1108(7) \\
       & 0 & 18 & \multicolumn{2}{c|}{}
                             &      & 0 & 02161(7) &
      \multicolumn{2}{c|}{}
                 &     & 0 & 1052(5) \\
       & 0 & 22 & \multicolumn{2}{c|}{}
                             &      & 0 & 02443(5) &
      \multicolumn{2}{c|}{}
                 &     & 0 & 1005(4) \\
    \hline
    32 & 0 & 02 & 300 & 2000 & 1700 & 0 & 00835(7) &
      800 & 2000 & 240 & 0 & 155(3)  \\
    \cline{6-8}
       & 0 & 06 & \multicolumn{2}{c|}{}
                             &  340 & 0 & 01169(8) &
      \multicolumn{2}{c|}{}
                 &     & 0 & 135(2)  \\
       & 0 & 1  & \multicolumn{2}{c|}{}
                             &      & 0 & 01504(8) &
      \multicolumn{2}{c|}{}
                 &     & 0 & 123(2)  \\
       & 0 & 14 & \multicolumn{2}{c|}{}
                             &      & 0 & 01822(7) &
      \multicolumn{2}{c|}{}
                 &     & 0 & 114(1)  \\
       & 0 & 18 & \multicolumn{2}{c|}{}
                             &      & 0 & 02124(6) &
      \multicolumn{2}{c|}{}
                 &     & 0 & 108(1)  \\
       & 0 & 22 & \multicolumn{2}{c|}{}
                             &      & 0 & 02410(6) &
      \multicolumn{2}{c|}{}
                 &     & 0 & 103(1)  \\
    \hline
    40 & 0 & 02 & 300 & 1350 & 1050 & 0 & 00772(8) &
      300 & 1350 & 210 & 0 & 157(3)  \\
    \cline{6-8}
       & 0 & 06 & \multicolumn{2}{c|}{}
                             &  210 & 0 & 0113(1)  &
      \multicolumn{2}{c|}{}
                 &     & 0 & 135(2)  \\
       & 0 & 1  & \multicolumn{2}{c|}{}
                             &      & 0 & 0147(1)  &
      \multicolumn{2}{c|}{}
                 &     & 0 & 123(1)  \\
       & 0 & 14 & \multicolumn{2}{c|}{}
                             &      & 0 & 01799(7) &
      \multicolumn{2}{c|}{}
                 &     & 0 & 114(1)  \\
       & 0 & 18 & \multicolumn{2}{c|}{}
                             &      & 0 & 02106(6) &
      \multicolumn{2}{c|}{}
                 &     & 0 & 1075(9) \\
       & 0 & 22 & \multicolumn{2}{c|}{}
                             &      & 0 & 02396(6) &
      \multicolumn{2}{c|}{}
                 &     & 0 & 1022(7) \\
    \hline
  \end{tabular}
\end{table}


\begin{table}
  \centering
  \caption{
    Valence $m_{\rm res}$ and $b_0$:
    $8^3\times 4, \beta=5.2, m_0=1.9, m_f=0.02$
  }
  \label{tab:GMOR_wd_8nt4_b5.2_h1.9_m0.02}
  \begin{tabular}{|r|r|r@{.}l|r@{.}l|r@{.}l|}
    \hline
    &
      \multicolumn{1}{c|}{jknf} & 
      \multicolumn{2}{c|}{} & 
      \multicolumn{2}{c|}{} & 
      \multicolumn{2}{c|}{} \\
    \multicolumn{1}{|c|}{$L_s$} &
      \multicolumn{1}{c|}{blks} &
      \multicolumn{2}{c|}{$m_{\rm res}$} &
      \multicolumn{2}{c|}{$-b_0$} &
      \multicolumn{2}{c|}{$\chi^2/{\rm dof}$} \\
    \hline\hline
    10 & 10 & 0 & 149(5) & 0 & 0094(5) & 0 & 75(42)   \\
    \cline{2-8}
       & 15 & 0 & 149(5) & 0 & 0094(5) & 0 & 78(41)   \\
    \cline{2-8}
       & 20 & 0 & 149(4) & 0 & 0094(4) & 0 & 78(44)   \\
    \hline
    12 & 10 & 0 & 129(2) & 0 & 0080(2) & 1 & 63(44)   \\
    \cline{2-8}
       & 15 & 0 & 129(2) & 0 & 0080(4) & 1 & 68(59)   \\
    \cline{2-8}
       & 20 & 0 & 129(2) & 0 & 0080(3) & 1 & 70(70)   \\
    \hline
    16 & 10 & 0 & 113(3) & 0 & 0080(4) & 1 & 14(45)   \\
    \cline{2-8}
       & 15 & 0 & 113(3) & 0 & 0080(4) & 1 & 20(37)   \\
    \cline{2-8}
       & 20 & 0 & 113(4) & 0 & 0080(5) & 1 & 21(46)   \\
    \hline
    24 & 10 & 0 & 095(2) & 0 & 0075(3) & 1 & 52(74)   \\
    \cline{2-8}
       & 15 & 0 & 095(3) & 0 & 0075(3) & 1 & 57(68)   \\
    \cline{2-8}
       & 20 & 0 & 095(3) & 0 & 0075(3) & 1 & 61(73)   \\
    \hline
    32 & 10 & 0 & 078(2) & 0 & 0068(5) & 0 & 71(38)   \\
    \cline{2-8}
       & 15 & 0 & 078(2) & 0 & 0069(4) & 0 & 72(39)   \\
    \cline{2-8}
       & 20 & 0 & 078(2) & 0 & 0069(4) & 0 & 72(42)   \\
    \hline
    40 & 10 & 0 & 059(3) & 0 & 0048(3) & 1 & 70(93)   \\
    \cline{2-8}
       & 15 & 0 & 059(2) & 0 & 0048(3) & 1 & 74(1.47) \\
    \cline{2-8}
       & 20 & 0 & 060(2) & 0 & 0048(3) & 1 & 74(1.56) \\
    \hline
  \end{tabular}
\end{table}

\begin{table}
  \centering
  \caption{
    Valence $m_{\rm res}$ and $b_0$
    for scale setting simulations, $m_0=1.9$
  }
  \label{tab:GMOR_val_scale_setting}
  \begin{tabular}
    {|c|r@{$\times$}l|r@{.}l|r|r@{.}l|r|r@{.}l|r@{.}l|r@{.}l|}
    \hline
    gauge &
      \multicolumn{2}{c|}{} & 
      \multicolumn{2}{c|}{} & 
      \multicolumn{1}{c|}{} & 
      \multicolumn{2}{c|}{} & 
      \multicolumn{1}{c|}{jknf} & 
      \multicolumn{2}{c|}{} & 
      \multicolumn{2}{c|}{} & 
      \multicolumn{2}{c|}{} \\
    action &
      \multicolumn{2}{c|}{Vol} & 
      \multicolumn{2}{c|}{$\beta$} & 
      \multicolumn{1}{|c|}{$L_s$} &
      \multicolumn{2}{c|}{$m_f$} & 
      \multicolumn{1}{c|}{blks} &
      \multicolumn{2}{c|}{$m_{\rm res}$} &
      \multicolumn{2}{c|}{$-b_0$} &
      \multicolumn{2}{c|}{$\chi^2/{\rm dof}$} \\
    \hline\hline
    W & 8 & 32 & 5 & 325 & 24 & 0 & 02 & 10 & 0 & 057(2) & 0 & 0045(2) &
      8 & 4(3.9) \\
    \cline{9-15}
      & \multicolumn{2}{c|}{}
               & \multicolumn{2}{c|}{}
                         &    & \multicolumn{2}{c|}{}
                                       & 20 & 0 & 057(2) & 0 & 0046(2) &
      8 & 8(3.9) \\
    \cline{7-15}
      & \multicolumn{2}{c|}{}
               & \multicolumn{2}{c|}{}
                         &    & 0 & 06 & 10 & 0 & 059(3) & 0 & 0045(3) &
      2 & 1(6)   \\
    \cline{9-15}
      & \multicolumn{2}{c|}{}
               & \multicolumn{2}{c|}{}
                         &    & \multicolumn{2}{c|}{}
                                       & 20 & 0 & 059(2) & 0 & 0045(2) &
      2 & 2(1.0) \\
    \hline
    R & 8 & 32 & 1 & 9   & 24 & 0 & 02 & 10 & 0 & 038(2) & 0 & 0027(1) &
      8 & 6(2.2) \\
    \cline{9-15}
      & \multicolumn{2}{c|}{}
               & \multicolumn{2}{c|}{}
                         &    & \multicolumn{2}{c|}{}
                                       & 20 & 0 & 038(2) & 0 & 0027(2) &
      9 & 1(2.8) \\
    \cline{4-15}
      & \multicolumn{2}{c|}{}
               & 2 & 0   & 24 & 0 & 02 & 10 & 0 & 016(2) & 0 & 0014(2) &
      6 & 8(2.6) \\
    \cline{9-15}
      & \multicolumn{2}{c|}{}
               & \multicolumn{2}{c|}{}
                         &    & \multicolumn{2}{c|}{}
                                       & 20 & 0 & 016(3) & 0 & 0015(4) &
      7 & 0(5.1) \\
    \cline{7-15}
      & \multicolumn{2}{c|}{}
               & \multicolumn{2}{c|}{}
                         &    & 0 & 06 & 10 & 0 & 017(2) & 0 & 0015(3) &
      1 & 6(8)   \\
    \cline{9-15}
      & \multicolumn{2}{c|}{}
               & \multicolumn{2}{c|}{}
                         &    & \multicolumn{2}{c|}{}
                                       & 20 & 0 & 017(2) & 0 & 0015(3) &
      1 & 6(8)   \\
    \hline
  \end{tabular}
\end{table}

\clearpage


\chapter*{Figures}
\addcontentsline{toc}{chapter}{Figures}
\thispagestyle{myheadings}
\markright{}

\clearpage


\clearpage

\clearpage


\clearpage

\clearpage


\clearpage

\clearpage


\clearpage

\clearpage


\clearpage

\clearpage


\clearpage

\clearpage


\clearpage

\clearpage


\clearpage

\clearpage


\clearpage

\clearpage

\clearpage


\clearpage


\clearpage


\clearpage


\clearpage
\begin{figure}
  \centering
  \includegraphics
    [height=0.9\textheight,width=\textwidth]
    {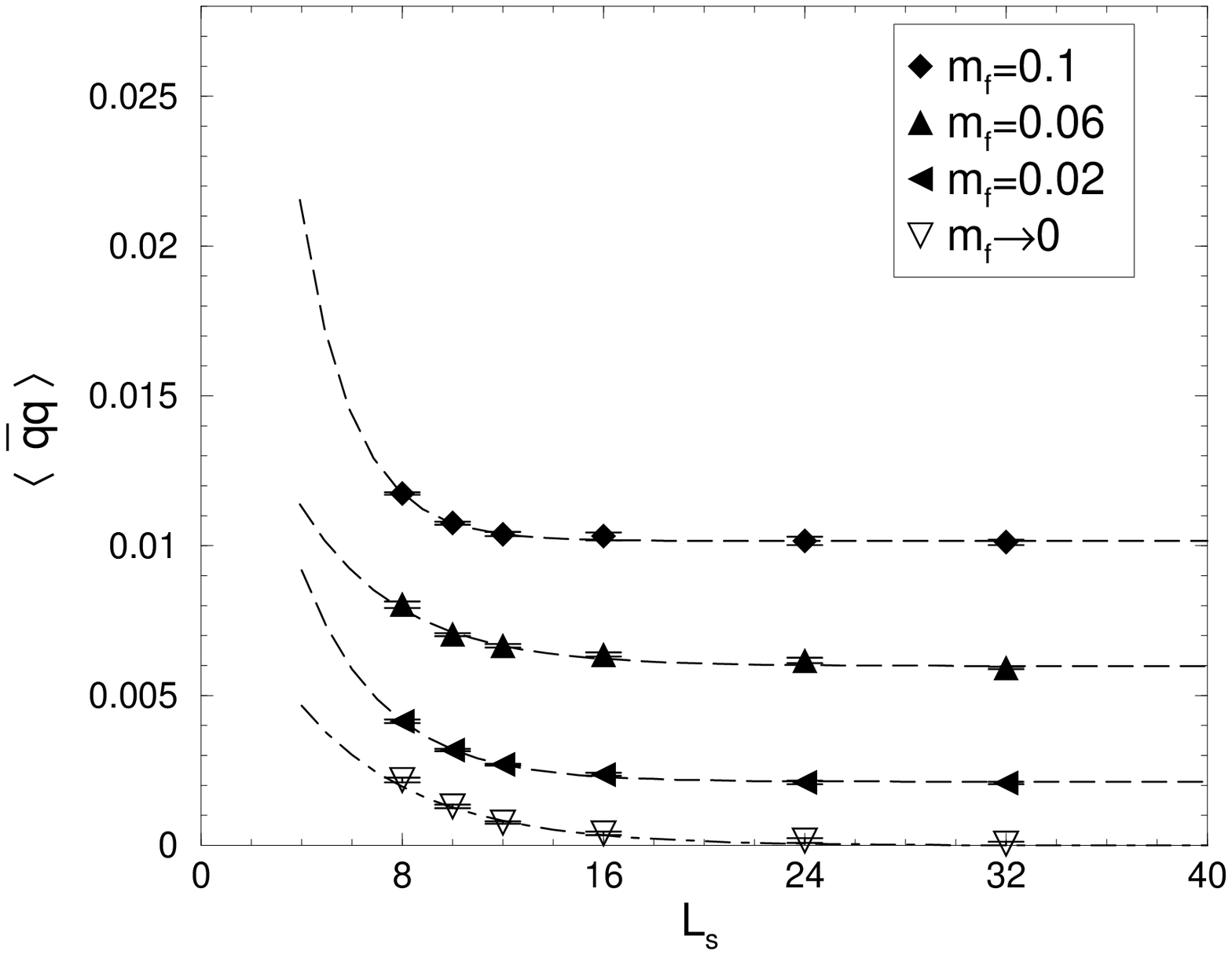}
  \caption{
    Fit of $\qbq = c_0 + c_1 \exp \left( -c_2 L_s \right)$:
    $8^3\times 4, \beta=5.45, m_0=1.9$
  }
  \label{fig:pbp_ls_fit_wd_8nt4_b5.45_h1.9}
\end{figure}

\clearpage


\begin{figure}
  \centering
  \includegraphics
    [height=0.9\textheight,width=\textwidth]
    {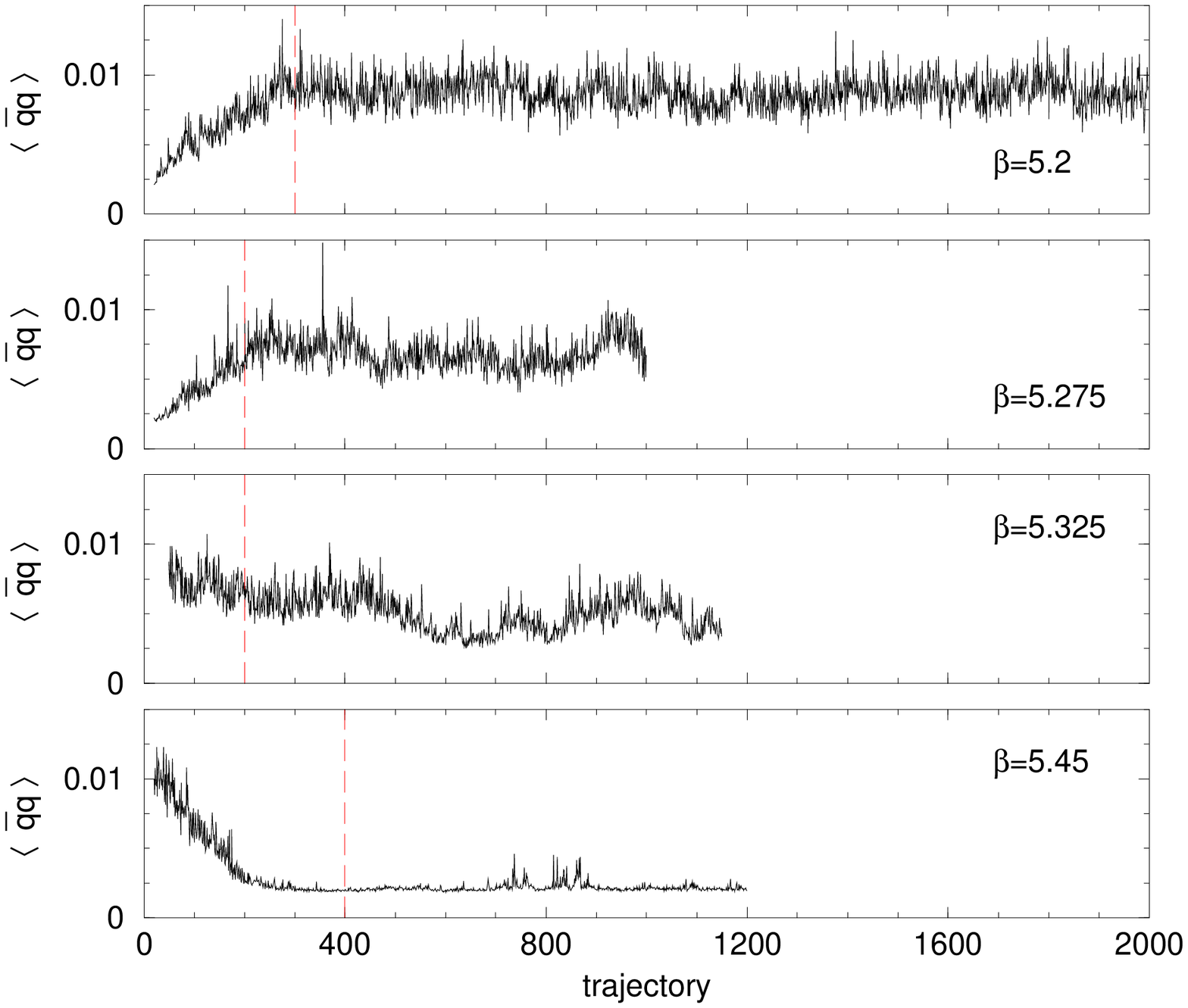}
  \caption{
    $\qbq$ evol: $8^3\times 4, m_0=1.9, L_s=24, m_f=0.02$
  }
  \label{fig:pbp_wd_8nt4_h1.9_l24_m0.02}
\end{figure}

\begin{figure}
  \centering
  \includegraphics
    [height=0.9\textheight,width=\textwidth]
    {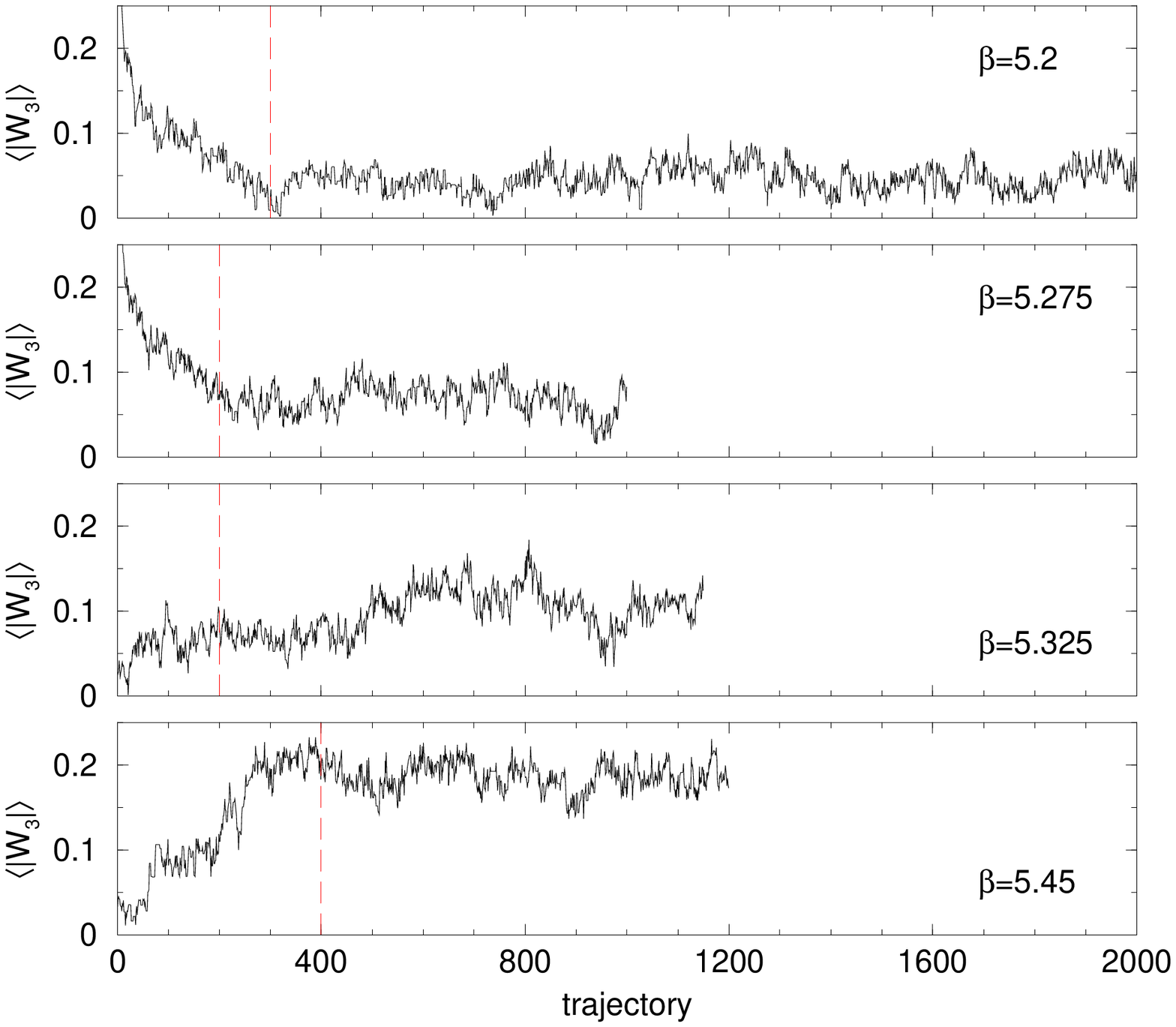}
  \caption{
    $\w3m$ evol: $8^3\times 4, m_0=1.9, L_s=24, m_f=0.02$
  }
  \label{fig:w3m_wd_8nt4_h1.9_l24_m0.02}
\end{figure}

\begin{figure}
  \centering
  \includegraphics
    [height=0.9\textheight,width=\textwidth]
    {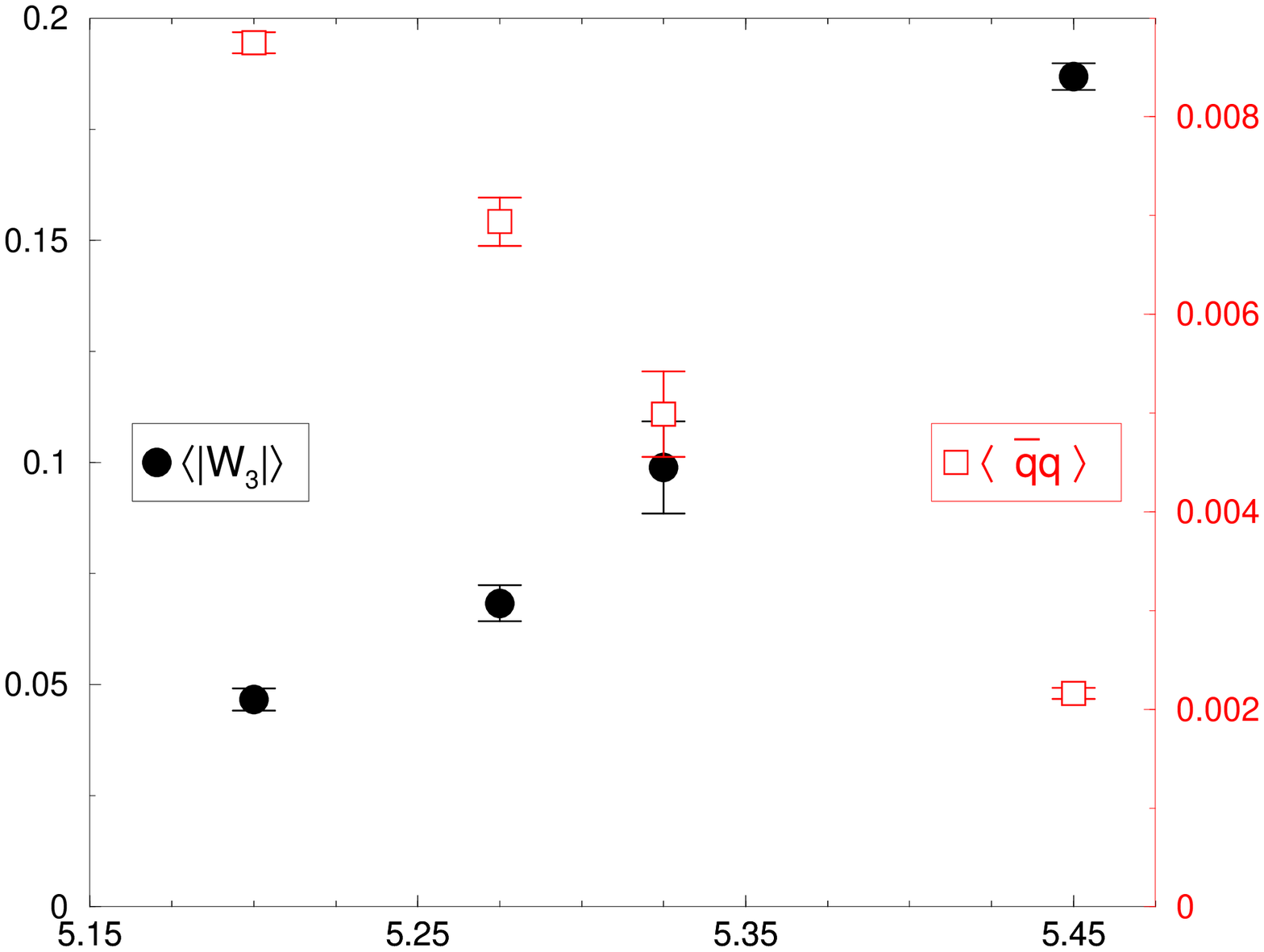}
  \caption{
    $\qbq$, $\w3m$ {\it vs}.~$\beta$: $8^3\times 4, m_0=1.9, L_s=24, m_f=0.02$
  }
  \label{fig:pbp_w3m_wd_8nt4_h1.9_l24_m0.02}
\end{figure}


\begin{figure}
  \centering
  \includegraphics
    [height=0.9\textheight,width=\textwidth]
    {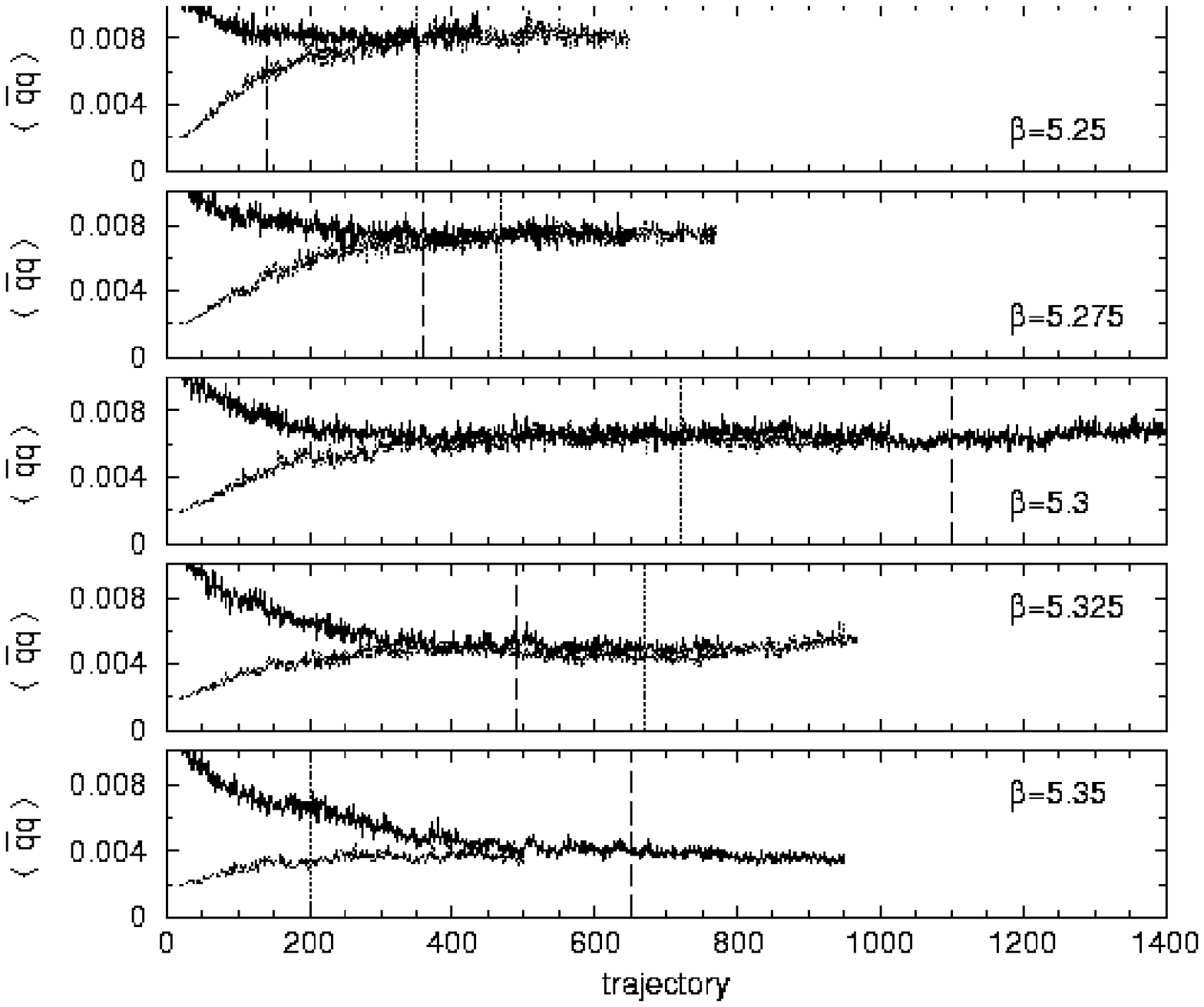}
  \caption{
    $\qbq$ evol: $16^3\times 4, m_0=1.9, L_s=24, m_f=0.02$
  }
  \label{fig:pbp_wd_16nt4_h1.9_l24_m0.02}
\end{figure}

\begin{figure}
  \centering
  \includegraphics
    [height=0.9\textheight,width=\textwidth]
    {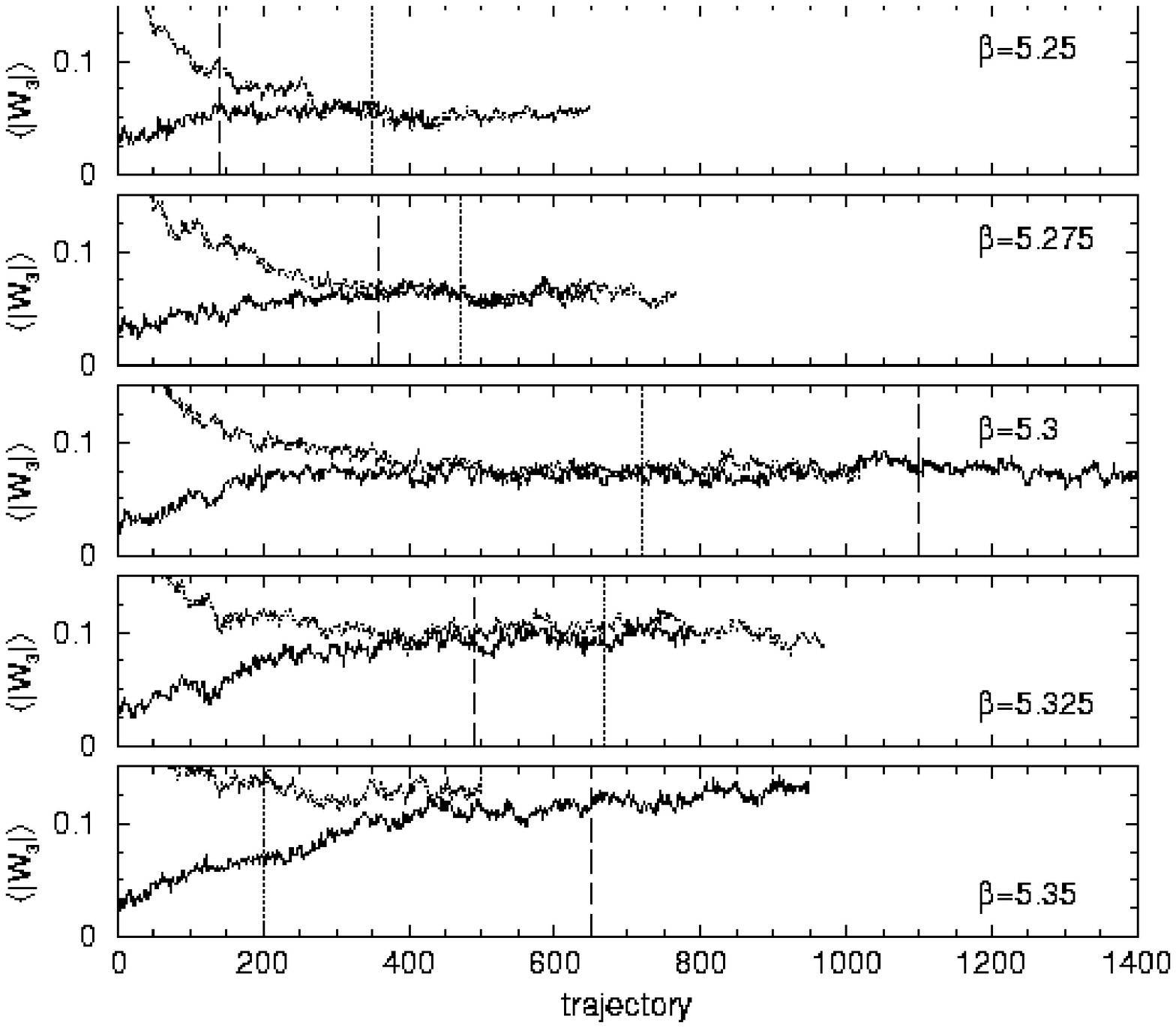}
  \caption{
    $\w3m$ evol: $16^3\times 4, m_0=1.9, L_s=24, m_f=0.02$
  }
  \label{fig:w3m_wd_16nt4_h1.9_l24_m0.02}
\end{figure}

\clearpage


\begin{figure}
  \centering
  \includegraphics
    [height=0.9\textheight,width=\textwidth]
    {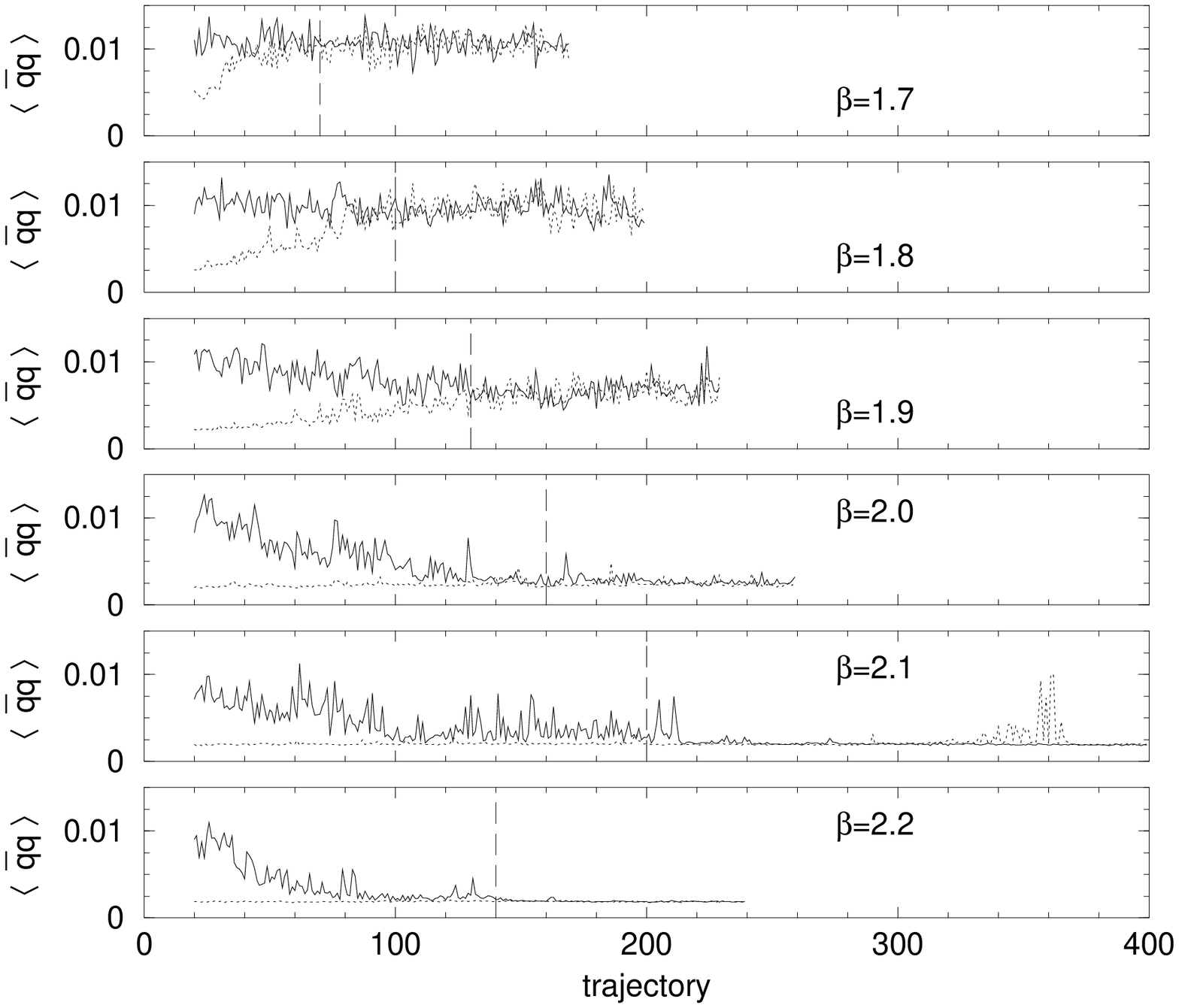}
  \caption{
    $\qbq$ evol: $8^3\times 4, c_1=-0.331, m_0=1.9, L_s=24, m_f=0.02$
  }
  \label{fig:pbp_rd_8nt4_h1.9_l24_m0.02}
\end{figure}

\begin{figure}
  \centering
  \includegraphics
    [height=0.9\textheight,width=\textwidth]
    {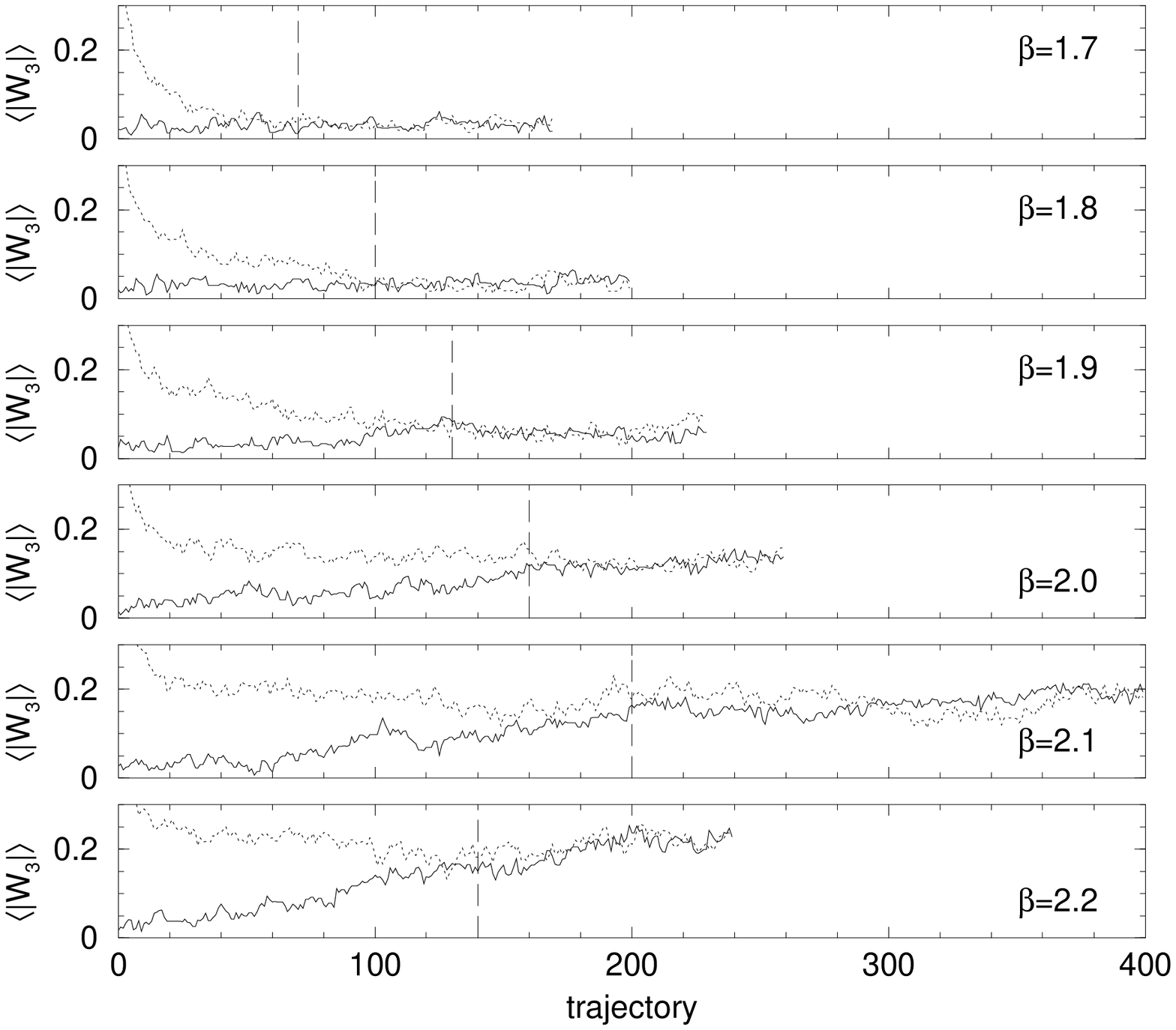}
  \caption{
    $\w3m$ evol: $8^3\times 4, c_1=-0.331, m_0=1.9, L_s=24, m_f=0.02$
  }
  \label{fig:w3m_rd_8nt4_h1.9_l24_m0.02}
\end{figure}


\begin{figure}
  \centering
  \includegraphics
    [height=0.9\textheight,width=\textwidth]
    {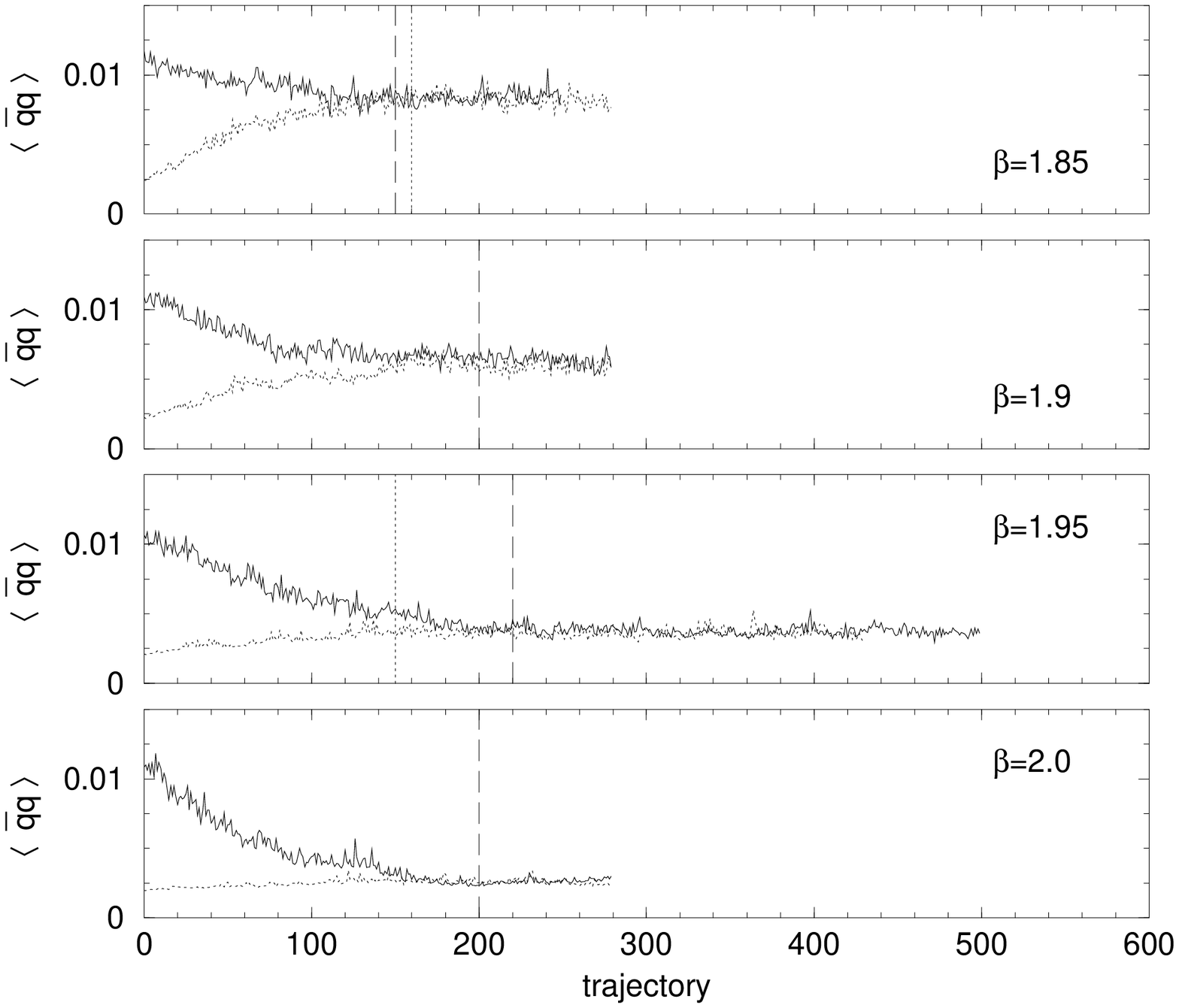}
  \caption{
    $\qbq$ evol: $16^3\times 4, c_1=-0.331, m_0=1.9, L_s=24, m_f=0.02$
  }
  \label{fig:pbp_rd_16nt4_h1.9_l24_m0.02}
\end{figure}

\begin{figure}
  \centering
  \includegraphics
    [height=0.9\textheight,width=\textwidth]
    {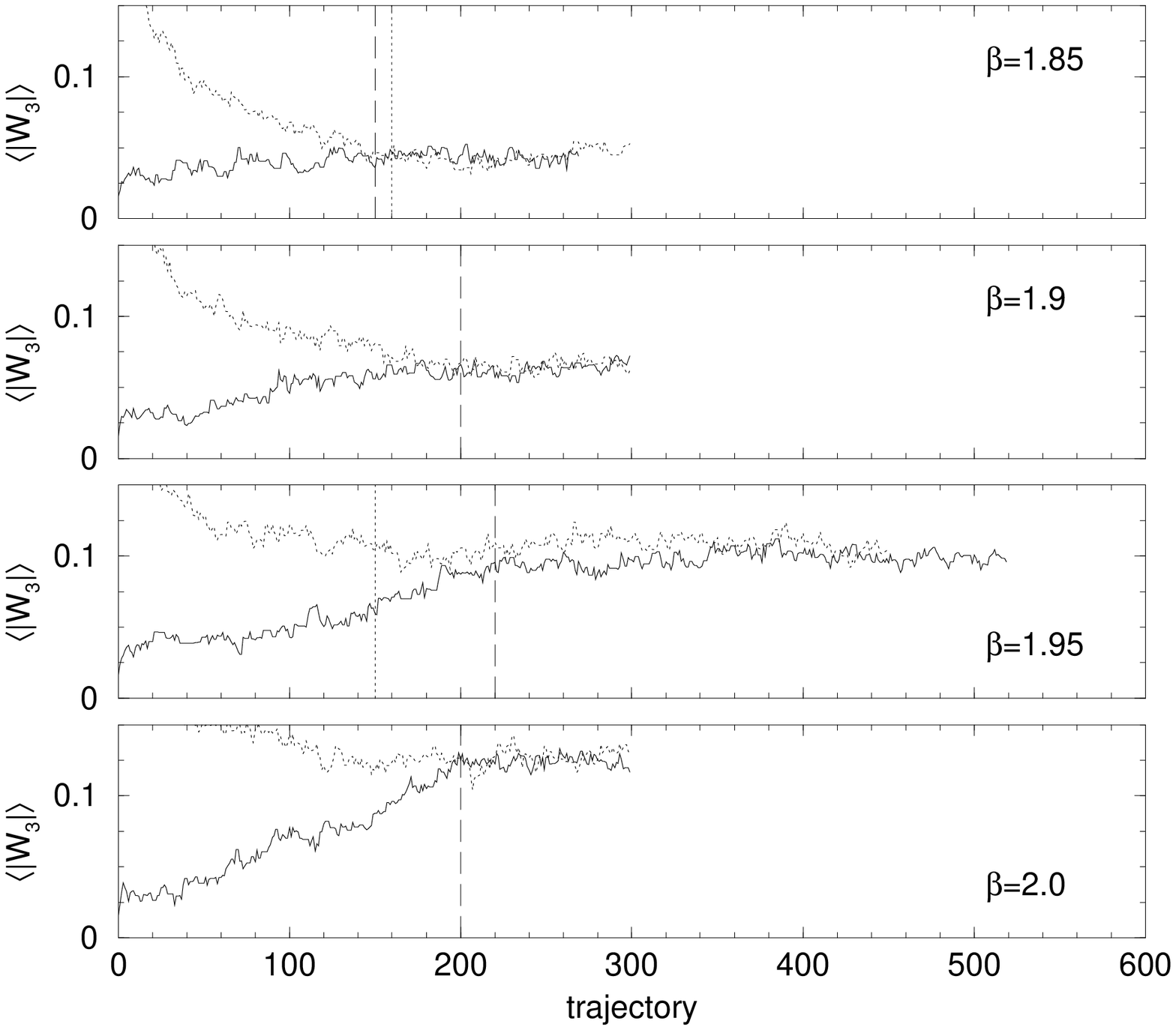}
  \caption{
    $\w3m$ evol: $16^3\times 4, c_1=-0.331, m_0=1.9, L_s=24, m_f=0.02$
  }
  \label{fig:w3m_rd_16nt4_h1.9_l24_m0.02}
\end{figure}


\clearpage
\begin{figure}
  \centering
  \includegraphics
    [height=0.9\textheight,width=\textwidth]
    {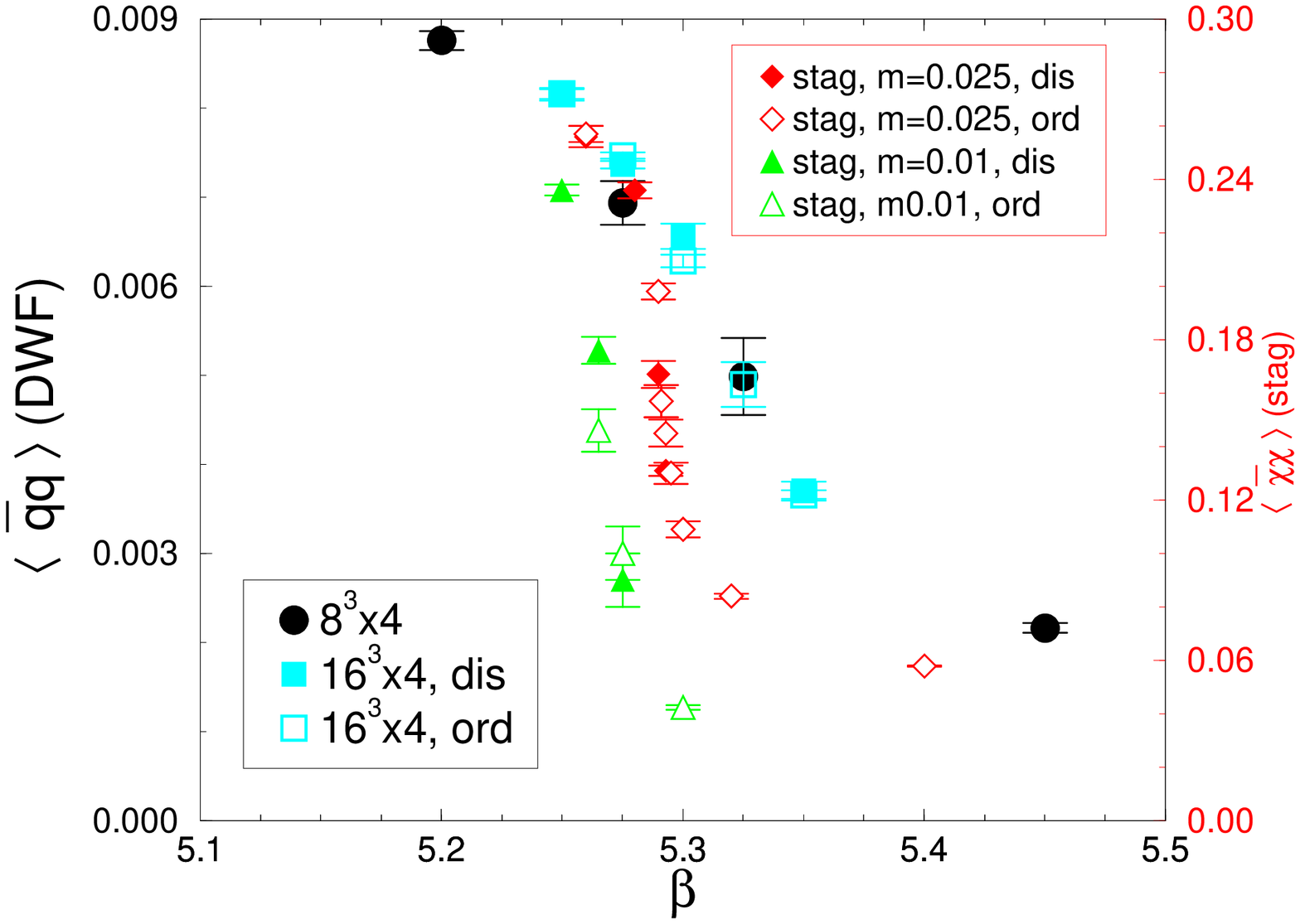}
  \caption{
    $\qbq$ {\it vs.}~$\beta$: $16^3\times 4, m_0=1.9, L_s=24, m_f=0.02$
  }
  \label{fig:pbp_wd_h1.9_l24_m0.02}
\end{figure}

\begin{figure}
  \centering
  \includegraphics
    [height=0.9\textheight,width=\textwidth]
    {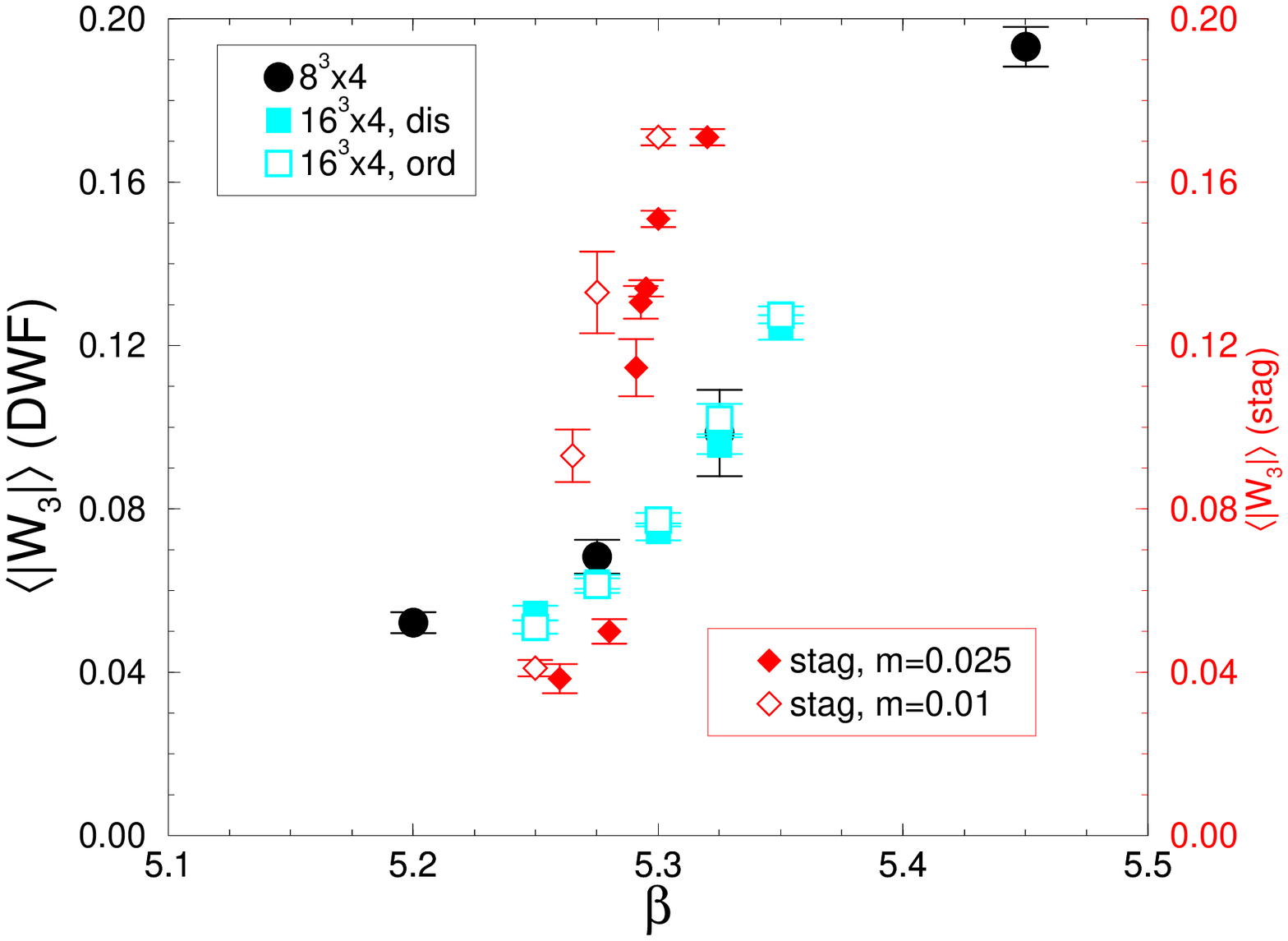}
  \caption{
    $\w3m$ {\it vs.}~$\beta$: $16^3\times 4, m_0=1.9, L_s=24, m_f=0.02$
  }
  \label{fig:w3m_wd_h1.9_l24_m0.02}
\end{figure}

\begin{figure}
  \centering
  \includegraphics
    [height=0.9\textheight,width=\textwidth]
    {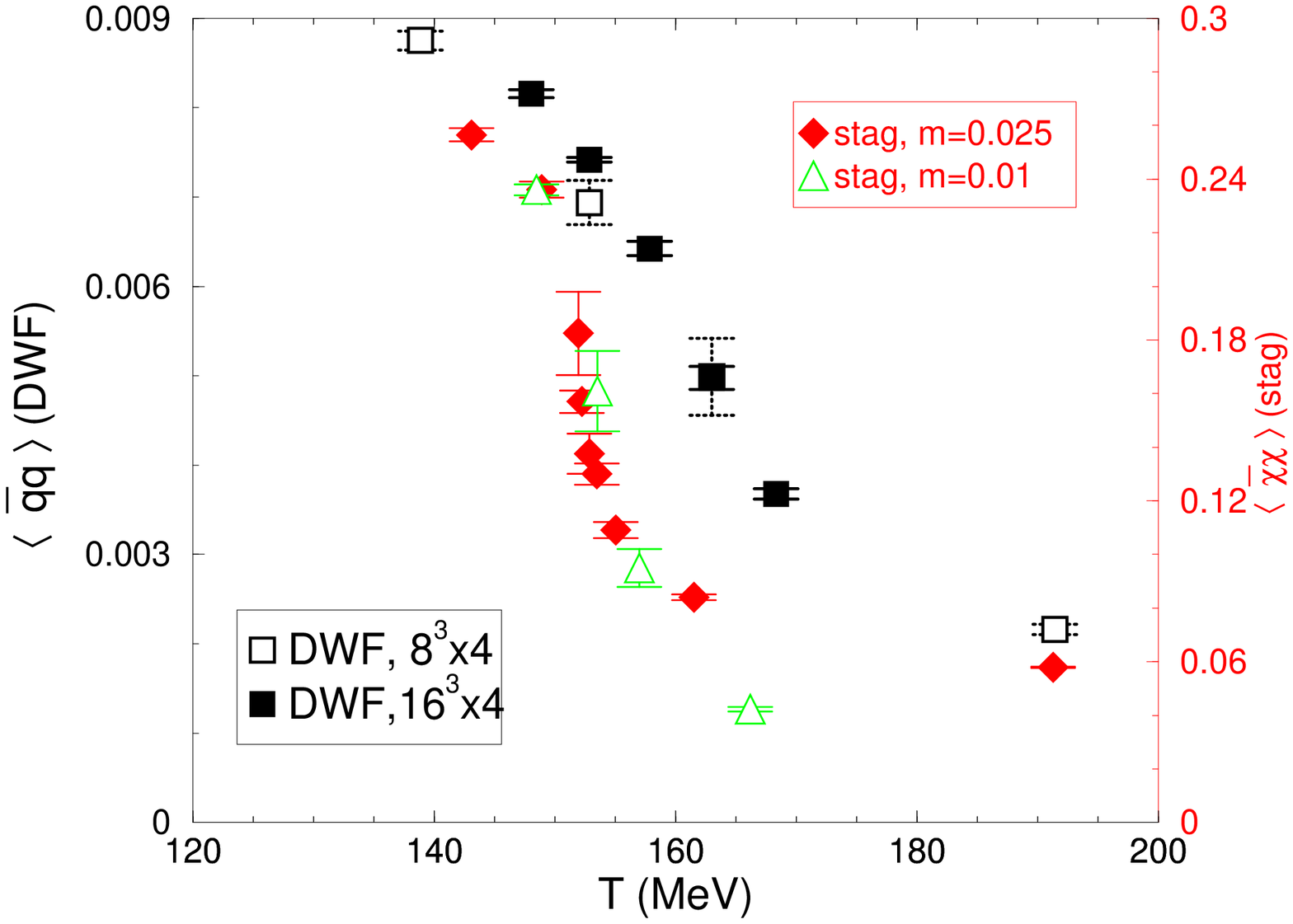}
  \caption{
    $\qbq$ {\it vs.}~$T$ (MeV): $16^3\times 4, m_0=1.9, L_s=24, m_f=0.02$
  }
  \label{fig:pbp_wd_vs_temp}
\end{figure}

\begin{figure}
  \centering
  \includegraphics
    [height=0.9\textheight,width=\textwidth]
    {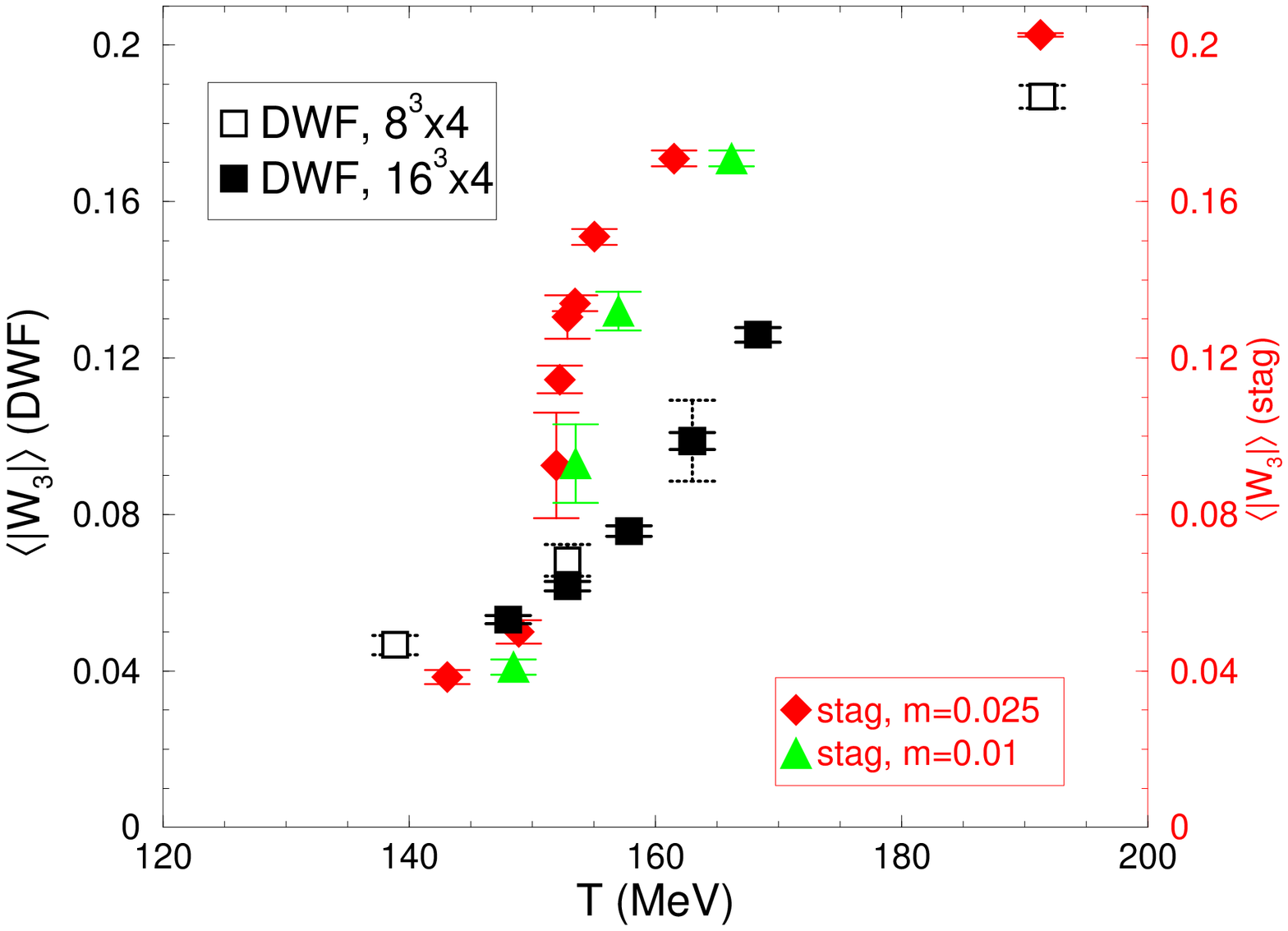}
  \caption{
    $\w3m$ {\it vs.}~$T$ (MeV): $16^3\times 4, m_0=1.9, L_s=24, m_f=0.02$
  }
  \label{fig:w3m_wd_vs_temp}
\end{figure}

\begin{figure}
  \centering
  \includegraphics
    [height=0.9\textheight,width=\textwidth]
    {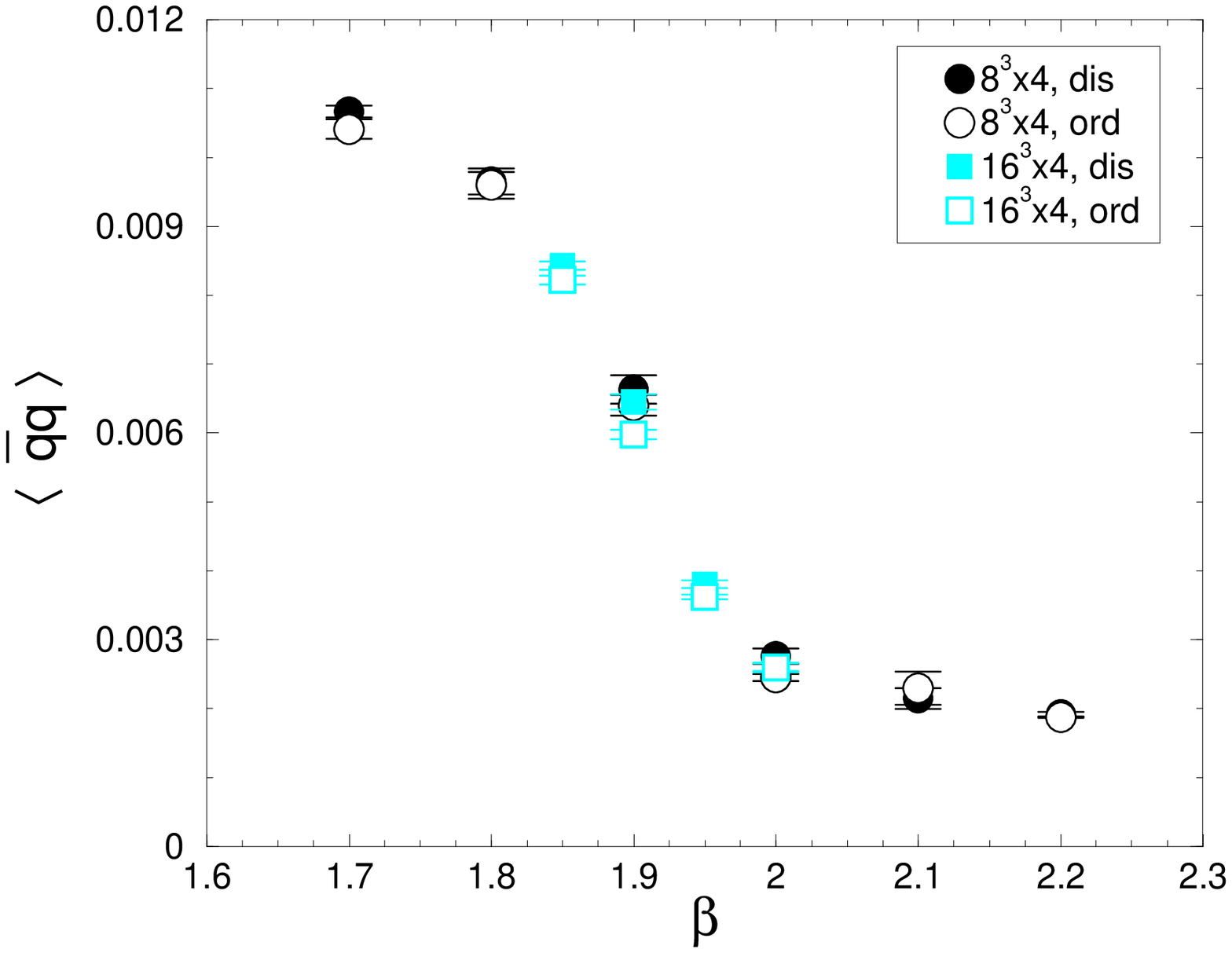}
  \caption{
    $\qbq$ {\it vs.}~$\beta$: $16^3\times 4$, $c_1$=-0.331, $m_0$=1.9,              $L_s$=24, $m_f$=0.02
  }
  \label{fig:pbp_rd_h1.9_l24_m0.02}
\end{figure}

\begin{figure}
  \centering
  \includegraphics
    [height=0.9\textheight,width=\textwidth]
    {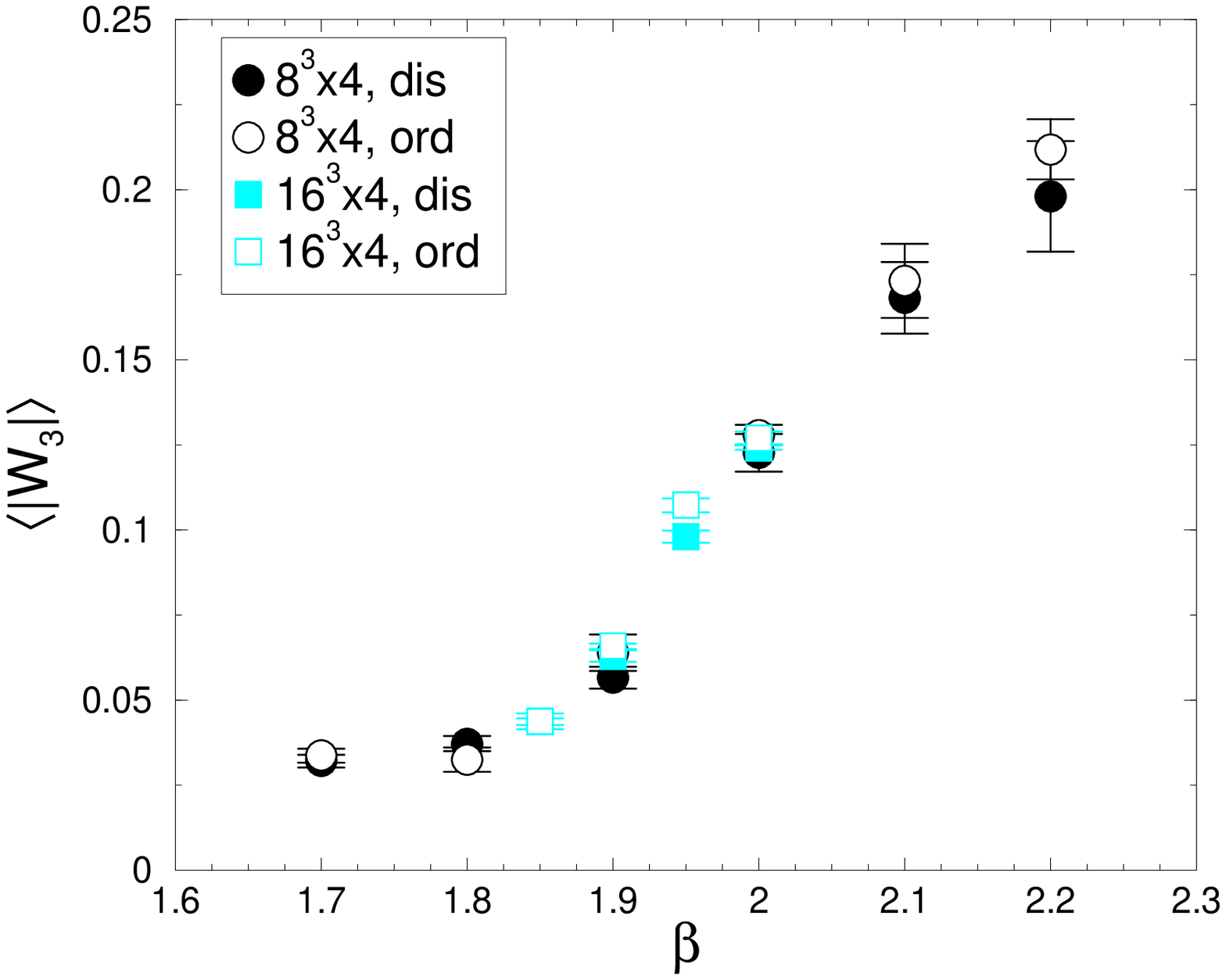}
  \caption{
    $\w3m$ {\it vs.}~$\beta$: $16^3\times 4$, $c_1$=-0.331, $m_0$=1.9,
    $L_s$=24, $m_f$=0.02
  }
  \label{fig:w3m_rd_h1.9_l24_m0.02}
\end{figure}

\begin{figure}
  \centering
  \includegraphics
    [height=0.9\textheight,width=\textwidth]
    {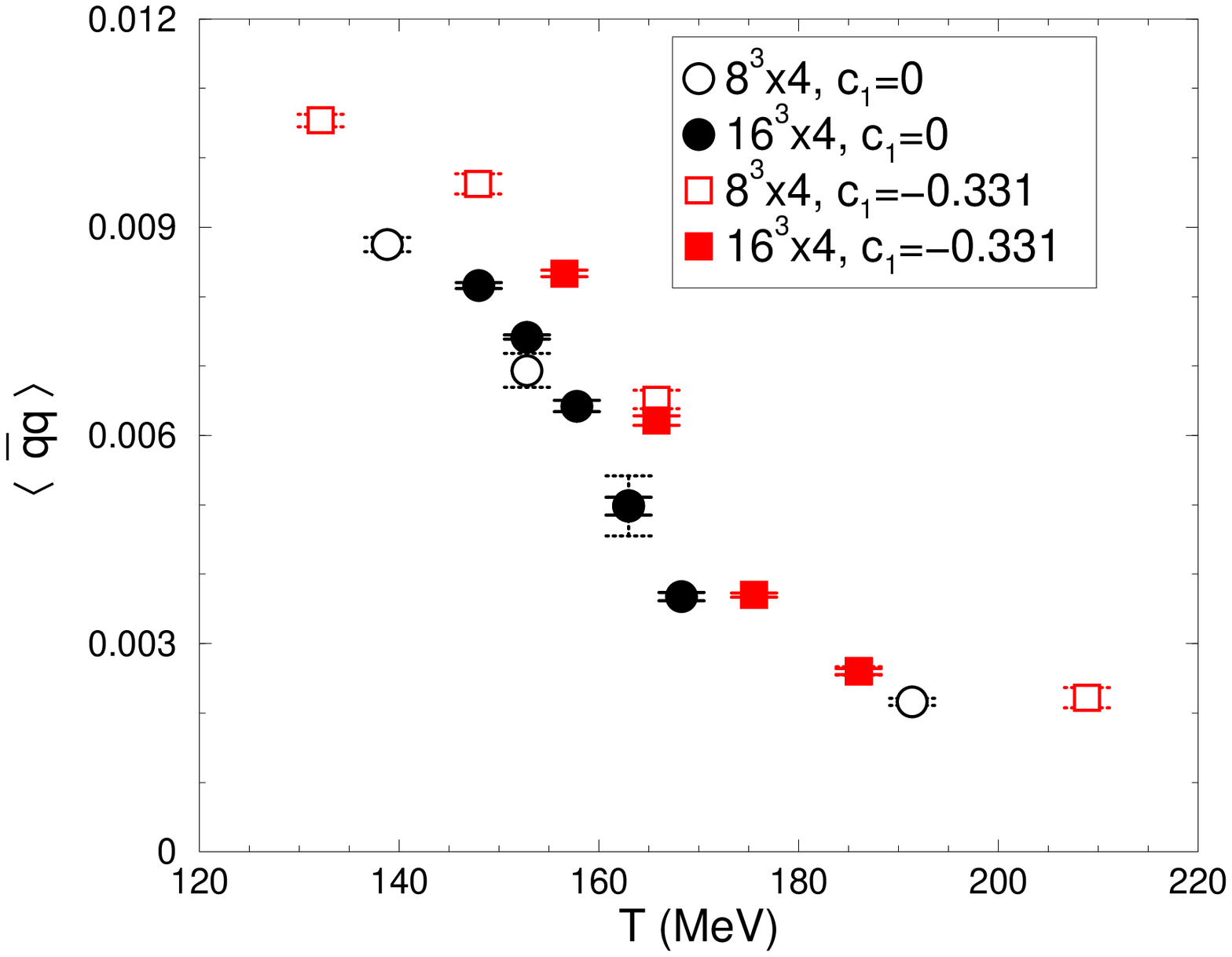}
  \caption{
    $\qbq$ {\it vs.}~$T$ (MeV): $16^3\times 4$, $c_1$=-0.331, $m_0$=1.9,
    $L_s$=24, $m_f$=0.02
  }
  \label{fig:pbp_rd_vs_temp}
\end{figure}

\begin{figure}
  \centering
  \includegraphics
    [height=0.9\textheight,width=\textwidth]
    {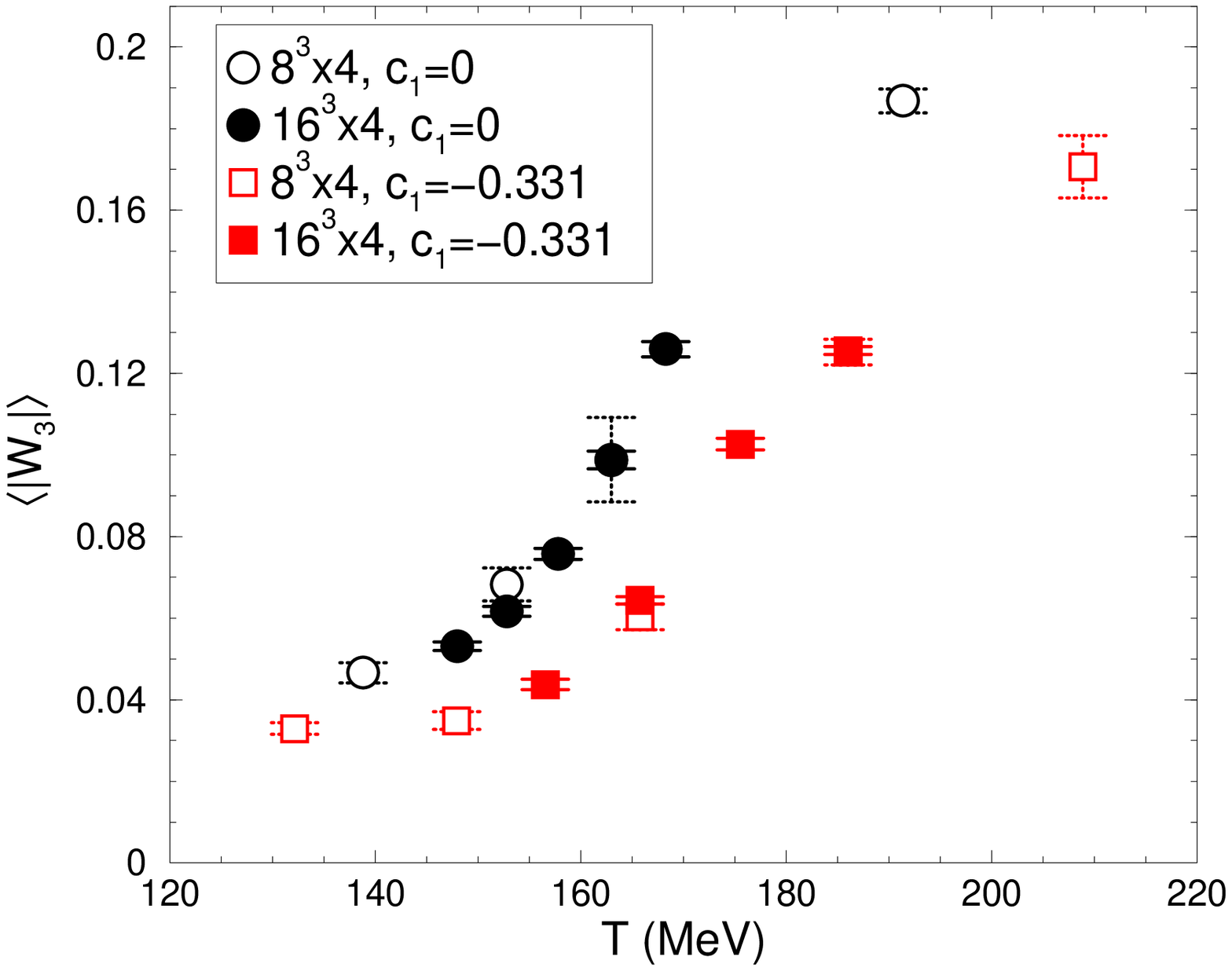}
  \caption{
    $\w3m$ {\it vs.}~$T$ (MeV): $16^3\times 4$, $c_1$=-0.331, $m_0$=1.9,            $L_s$=24, $m_f$=0.02
  }
  \label{fig:w3m_rd_vs_temp}
\end{figure}

\clearpage


\clearpage
\begin{figure}
  \centering
  \includegraphics
    [height=0.9\textheight,width=\textwidth]
    {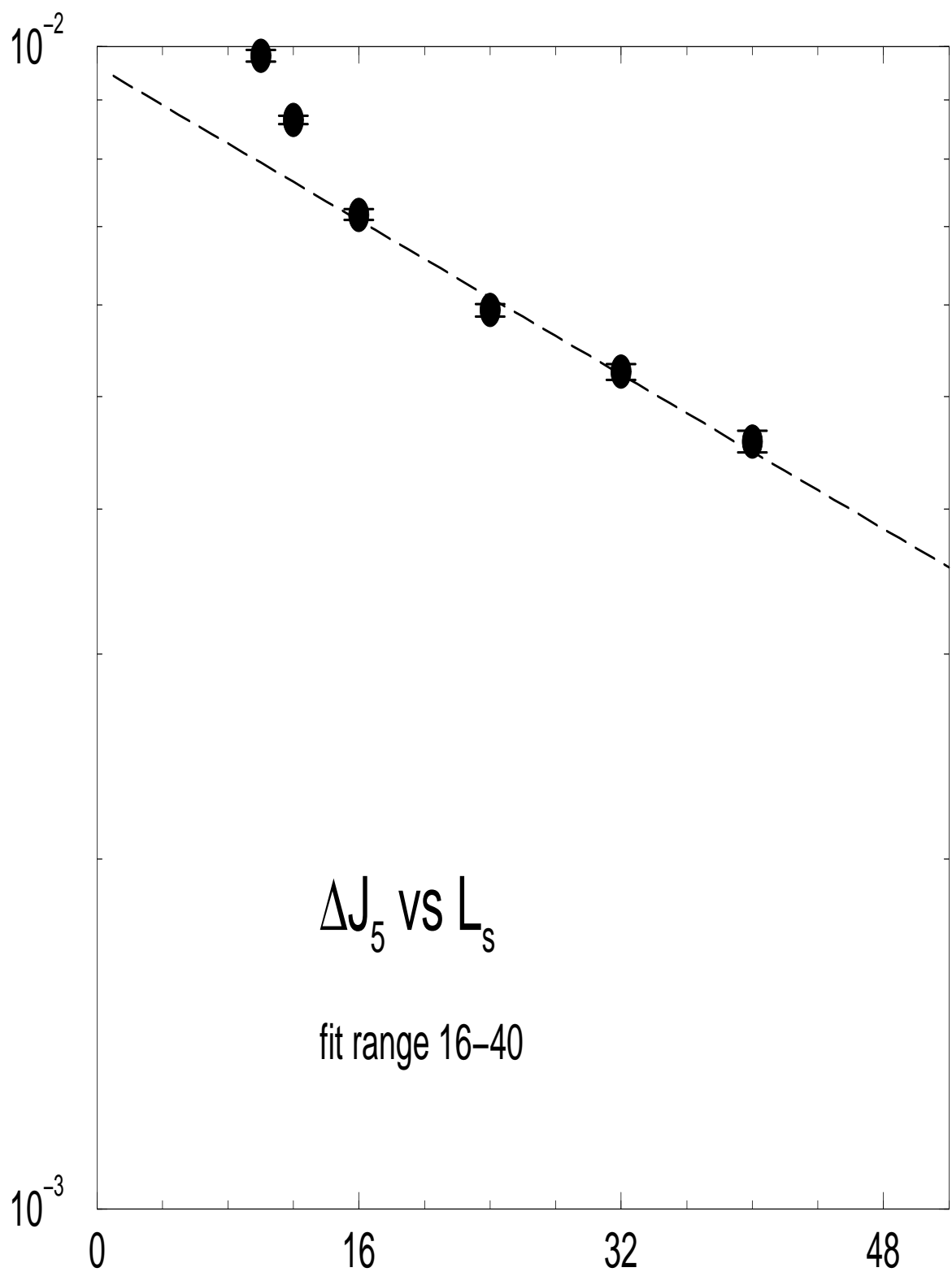}
  \caption{
    Fit of $\Delta J_5 = c_0 e^{-c_1 L_s}$;
    $8^3\times 4, \beta=5.2, m_0=1.9, m_f=0.02$
  }
  \label{fig:dj5_wd_8nt4_b5.2_h1.9_m0.02}
\end{figure}


\clearpage
\begin{figure}
  \centering
  \includegraphics
    [height=0.9\textheight,width=\textwidth]
    {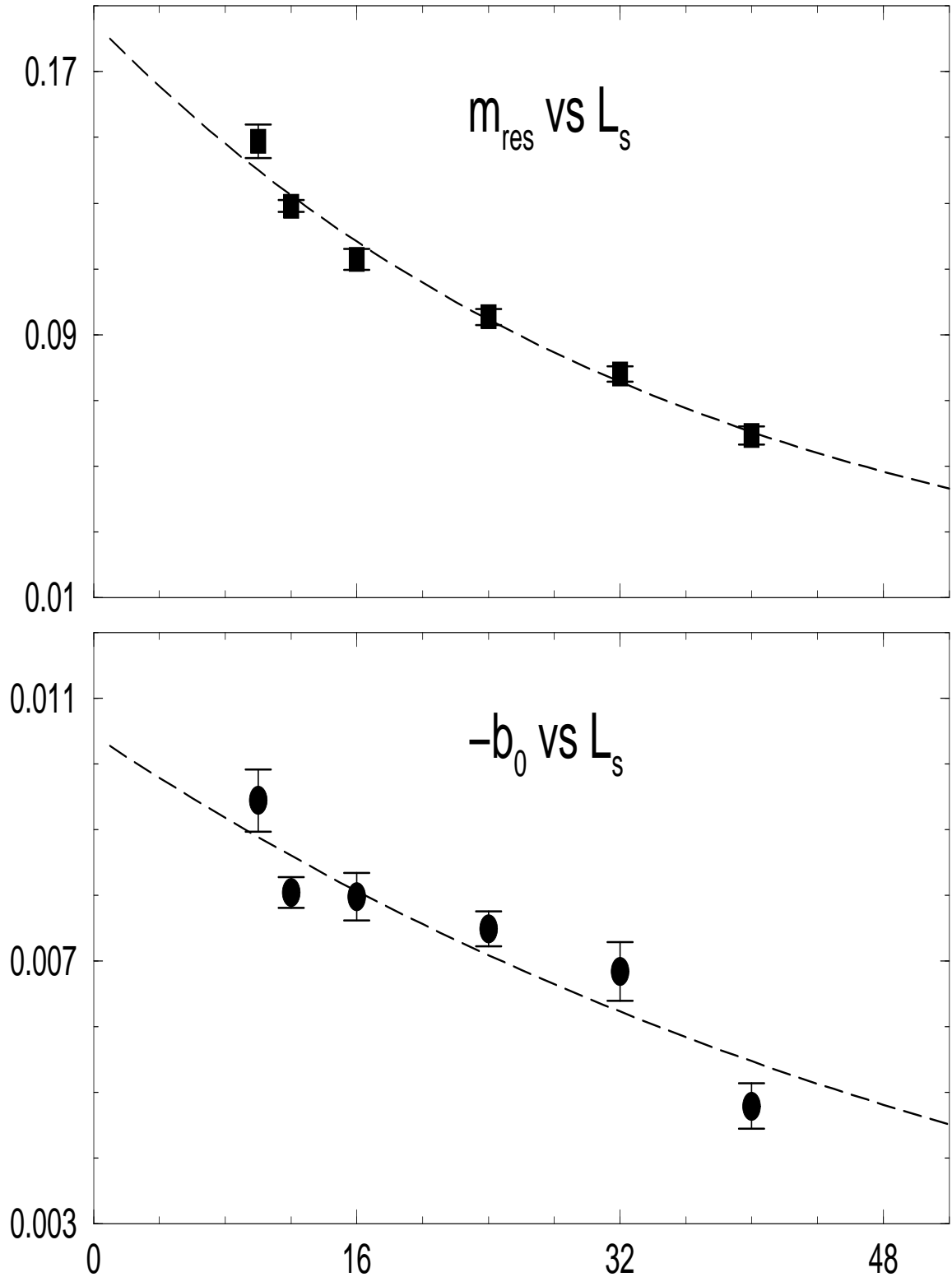}
  \caption{
    $L_s$ dependence of $m_{\rm res}$ and $b_0$:
    $8^3\times 4, \beta=5.2, m_0=1.9, m_f=0.02$
  }
  \label{fig:GMOR_wd_8nt4_b5.2_h1.9_m0.02}
\end{figure}

\end{document}